\pdfoutput=1
\documentclass{iopart}
\usepackage{iopams,setstack}
\usepackage{graphicx,bbm}
\usepackage{multirow,subfigure}
\usepackage[utf8]{inputenc}

\begin{document}

\title{Periodic boundary conditions on the pseudosphere}
	\author{F Sausset and G Tarjus}
	\address{Laboratoire de Physique Théorique de la Matière Condensée, Université Pierre et Marie Curie - Paris 6, UMR CNRS 7600, 4 place Jussieu, 75252 Paris Cedex 05, France}
	\ead{sausset@lptmc.jussieu.fr, tarjus@lptmc.jussieu.fr}
	\begin{abstract}
	We provide a framework to build periodic boundary conditions on the pseudosphere (or hyperbolic plane), the infinite two-dimensional Riemannian space of constant negative curvature. Starting from the common case of periodic boundary conditions in the Euclidean plane, we introduce all the needed mathematical notions and sketch a classification of periodic boundary conditions on the hyperbolic plane. We stress the possible applications in statistical mechanics for studying the bulk behavior of physical systems and we illustrate how to implement such periodic boundary conditions in two examples, the dynamics of particles on the pseudosphere and the study of classical spins on hyperbolic lattices.
	\end{abstract}

\section{Introduction}

The use of periodic boundary conditions has become a standard tool in statistical physics, and more particularly in computer simulations, to extract from the behavior of finite systems relevant information on the properties in the thermodynamic limit. In Molecular Dynamics and Monte Carlo simulations, the number of particles, or more generally degrees of freedom, that can be studied with present-time computers remains limited (say, of the order of $10^6$ or less) and extrapolating to the ``bulk'' behavior of a macroscopic system requires to somewhat minimize the influence of the system's boundaries. This is commonly done by employing periodic boundary conditions \cite{Hansen:1986,Frenkel:1996,Landau:2000}.

The practical implementation of such boundary conditions is well developed in Euclidean, ``flat'' spaces. However, much less has been achieved with regard to curved spaces. We do not consider here spherical-like spaces characterized by a positive Gaussian curvature, such as the surface of a sphere in three-dimensional Euclidean space, because such spaces have a finite extent; as a result, the thermodynamic limit is only attained by letting the curvature go to zero. We rather focus on hyperbolic spaces, characterized by a negative curvature, which can indeed be infinite.  Due to the peculiar character of the hyperbolic metric, the thermodynamic limit in the case of an open space may crucially depend on the nature of the boundary conditions: for a homogeneous, simply-connected hyperbolic space, the perimeter (or area in $3$ dimensions, etc) of the boundary of a disk of radius $r$ is proportional to the area of the disk (volume in $3D$, etc), when $r$ is much larger than the radius of curvature.

Negatively curved spaces show up in different areas of physics. The more prominent one is general relativity and cosmology where the curvature of spacetime is a fundamental ingredient. For the same reason, and usually motivated by quantum gravity and string theories, quantum field theories have been studied in curved spacetime \cite{Birrel:1982}, in particular spacetimes with a hyperbolic spatial section \cite{Birrel:1982,Callan:1990,Camporesi:1991,Cognola:1993,Miele:1997,Doyon:2003}. Negatively curved spaces have also recently been considered in the context of statistical physics and condensed matter theory. There, the motivations are both practical (describing the behavior of newly designed mesoscopic and nanoscopic objects with exotic shapes, curvatures, and topological properties \cite{Avron:1992,Alimohammadi:1999,Bulaev:2003,Travesset:2005,Vitelli:2006,Giomi:2007}) and theoretical (understanding how curvature influences the critical behavior and the phase transitions of classical statistical systems \cite{Callan:1990,Miele:1997,Nelson:1983,Rietman:1992,Wu:1996,Jancovici:1998,Wu:2000,Angles-dAuriac:2001,Benjamini:2001,Doyon:2004,Belo:2006,Shima:2006a,Shima:2006, Hasegawa:2006, Ueda:2007, Baek:2007}). Still in condensed-matter theory, negatively curved spaces appear in studies of the Quantum Hall effect \cite{Avron:1992,Alimohammadi:1999,Bulaev:2003} and in the framework of ``geometrical frustration'' \cite{Sadoc:1999,Nelson:2002,Tarjus:2005}. In the latter case, the hyperbolic reference space serves either for providing crystal-like templates to amorphous solids \cite{Kleman:1982} or exotic liquid crystalline structures (such as the ``infinite periodic minimal surfaces'' associated with the bicontinuous cubic phases of liquid crystals formed by amphiphilic molecules and water \cite{Sadoc:1988}), or for building toy models of supercooled liquids and metallic glasses \cite{Nelson:1983,Nelson:2002,Rubinstein:1983}. Finally, the free motion of mass points on hyperbolic surfaces has been used as a prototype to study classical and quantum chaos \cite{Balazs:1986}, and properties of random walks have also been investigated \cite{Comtet:1996,Nechaev:2002}.

In the above listed studies, need for periodic boundary conditions may arise for different reasons. (i) First, and as mentioned above, they may be useful to simulate, as closely as possible, the bulk behavior of a macroscopic sample through the study of finite-size models. (ii) Secondly, periodic boundary conditions, as will be further detailed in the following, are intimately connected with \textit{tessellations} of space, \textit{i.e.}, tilings of space by identical replicas of a unit cell; as a consequence, their study has a direct bearing on the characterization of crystalline structures, and thus, in the present case, on ``hyperbolic crystallography''. (iii) From a more topological point of view, periodic boundary conditions are also related to the generation of \textit{compact manifolds with nontrivial topology}. Such multi-connected hyperbolic manifolds are for instance relevant to ``cosmic topology", which investigates the possibility that the universe is not simply connected \cite{Lachieze-Rey:1995,Levin:2002}, and to studies on classical and quantum chaos \cite{Balazs:1986}.

In the following, we restrict ourselves to the consideration of periodic boundary conditions on the ``hyperbolic plane" or ``pseudosphere" (also called ``Bolyai-Lobachevsky plane"), which is the infinite, simply-connected two-dimensional space with constant negative curvature \cite{Coxeter:1961,Hilbert:1983}. Our goal is to provide a framework for implementing periodic boundary conditions with \textit{chosen} properties. Such properties will generally follow from physical requirements and may be specific to the problem under study.

The rest of the article is organized as follows. In section II we revisit the case of periodic boundary conditions in the Euclidean plane in order to introduce the two-fold approach, geometrical and topological, to such conditions. We provide in the next section the basic mathematical formalism for studying periodic boundary conditions in the hyperbolic plane. In section III, we address the classification of the possible periodic boundary conditions in the hyperbolic plane and give a guideline for building the proper conditions adapted to a given physical problem. Finally, in section IV, we illustrate how to implement the proposed formalism by considering two examples of statistical mechanical models: the dynamics of particles on the pseudosphere and spin models on hyperbolic lattices. We complement the presentation by appendices in which we summarize basic elements on hyperbolic geometry (\ref{ap:geo}), on isometries and Fuchsian groups (\ref{ap:isometries}), and on hyperbolic trigonometry and tessellations (\ref{ap:tilings}).

\section{The Euclidean case revisited}

We first consider the well-known case of the Euclidean plane $E^2$. Implementing periodic boundary conditions in $E^2$ consists in choosing a primitive cell (which contains the physical system) such that it can be infinitely replicated to tile the whole plane with no overlaps between copies of the cell and no voids. To ensure smoothness and consistency, the edges of the primitive cell have to be paired in a specific way: leaving the cell through one edge implies to come back by another edge, which should be equivalent to exploring the whole tiling of the plane.

\begin{figure}
	\subfigure[]{\label{fig:rep-car} \includegraphics[scale=0.4]{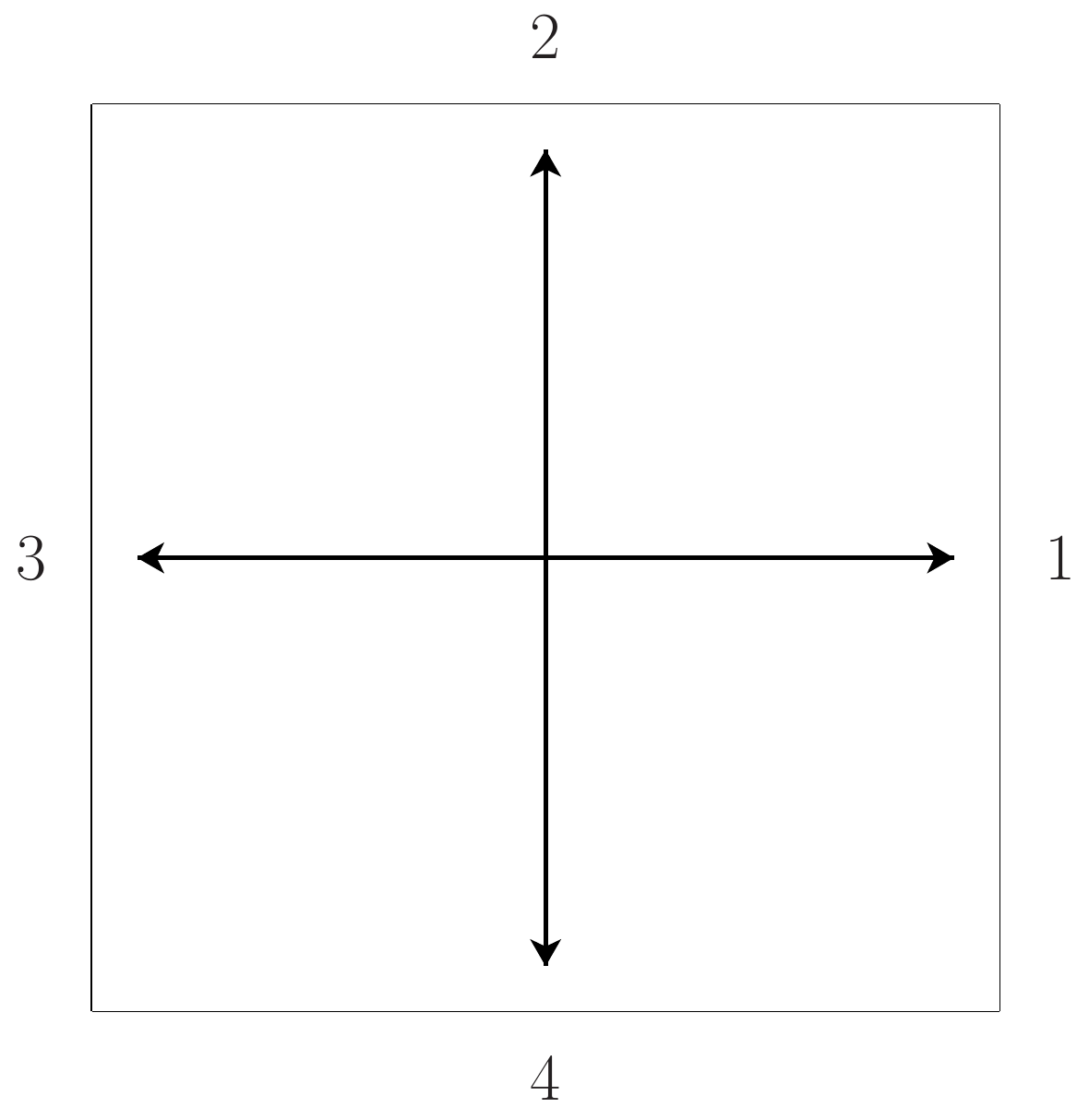}}
	\subfigure[]{\label{fig:tor-car} \includegraphics[scale=0.4]{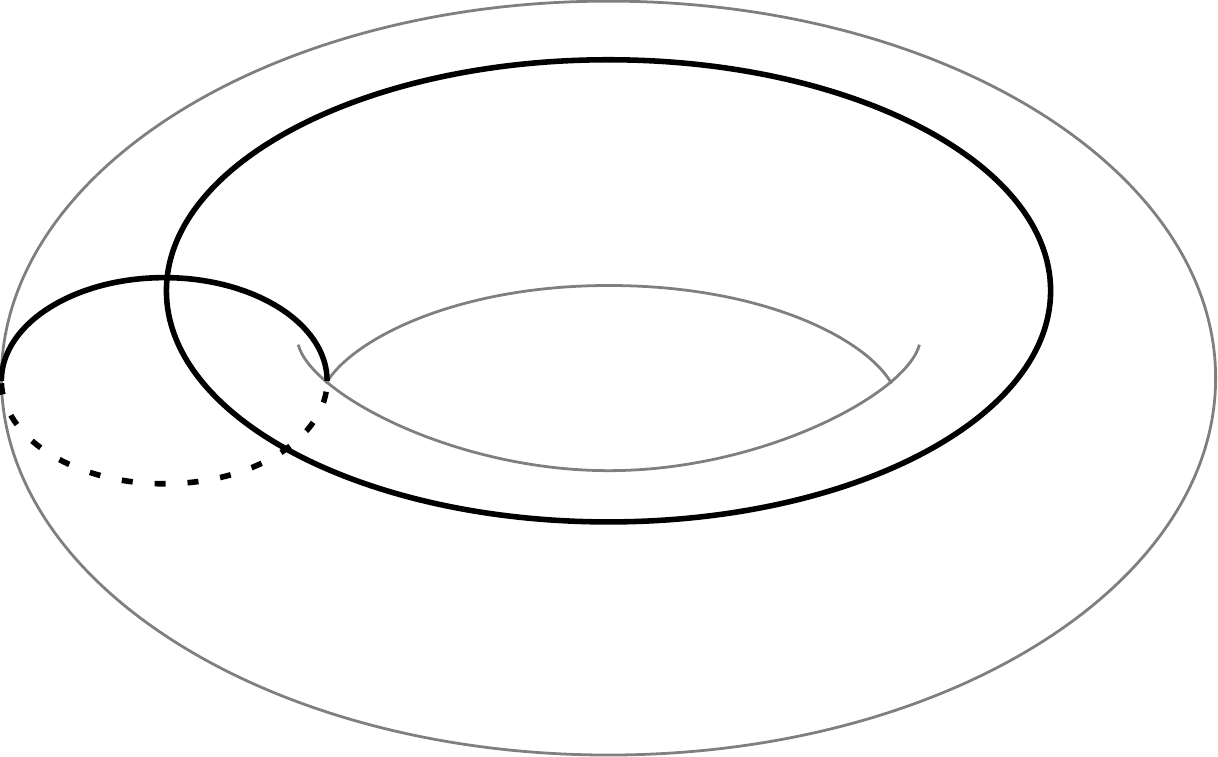}}
	\caption{\label{fig:car} Square periodic boundary conditions in the Euclidean plane. (a) Square with pairing of the facing edges. (b) One-hole torus obtained by gluing paired edges of the square. The embedded graph (black and dotted lines) corresponds to the paired edges.}
\end{figure}

The simplest periodic boundary conditions in $E^2$ are the ``square" ones: taking a square as the primitive cell, one can obviously replicate it to form a square lattice, which corresponds to a particular tiling of the plane, and one can pair facing sides of this primitive square (see \fref{fig:rep-car}).
This operation can be framed in a more mathematical setting {\cite{Nakahara:1990}}. Indeed, in the language of group theory, periodic boundary conditions are just an equivalence relation. In the present case, choosing any two points in $E^2$, \textit{i.e.}, any two pairs $(x_1, y_1)$ and $(x_2, y_2)$ in $\mathbb{R}^2$, one can introduce the following equivalence relation:
\begin{equation}
	\label{eq:rel-eq}
	(x_1, y_1) \sim (x_2, y_2) \qquad \mathrm{if} \;
	\cases{x_2 = x_1 + a \, n_{\mathrm{x}}\\
	y_2 = y_1 + a \, n_{\mathrm{y}}},
\end{equation}
with $n_{\mathrm{x}}, n_{\mathrm{y}} \in \mathbb{Z}$ and $a$ the side length of the square. The quotient space $\mathbb{R}^2 / \sim$ associated with this equivalence relation (in geometric context, ``orbit space" is sometimes used in place of  ``quotient space") is then the one-hole torus, noted $T^2$. As can be seen in \fref{fig:tor-car}, $T^2$ can be represented by the primitive cell with \textit{identified} edges (which is then called in two dimensions the \textit{fundamental polygon}): by ``gluing" paired edges together, one can indeed obtain a torus which is topologically equivalent to the one-hole torus $T^2$. Note that the torus shown in \fref{fig:tor-car} is represented as the projection of an object that seems to be physically embedded in the $3D$ Euclidean space $E^3$ and appears, with the natural Euclidean metric, curved like a doughnut; this is not the case of the ``flat" two-dimensional torus $T^2$, whose actual visualization is then more elusive.

The way to pair the edges is crucial. Other pairings of the edges of the primitive square are in fact possible. They lead to different $2D$ manifolds, which can be considered as more ''exotic``  than the torus, being nonorientable and intersecting with themselves in $E^3$, such as the Klein bottle and the projective plane (this latter, in addition, having a positive curvature)\cite{Hilbert:1983,Nakahara:1990}. Actually, the way to pair edges reflects the equivalence relation underlying the periodic boundary conditions.

As in the simple example above, we will use in this article, depending on the context, the two views concerning periodic boundary conditions: 
\begin{itemize}
	\item[(1)] the ``geometrical'' one, based on the tiling of the infinite and simply connected space $X$ endowed with its metric ($E^2$ in the above example) by identical (congruent) replicas of the fundamental cell; such a tiling or tessellation involves a discrete subgroup $\Gamma$ of isometries of the space $X$ (\textit{i.e.}, displacements preserving the metric), whose elements transform the replicas of the cell one into another.
	\item[(2)] The ``topological'' one, using multi-connected manifolds (here, surfaces) to represent the quotient space, $X / \Gamma$. The link between $\Gamma$ and the equivalence relation $\sim$ defined in equation \eref{eq:rel-eq} is that two points $x, y$ of $X$ are equivalent if and only if there exists an element $\gamma$ of the group $\Gamma$ such that $\gamma(x) = y$. For example, in the case of equation \ref{eq:rel-eq}, $\Gamma$ is generated by the two translations $T_{a \vec{x}}$ and $T_{a \vec{y}}$.
\end{itemize}
These two points of view are fully compatible and closely connected. A ``visual'' link can be made between them by gluing paired edges of the fundamental polygon (here, the square with identified edges) to obtain a manifold representing the quotient space (here, the one-hole torus), with the glued edges forming a \textit{graph embedded in the manifold} as shown in \fref{fig:tor-car}. As we shall see below, this latter aspect is essential for more general geometries.

\begin{figure}
	\subfigure[]{\label{fig:rep-hex} \includegraphics[scale=0.4]{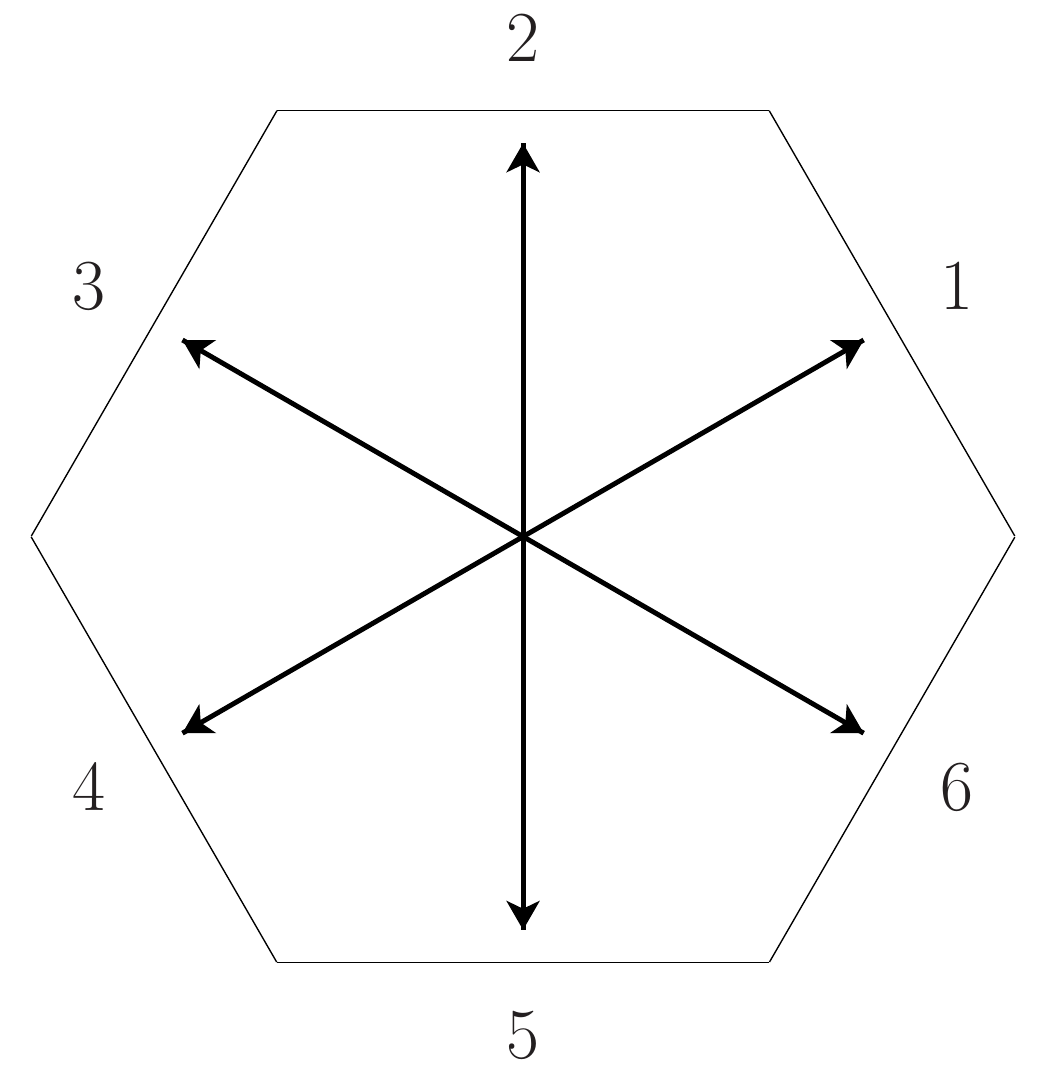}}
	\subfigure[]{\label{fig:tor-hex} \includegraphics[scale=0.4]{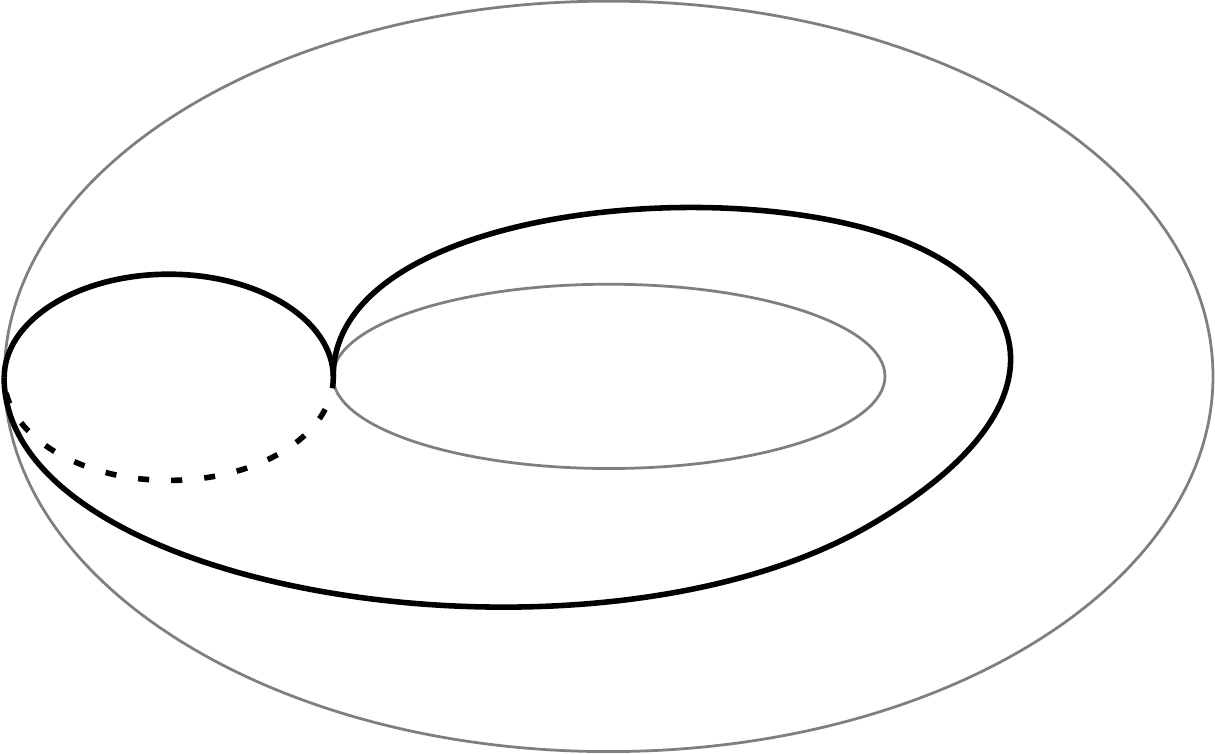}}
	\caption{Hexagonal periodic boundary conditions in the Euclidean plane. (a) Hexagon with pairing of the facing edges. (b) One-hole torus obtained by gluing paired edges of the hexagon. The embedded graph (black and dotted lines) corresponds to the paired edges. Note the difference with \fref{fig:tor-car}.}
\end{figure}

In the $2D$ flat space, alternatives exist to build periodic boundary conditions. One can consider a hexagonal cell, also with paired opposite edges (see \fref{fig:rep-hex}). In this case, the underlying equivalence relation for two pairs $(x_1, y_1)$ and $(x_2, y_2)$ in $\mathbb{R}^2$ is:
\begin{equation}
	\label{eq:rel-eq-hex}
	(x_1, y_1) \sim (x_2, y_2) \qquad \mathrm{if} \;
	\cases{x_2 = x_1 + \frac{3 \, a}{2} \, n_{\mathrm{x}}\\
	y_2 = y_1 + \frac{\sqrt{3} \, a}{2} \, n_{\mathrm{y}}},
\end{equation}
where $n_{\mathrm{x}}, n_{\mathrm{y}}$ and $a$ are defined as below \eref{eq:rel-eq}.
The associated quotient space is also the one-hole torus $T^2$, but the graph formed by glued edges of the hexagon is embedded in $T^2$ in a totally different way than that used before, as illustrated in \fref{fig:tor-hex}. As in the case of the square, other pairings of the edges are possible. However, the resulting manifolds associated with the quotient space are again exotic, nonorientable surfaces.

In what follows, we generalize the present considerations to the hyperbolic plane. As they are simpler and usually more physical, we only consider periodic boundary conditions that can be built from a primitive cell which is a regular polygon with a finite area (due to the pairing requirement, the number of edges must in addition be even). The associated regular tessellation of the plane is generically noted $\{p,q\}$, where $p$ is the (even) number of polygon edges and $q$ the number of polygons that meet at a vertex. (We shall discuss the constraints on $p$ and $q$ below.) We also avoid ``twisted'' boundary conditions and therefore restrict  pairings of the sides of the primitive cell to schemes that lead to quotient spaces representable as orientable surfaces. With the above restrictions, the surface is then closed and of finite extent with respect to the metric under consideration, namely, \textit{compact}.

\section{Generalization to the hyperbolic plane: fundamentals}

\subsection{Tessellations, quotient spaces, and fundamental polygons}

Going back for a moment to the case of the Euclidean plane, we recall that well known geometric constraints limit the possible regular tilings $\{p,q\}$ to those verifying $(p - 2)(q - 2) = 4$. This includes the tilings we have already considered, $\{4,4\}$ (tiling by squares) and $\{6,3\}$ (tiling by hexagons), plus a third one, the tiling $\{3,6\}$ by triangles which is dual to the $\{6,3\}$. This $\{3,6\}$ tiling however involves a regular polygon with an odd number of sides, which cannot be paired to generate periodic boundary conditions. (It is nonetheless possible to build periodic boundary conditions by using the $\{3,6\}$ tiling, but in that case the fundamental polygon is made of two triangles joined to form a parallelogram with paired opposite sides; but as we have stressed above, we only consider in this article cases where there exists a regular fundamental polygon.)

In the hyperbolic plane, which we shall denote $H^2$ irrespective of the model used for its representation (see below and \ref{ap:geo}), regular tilings exist provided
\begin{equation}
	\label{eq:tiling}
	(p - 2) (q - 2) > 4,
\end{equation}
which now leads to an infinite number of possibilities \cite{Coxeter:1961,Hilbert:1983,Coxeter:1965}. The richness of the tessellations in hyperbolic space, compared to that in Euclidean space, can be traced back to the existence of an intrinsic lengthscale, the radius of curvature $\kappa^{-1}$, where $- \kappa^2$ is the Gaussian curvature of $H^2$ \cite{Balazs:1986}.

Another property brought about by hyperbolic geometry is that quotient spaces associated with periodic boundary conditions have a greater variety than in Euclidean space. In the latter, we have seen that $E^2 / \Gamma$ is a one-hole torus, which is a closed orientable surface of genus $1$. (The genus of a connected, orientable surface is the number of holes, or handles, on it: for instance, a sphere and a disk have genus zero, whereas a one-hole torus has genus one.) Using the Gauss-Bonnet theorem which links the area $A$ of a surface to its genus $g$, one finds, for any surface embedded in $H^2$ with negative curvature $- \kappa^2$,
\begin{equation}
	\label{eq:gauss-bonnet}
	A = 4 \pi \kappa^{-2}(g - 1),
\end{equation}
which implies that any such surface has a genus $g \geqslant 2$. As a result a (finite-sized) fundamental polygon in $H^{2}$ must have a genus $g \geqslant 2$ and so does the associated (compact) quotient space. As we shall see, there is no other restriction on the value of the genus in this case.

To proceed further in our investigation of periodic boundary conditions in the hyperbolic plane, we focus on the way to obtain fundamental polygons. These polygons are directly connected to the discrete subgroups $\Gamma$ of the group of isometries of $H^2$ (see \footnote{Two points are images one of each other by a transformation of a subgroup $\Gamma$ if and only if they satisfy the equivalence relation associated with the quotient space $H^2 / \Gamma$.} and \ref{ap:isometries}). There are a number of restrictions on $\Gamma$. First, being interested in quotient spaces $H^2 / \Gamma$ that are orientable manifolds, we only consider groups containing orientation-preserving transformations. Such groups  $\Gamma$ are called Fuchsian groups \cite{Poincare:1882,Beardon:1983} (see \ref{ap:isometries}). Fuchsian groups play for the hyperbolic geometry a role similar to that of crystallographic groups for the Euclidean geometry. Note however that a Fuchsian group is non-Abelian. (Discrete subgroups containing elements which do not preserve orientation also exist and are generically called non-Euclidean crystallographic groups.) Actually, we will be interested only by ``purely hyperbolic'' Fuchsian groups which only contain transformations acting on the hyperbolic plane without fixed points and which do not lead to infinite polygons, \textit{i.e.}, generalizations of the Euclidean translations used in standard crystallography \cite{Poincare:1882,Beardon:1983}.

To each Fuchsian group $\Gamma$ is associated one or more fundamental polygons. In general indeed, the fundamental polygon is not unique. Among the possible ones, two types are worth noticing: the \textit{standard} or \textit{canonical} fundamental polygon and the \textit{metric} fundamental polygon. The former is defined by a side pairing which is given by a specific sequence of the generators of the associated Fuchsian group \footnote{The specific sequence of generators can be written $\gamma_1 \, \gamma_2 \, \gamma_1^{-1} \, \gamma_2^{-1} \ldots \gamma_{2g-1} \, \gamma_{2g} \, \gamma_{2g-1}^{-1} \, \gamma_{2g}^{-1} = \mathbbm{1}$, where $g$ is the genus and the $\gamma_i$'s and their inverses are the generators of the Fuchsian group.}. The latter is defined as the Voronoi cell (in physicists' language) or Dirichlet polygon (in mathematicians' language) built around a suitable base point and endowed with a proper side pairing; more precisely, the Voronoi construction is applied around the base point by considering all the images of the point obtained by applying the generators of the relevant Fuchsian group. It has been proven that for each $\Gamma$, there exists a unique standard polygon \cite{Klein:1890,Keen:1966}, but a metric one may not necessarily exist. (On the other hand, once the metric fundamental polygon is known, one can construct the associated standard polygon \cite{Poincare:1882}.) Note also that these two polygons usually have a different number of sides.

The difference between the two types of fundamental polygons can be illustrated by taking again the example of the Euclidean plane. In the case of the $\{6,3\}$ tiling, the metric fundamental polygon is a hexagon (the corresponding Voronoi polygon is built around the center of the hexagon and not, as often in crystallography, around one of the vertices), whereas the standard fundamental polygon is a parallelogram. In  the latter case, the pairing of the sides and the associated graph are similar to those of the square periodic boundary conditions shown in \fref{fig:car}. The equivalence between the two fundamental polygons can be seen in terms of homotopy, as curves formed by paired edges of the hexagon (see \fref{fig:tor-hex}) and embedded in the torus are homotopic to the ones formed by the parallelogram (see \fref{fig:tor-car}). The two nodes of the graph of \fref{fig:tor-hex} can indeed be assembled in one node to give the graph of \fref{fig:tor-car}.
If one now considers the $\{3,6\}$ tiling, dual to the $\{6,3\}$, the metric fundamental polygon does not exist, as the unit cell is a triangle and has an odd number of edges. However, as seen above, the standard fundamental polygon exists and is a parallelogram.
Finally, for the $\{4,4\}$ tiling, the metric and the standard polygon are identical and both realized by a square with paired sides.

The convenience of studying metric polygons is that the link between the polygon and the associated tiling is clear from a geometrical point of view, whereas standard fundamental polygons are not trivially related to their corresponding tiling. Moreover, for a regular tiling, the associated Voronoi polygon is always regular, whereas standard polygons are regular only in some specific cases. In any case, as we are interested by periodic boundary conditions built from a primitive cell which is a regular polygon, we only consider metric fundamental polygons in the following.

\subsection{Metric fundamental polygon and associated graph}

Like any fundamental polygon, a metric one is the unit cell of a tessellation of $H^2$ by the action of the associated Fuchsian group $\Gamma$. This unit cell is usually called ``fundamental domain'', but we shall not use this terminology to avoid confusion with the concept of ``fundamental polygon" which, contrary to that of ``fundamental domain'', implies a specific pairing of the edges. Being a Dirichlet domain, the metric fundamental polygon is convex, \textit{i.e}, the geodesic joining any two points of the polygon is in the interior of the polygon and the edges are geodesic arcs. The edges are paired together, which means that each edge of the polygon can be transformed in the associated one by the action of some element $\gamma$ of the group $\Gamma$. As a result, the number of edges must be even. These transformations $\gamma$ actually form a set of generators of the group $\Gamma$ \cite{Coxeter:1965,Poincare:1882,Beardon:1983}. Since we restrict ourselves to orientation-preserving groups that act without fixed points, the generators are (hyperbolic) translations. If $2N$ is the number of edges of the metric fundamental polygon, there are $2N$ translations $\gamma_{i}$ by which edge $i$ is transformed into its associated counterpart. (Note that the labeling of the sides can be taken either clockwise or counterclockwise.) These generators must then satisfy the constraint
\begin{equation*}
\gamma_1\gamma_2\dots \gamma_i\dots \gamma_{2N}=\mathbbm{1}.
\end{equation*}
Note that for each $\gamma_i$, its inverse $\gamma_i^{-1}$ is also present in the $2N$ generators. In the hyperbolic plane, these generators do not commute, and their associated Fuchsian group is non-Abelian. Concrete examples will be given below.

In an attempt to classify the possible periodic boundary conditions, one may first consider the properties of the primitive cell (which we take as a regular polygon) in which the physical system will be placed: more specifically, its surface area and the number of its sides. From the Gauss-Bonnet theorem, see equation \eref{eq:gauss-bonnet}, the area is fixed by the curvature of the embedding hyperbolic space $H^2$ and by the genus $g$ of the quotient space $H^2/\Gamma$ associated with the fundamental polygon. Notice that, the genus being an integer, the accessible values of the fundamental polygon areas form a discrete set. To make progress in building a classification, it is now convenient to focus on the graph formed by the identifying (or in more pictorial terms, gluing together) the edges of the fundamental polygon. This graph, which is embedded in the quotient space, contains all the information about the fundamental polygon, and a closed walk on the graph is equivalent to going around the boundary of the corresponding fundamental polygon. The constraints on such a closed walk will be detailed in the next subsection.

The graphs can be first classified by their ``genus". The genus of a graph is defined  as the minimum integer $g$ such that there exists a closed orientable surface of genus equal to $g$ in which the graph can be embedded without crossing itself. For instance, the graphs of figures \ref{fig:tor-car} and \ref{fig:tor-hex} have genus $1$ as they live on the one-hole torus without crossing themselves; on the other hand, a standard planar graph has genus $0$ as it can be embedded in the sphere without crossing itself. The genus of a graph and that of the associated fundamental polygon are equal by construction.

The graphs are also characterized by the number of their vertices $v$, the number of their edges $e$, and that of their faces $f$. Euler's formula relates $v$, $e$, $f$ and the genus $g$ of the graph:
\begin{equation}
	\label{eq:euler}
	v-e+f=-2 \, (g-1).
\end{equation}
Faces of a graph are regions bounded by edges. In the present case, there is one and only one face, corresponding to the interior of the polygon (which is a simply-connected surface). The quotient space is indeed entirely covered by the surface of the polygon whose edges form the graph. Stated otherwise, the compact manifold formed by the quotient space leads back to the fundamental polygon when cut along the edges of the graph.

Inserting this property, $f=1$, in the above equation gives the following relation between $e$ and $v$:
\begin{equation}
	\label{eq:edges}
	e=v+2\,g-1.
\end{equation}
The smallest number of vertices in the graph is $v=1$, which implies that the smallest number of edges is $e=2g$. 

To further investigate the constraints set on the characteristics of an acceptable graph, it is useful to relate the latter to those of the associated metric fundamental polygon. The number of sides of the polygon is $2N$ (see above) and the corresponding tiling of $H^2$ is $\{p,q\}$ with $p=2N$. Due to the pairing of polygon edges, each edge of the graph corresponds to two edges of the polygon, \textit{i.e.}, $p=2e$ or equivalently $N=e$. In addition, each vertex of the graph corresponds to $q$ vertices of the polygon. As the fundamental polygon has as many edges as vertices, $p=qv$. As a result, and using the fact that the coordinence $q$ of the vertices in any tiling is larger than $3$ (or equal), one finds that $2e\geqslant  3v$, which implies that $e\leqslant  3\,(2g-1)$. The only acceptable graphs must therefore satisfy, on top of $f=1$,
\begin{eqnarray*}
	2g & \leqslant e & \leqslant 3\,(2g-1) \\
	\ 1 & \leqslant v & \leqslant 2\,(2g-1)
\end{eqnarray*}
with $e$ and $v$ linked by equation \eref{eq:edges}, and $g=1$ for the Euclidean plane and $g\geqslant  2$ for the hyperbolic plane. The characteristics of the associated metric fundamental polygons can be inferred from the previous relations: for instance, one has
\begin{equation}
\label{eq:sides}
2g \leqslant  N \leqslant  3(2g-1).
\end{equation}
Note that the fact that $q$ is an integer imposes that $v$ must be a divider of $2e$.

\subsection{Edge pairings and closed walks on graphs}

\begin{figure}
	\begin{center}
		\subfigure[]{\label{fig:walk1} \includegraphics[scale=0.3,clip=true]{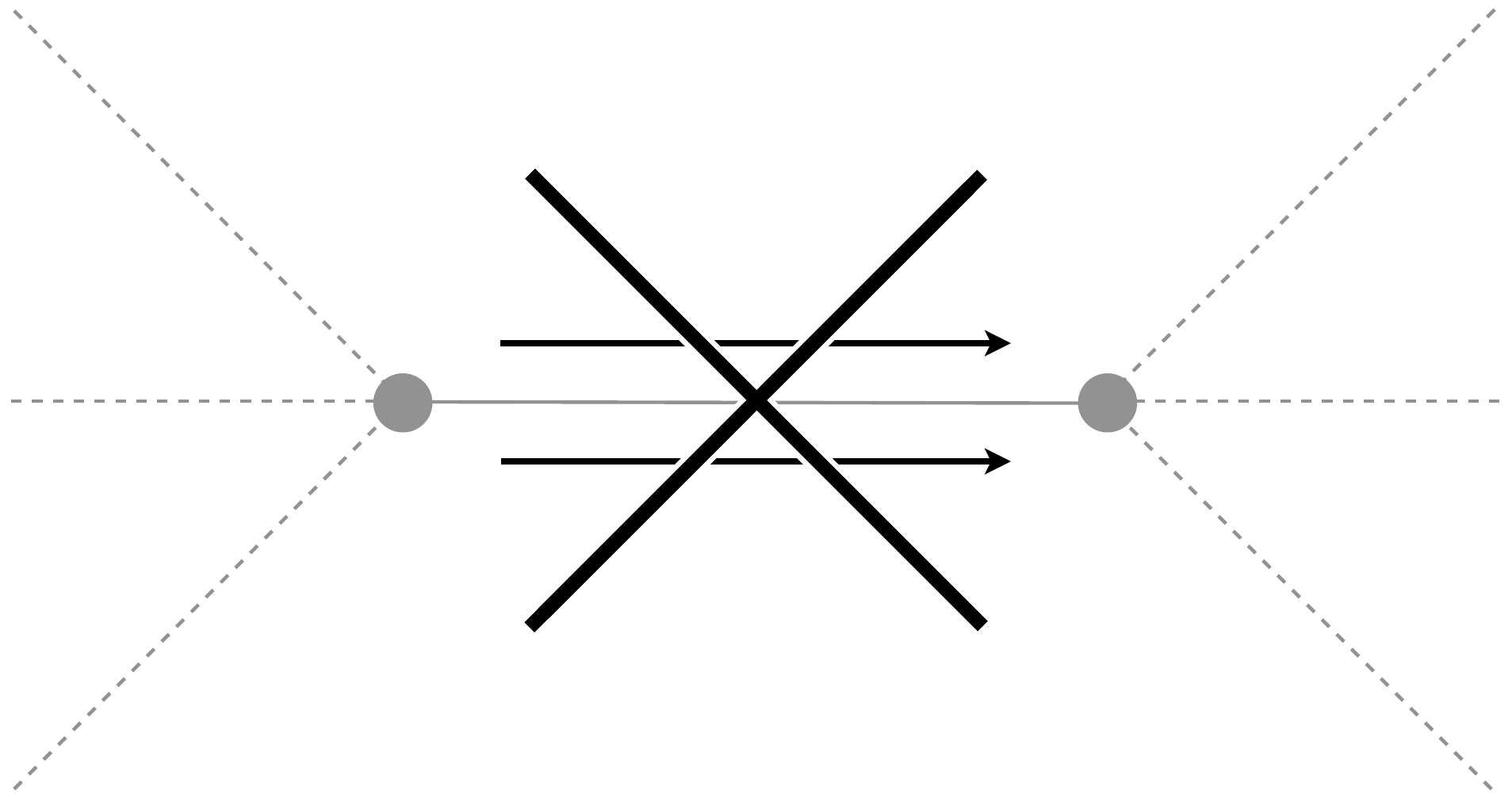}}
		\subfigure[]{\label{fig:walk2} \includegraphics[scale=0.3,clip=true]{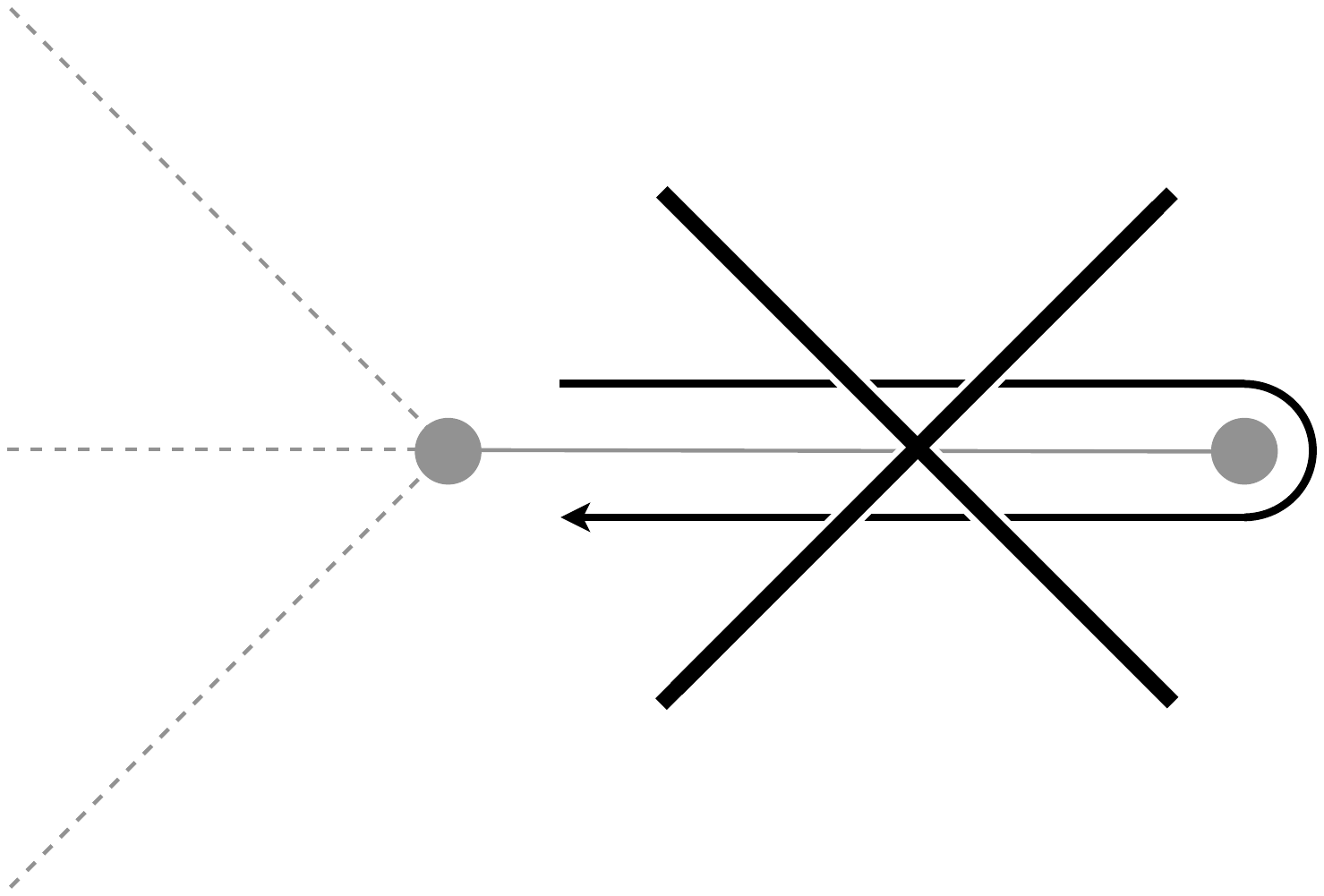}}
	\end{center}
	\caption{\label{fig:walk}Constraints on the allowed closed walks on graphs: (a) each edge of the graph must be followed exactly once in each direction; (b) two consecutive steps of the walk cannot pass over the same graph edge, even if their directions are opposite.}
\end{figure}

One more step is still required. Indeed, to fully determine a fundamental polygon, the pairing of the edges has to be specified in addition to its shape. This pairing is closely related to the structure of the associated graph. We have already addressed some properties of the graphs: genus, numbers of vertices, edges and faces. Edge pairing now amounts to finding a closed walk on the graph. As explained before, going around the polygon boundary once is equivalent to a closed walk. This latter is constrained by two conditions:
\begin{itemize}
	\item[-] the walk follows each edge of the graph exactly once in each direction (see \fref{fig:walk1}),
	\item[-] when an edge has been traversed in one direction, the walk cannot immediately go backward to follow the edge in the opposite direction (see \fref{fig:walk2}).
\end{itemize}
With these conditions, one can enumerate all possible graphs and closed walks for a given polygon. This has been done in the particular case of fundamental polygons with $18$ sides and genus $2$ \cite{Jorgensen:1982}.

Graphs can be represented in the Euclidean plane. However, one should remember that the graph does not live in this plane, but rather on the compact manifold of given genus $g$. In the following we use the term ``graph" to indicate such a representation without a specified closed walk. With this terminology, different closed walks may then exist for the same graph. However, two different closed walks on a given graph may correspond to the same way of pairing polygon edges. For example, in the case of the $18$-gons of genus $2$ \cite{Jorgensen:1982}, 5 graphs and 13 different closed walks were found, but this led to only 8 different side pairings. 

\begin{figure}
	\includegraphics[scale=0.45]{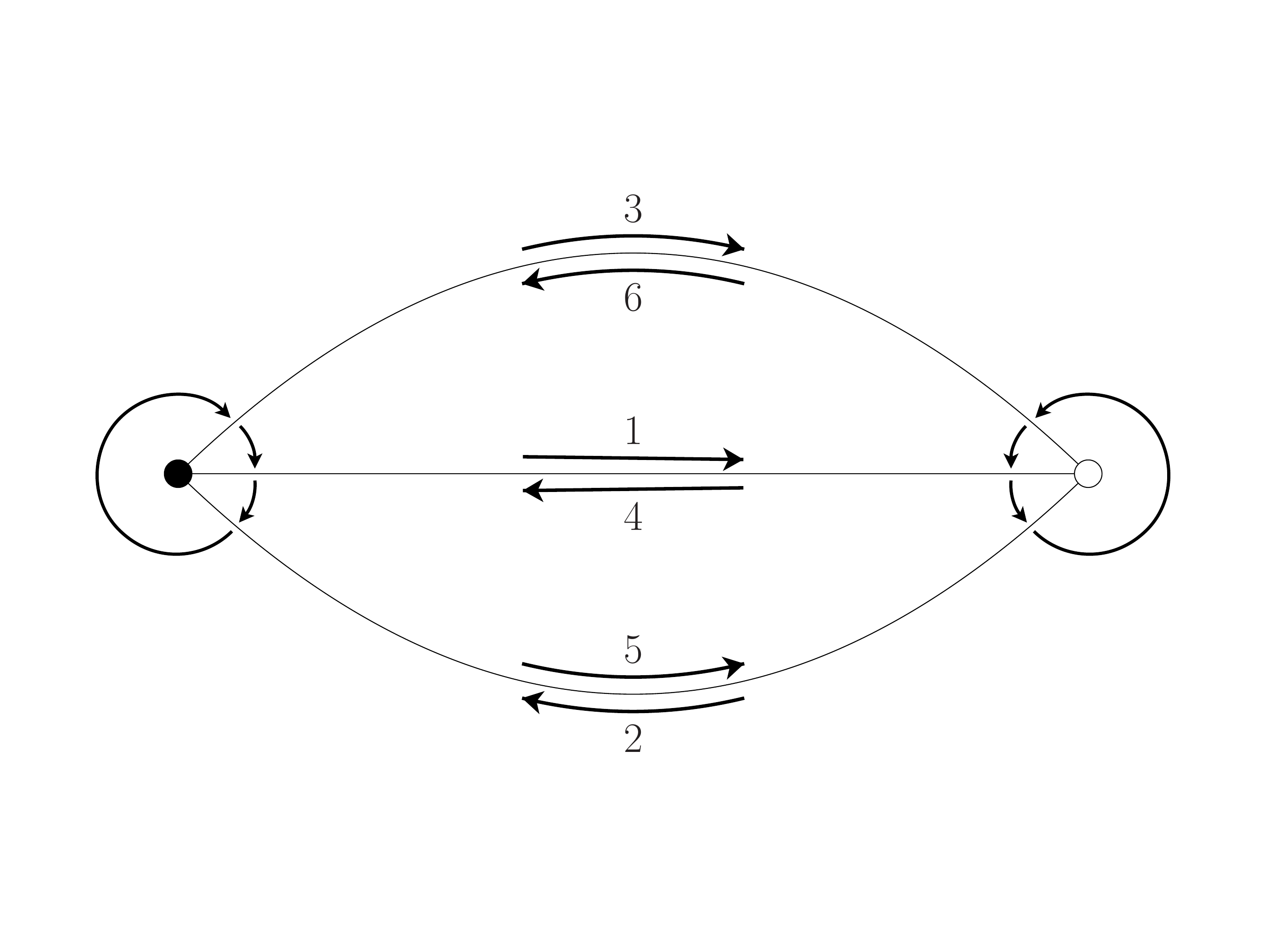}
	\caption{\label{fig:graphe-hex}Decorated graph and its associated closed walk for the hexagonal periodic boundary condition in $E^2$ shown in \fref{fig:rep-hex}. As indicated by the arrows, the vertex with a black circle is decorated by a clockwise rotation and that with a white circle with a counterclockwise rotation. The present graph is a planar representation of the graph embedded in the one-hole torus (see \fref{fig:tor-hex}).}
\end{figure}

To determine the allowed closed walks on a graph, it is convenient to use the concept of ``rotation" for vertices of the graph. For each closed walk and each vertex of the graph, one must specify the rules describing how one goes from one edge by which the walk arrives at the vertex to the subsequent one by which the walk next leaves the vertex, and this of course for all the edges connected to the vertex. Considering that for any two consecutive polygon edges their counterparts on the graph are adjacent edges, the only acceptable rules then correspond to clockwise and counterclockwise rotations around a vertex (see \fref{fig:graphe-hex} for illustration).

All possible closed walks for a given graph are then obtained after first decorating vertices with rotations. For each vertex there are two rotations, so that for a graph with $v$ vertices there will be $2^v$ possible decorations. In some cases, simple rules allow to reduce the number of possible decorations \cite{Jorgensen:1982}. Once all possible decorations have been performed, the corresponding edge pairings have to be found, knowing that different decorations can give the same edge pairing. A simple way to link edge pairings and closed walks on decorated graphs is to report on the graph the numbers labeling the polygon edges in the order in which they are encountered in the closed walk. Then, each graph edge is labeled by two numbers that indicate the pairing of two specific polygon sides. We now illustrate this procedure.

Let us first detail the hexagonal case on the Euclidean plane. The graph is represented in \fref{fig:graphe-hex}, where a black vertex indicates a clockwise rotation and a white one a counterclockwise rotation. The displayed closed walk, which can be easily checked to satisfy the conditions given above, corresponds to the following edge pairing:
\begin{equation*}
	1-4\;,\quad 2-5\;,\quad 3-6.
\end{equation*}
This notation must be read in this way: side 1 is paired with side 4, side 2 with side 5, side 3 with side 6. Another notation, more convenient for polygons with numerous sides and complex side-pairing pattern, is based on a visual description that gives, for each side of the polygon and when moving around in a chosen direction, the number of sides between two paired sides. The above pairing can then be noted
\begin{equation*}
	2-2-2-2-2-2,
\end{equation*}
which means, going counterclockwise around the hexagon and starting with side $1$: $1$ with $4$, $2$ with $5$, ..., $6$ with $3$. This notation will be preferred in the following.

\section{Classifying and building periodic boundary conditions}

\subsection{Elements of classification}

Consider first $g=1$, which corresponds to the Euclidean case. Setting $g=1$ in equation \eref{eq:sides} gives that the number of sides of the fundamental polygon can be 4 or 6, which confirms that the only two periodic boundary conditions with a regular polygon as primitive cell (and an associated compact quotient space) are the square and the hexagonal ones. We can reframe the side pairings in these two cases in the graph formalism of the preceding section. For the square, the graph obtained by gluing polygon edges and embedded in the one-hole torus has one vertex and two edges. So, edges form loops attached to the unique vertex. However, two arrangements are possible: interlaced loops and disjoint loops.
\begin{figure}
	\includegraphics[scale=1]{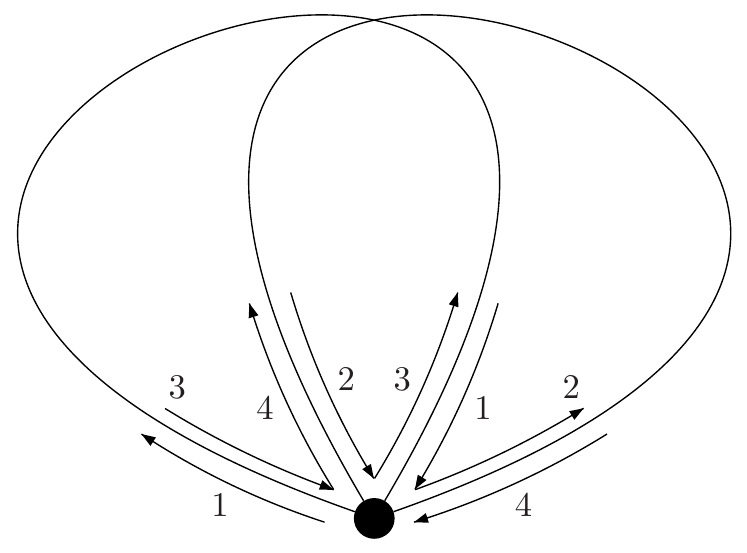}
	\caption{\label{fig:graphe-carre}Decorated graph and its associated closed walk for the square periodic boundary condition in $E^2$ (see  \fref{fig:rep-car}). The two edges form two interlaced closed loops; for clarity, they are labeled twice in each direction. The graph embedded in the one-hole torus is shown in \fref{fig:tor-car}.}
\end{figure}
By trying to perform a closed walk on the two possible graphs (the rotation of the unique vertex has no importance here), one can see that the only acceptable graph is the interlaced one shown in \fref{fig:graphe-carre}. This graph leads to the following edge pairing:
\begin{equation*}
	1-1-1-1.
\end{equation*}

For the hexagon, the graph has two vertices and three edges. Of all the possibilities with vertices decorated by ``rotations", the only one which provides a suitable closed walk is that represented in \fref{fig:graphe-hex}. As seen previously, it leads to the following edge pairing:
\begin{equation*}
	2-2-2-2-2-2.
\end{equation*}
The embedding of the graphs associated to the square and hexagonal periodic boundary conditions in the one-hole torus are displayed in figures \ref{fig:graphe-carre} and \ref{fig:graphe-hex}.

Taking now $g=2$, one has to consider the hyperbolic geometry. Equation \eref{eq:sides} for $g=2$ becomes
\begin{equation}
\label{eq:sides2}
	4 \leqslant N \leqslant 9
\end{equation}
with, we recall, $2N$ denoting the number of polygon edges. For all these fundamental polygons, the quotient space will be a two-hole torus (see \fref{fig:tor-8}). It is easier to first detail the case $N=4$ for which the graph as only one vertex, like the Euclidean square case. Due to this unique vertex, the 4 edges become 4 loops attached to the vertex.
\begin{figure}
	\begin{center}
		\subfigure[]{\label{fig:graphe-8-can} \includegraphics[scale=0.5]{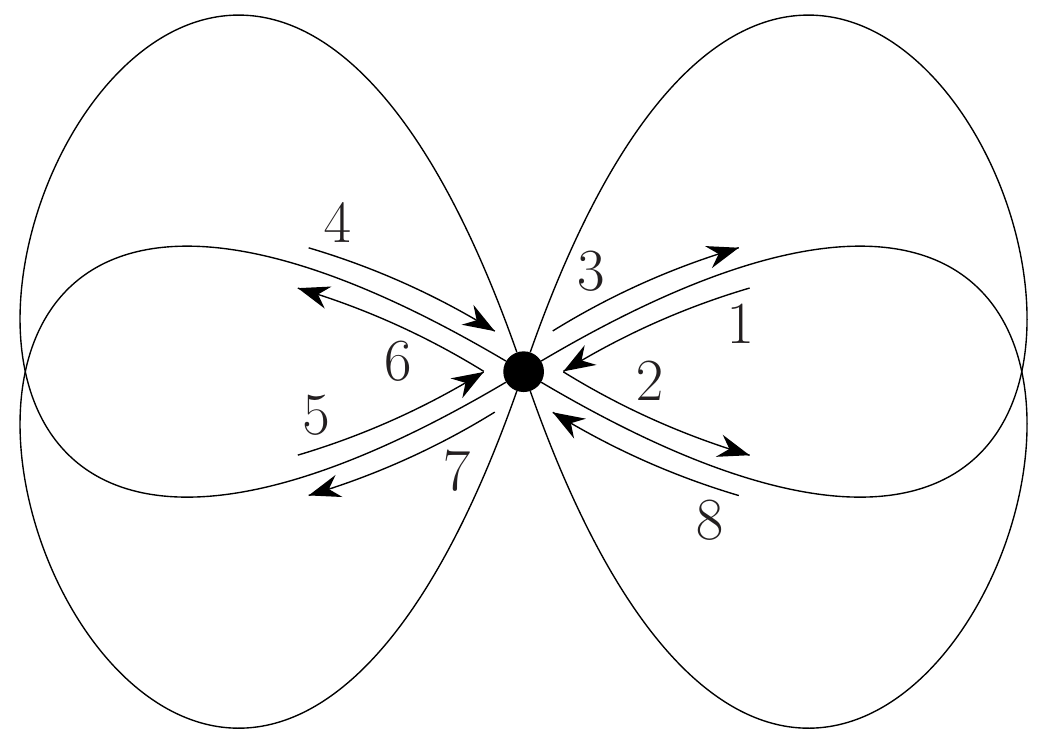}}
		\subfigure[]{\label{fig:graphe-8-simple} \includegraphics[scale=0.5]{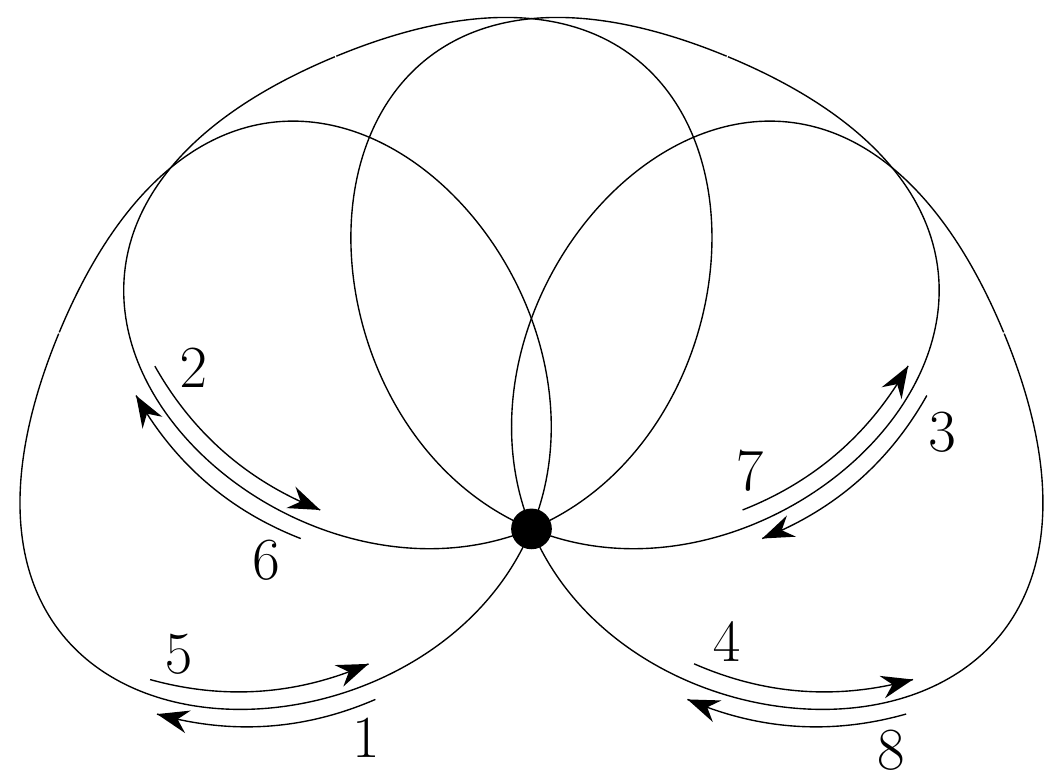}}
	\end{center}
	\caption{\label{fig:graphe-8}The only two possible decorated graphs (and their associated closed walk) with one vertex and four edges. The four edges form four interlaced closed loops (that appear in two disconnected sets in (a), but not in (b)). Contrary to \fref{fig:graphe-carre}, the loops are only labeled once in each direction. In both cases, the genus is equal to 2 and the associated fundamental polygon is an octagon.}
\end{figure}
In this case, only two graphs with their associated closed walk are possible (see \fref{fig:graphe-8}) and the corresponding side pairings are:
\begin{equation*}
	1-1-5-5-1-1-5-5 \quad \mathrm{and} \quad 3-3-3-3-3-3-3-3
\end{equation*}
which correspond, respectively, to the pairing patterns shown in figures \ref{fig:rep-8} and \ref{fig:rep-8-simple}.
\begin{figure}
	\begin{center}
		\subfigure[]{\label{fig:rep-8} \includegraphics[scale=0.5]{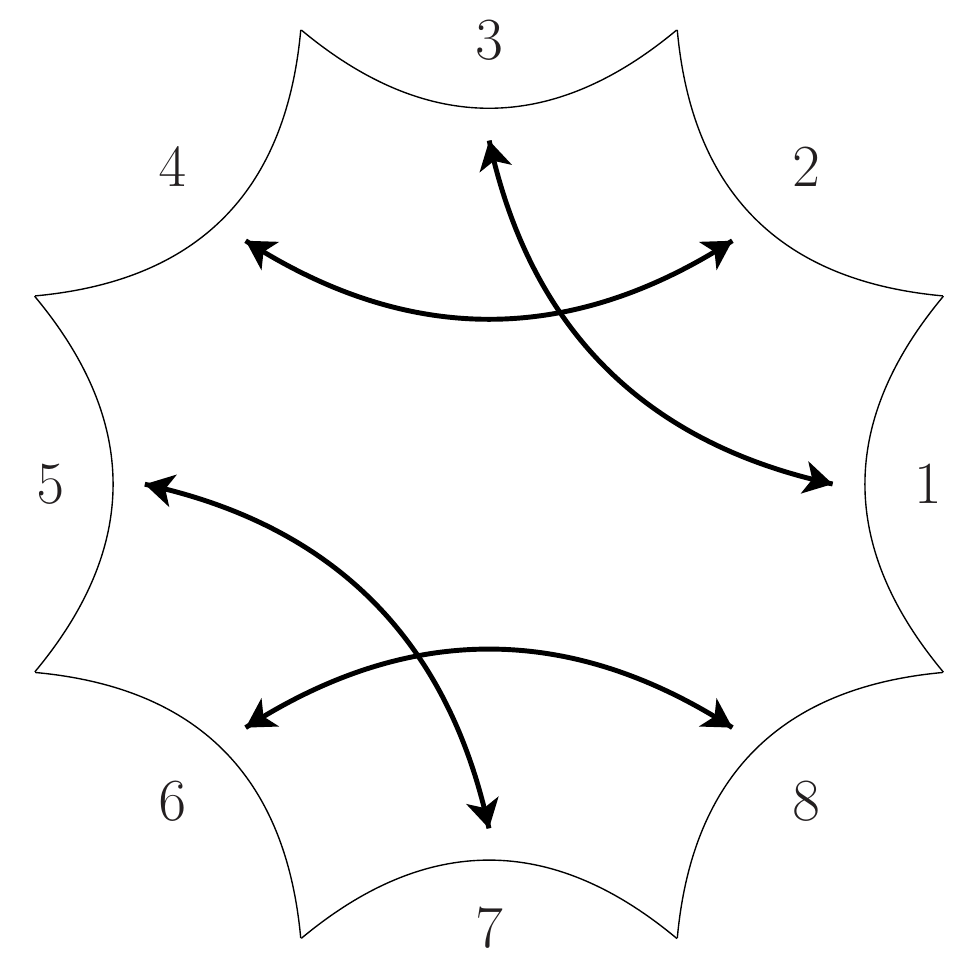}}
		\subfigure[]{\label{fig:rep-8-simple} \includegraphics[scale=0.5]{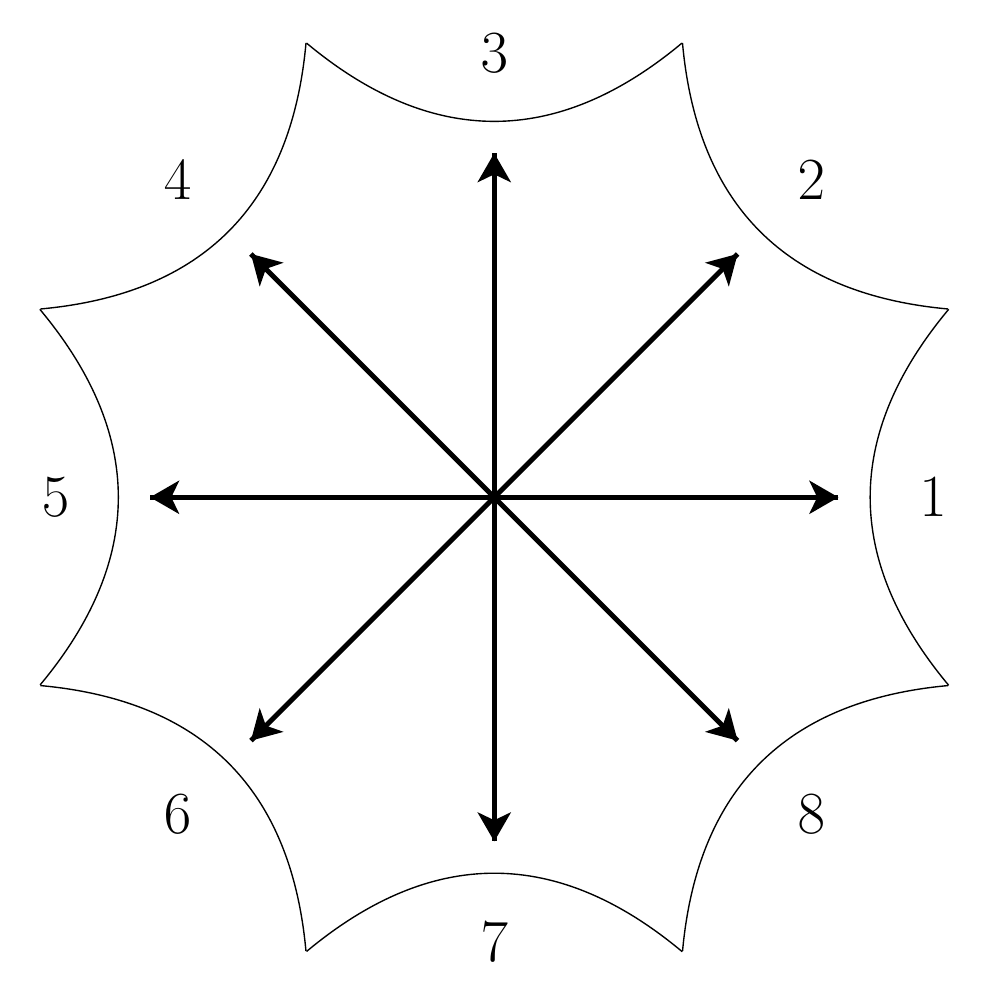}}
	\end{center}
	\caption{Fundamental polygons corresponding to the decorated graphs of \fref{fig:graphe-8}: (a) corresponds to \fref{fig:graphe-8-can} and (b) to \fref{fig:graphe-8-simple}. The pairings are the only ones allowed for the $\{8,8\}$ tiling of $H^2$.}
\end{figure}
In terms of the generators of the associated Fuchsian group introduced above ($\gamma_{i}$ is the hyperbolic translation that takes side $i$ into its paired counterpart), the two pairings can also be expressed as
\begin{equation*}
\gamma_1\gamma_2\gamma_1^{-1}\gamma_2^{-1}\gamma_5\gamma_6\gamma_5^{-1}\gamma_6^{-1}=\mathbbm{1}\quad \mathrm{and} \quad \gamma_1\gamma_2\gamma_3\gamma_4\gamma_1^{-1}\gamma_2^{-1}\gamma_3^{-1}\gamma_4^{-1}=\mathbbm{1},
\end{equation*}
where in the first case, $\gamma_3=\gamma_1^{-1}, \gamma_4=\gamma_2^{-1}, \gamma_7=\gamma_5^{-1}, \gamma_8=\gamma_6^{-1}$, and in the second one, $\gamma_5=\gamma_1^{-1}, \gamma_6=\gamma_2^{-1}, \gamma_7=\gamma_3^{-1}, \gamma_8=\gamma_4^{-1}$.

\begin{figure}
	\begin{center}
		\subfigure[]{\includegraphics[scale=0.15]{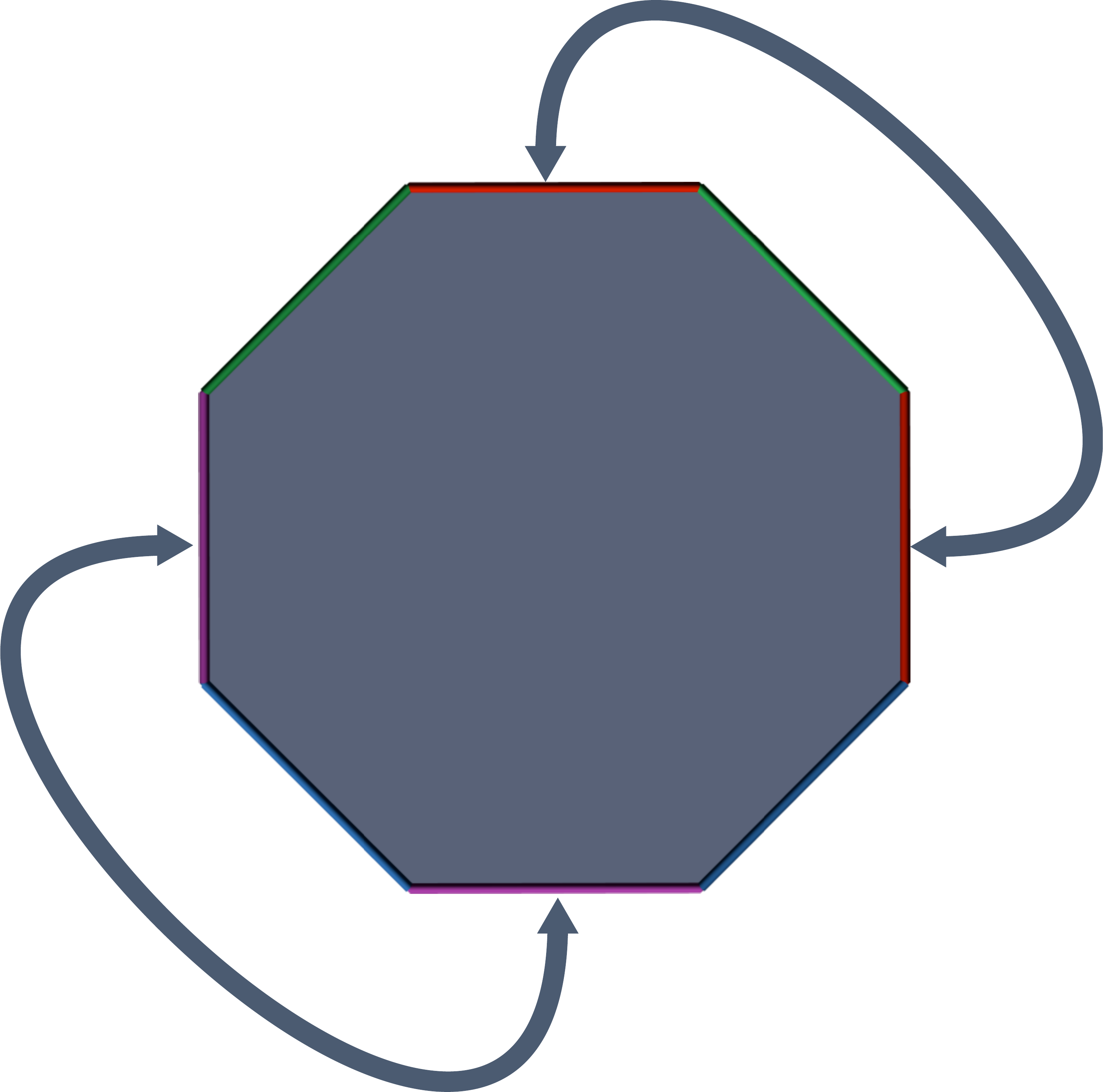}}
		\subfigure[]{\includegraphics[scale=0.2]{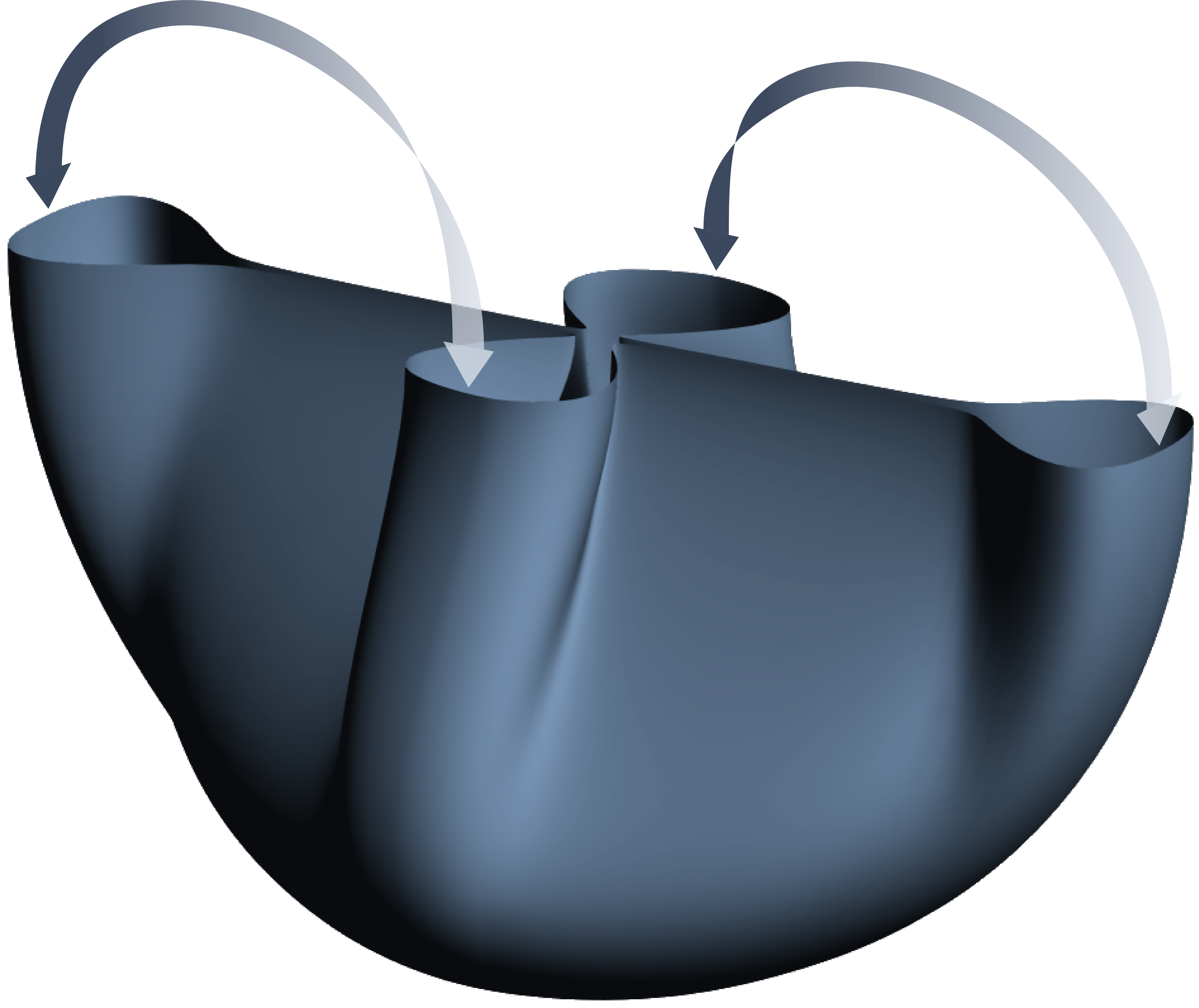}}
	\end{center}
	\caption{\label{fig:tor-8-rep}Schematic representation of the ``compactification" of the fundamental polygon shown in \fref{fig:rep-8}. The paired edges are glued together: 1 with 3 and 5 with 7 in (a); 2 with 4 and 6 with 8 in (b). The final compact manifold is a two-hole torus represented in \fref{fig:tor-8-can}.}
\end{figure}

Note that for $g=2$ and $N=4$, the fundamental polygon is an octagon of area equal to $4\pi \kappa^{-2}$, which is associated to the $\{8,8\}$ tiling of the hyperbolic plane.
The link between the graph and the fundamental polygon is made by identifying paired polygon edges (see \fref{fig:tor-8-rep}), which leads to the quotient space in which the graph is embedded (see \fref{fig:tor-8}) \cite{Balazs:1986,Nakahara:1990}.
\begin{figure}
	\begin{center}
		\subfigure[]{\label{fig:tor-8-can} \includegraphics[scale=0.25]{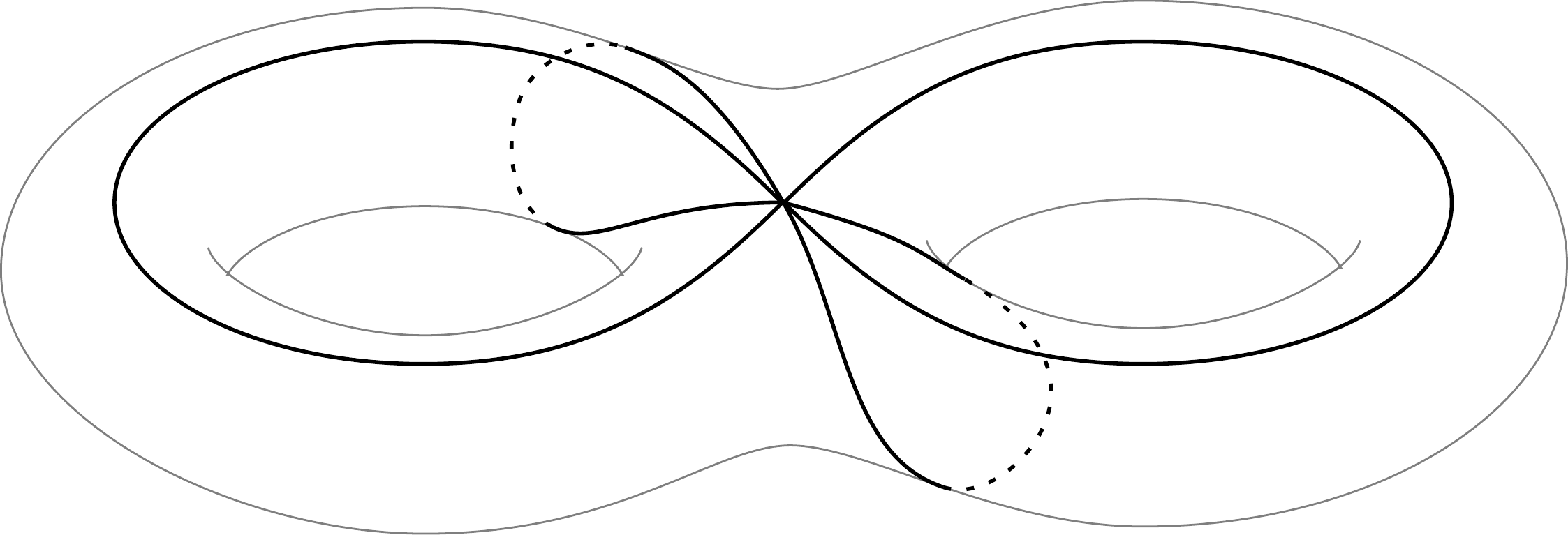}}
		\subfigure[]{\includegraphics[scale=0.25]{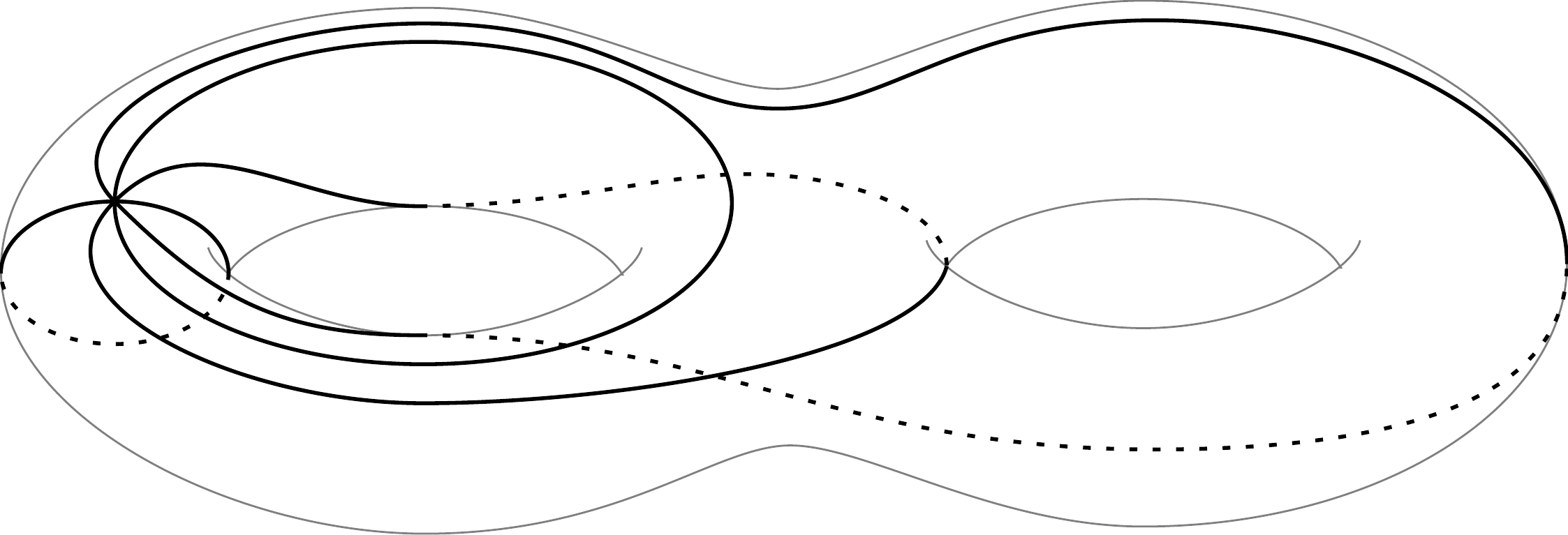}}
	\end{center}
	\caption{\label{fig:tor-8}Compact manifold and graphs obtained by gluing the edges of the fundamental polygons of \fref{fig:rep-8} (in(a)) and \fref{fig:rep-8-simple} (in (b)). The graphs embedded in the compact manifold (a two-hole torus) correspond to the planar representations of \fref{fig:graphe-8}.}
\end{figure}

Still for $g=2$, increasing $N$ changes the shape of the fundamental polygon and its edge pairing, while its area remains the same. We shall not detail all the possibilities, and we just recall that a complete enumeration of the fundamental polygons for $g=2$ and $N=9$, which corresponds to the upper bound for $N$ in that case (see equation \eref{eq:sides2}), has been given in \cite{Jorgensen:1982}.

What happens when $g$ increases? First, as the value of $g$ has no upper bound, one finds that  an infinity of periodic boundary conditions are possible in the hyperbolic plane. Secondly, for given $g$ and $N$, many different edge pairings are generally possible. There is no point in searching for an exhaustive description, and we rather focus on three specific ``families" of edge pairings which can be easily described for any value of the genus $g$. These three families can be thought of as generalizations of the two Euclidean periodic boundary conditions.

To explain the first two families, let us go back again to the Euclidean square case. The associated graph has only one vertex, which is related to the fact that $2N=4$ is the minimum value for  $g=1$. The analogous case for $g=2$, with the minimum value $2N=8$, has just been considered before: two different graphs, hence two different edge pairings, have been found. More generally, the minimum value of the number of polygon edges for a given genus $g$ is $2N=4g$ (see \tref{tab:classification}). Periodic boundary conditions built with such regular $4p$-gons are characterized by a graph with a unique vertex and $2g$ edges, so that the vertex has a coordinence of $4g$. Yet, multiple possibilities arise for constructing acceptable closed walks on graphs with such properties (only one actually for $g=1$, two for $g=2$, and more for higher $g$'s).
\begin{figure}
	\begin{center}
		\subfigure[]{\label{fig:graphe-4p-can} \includegraphics[scale=0.5]{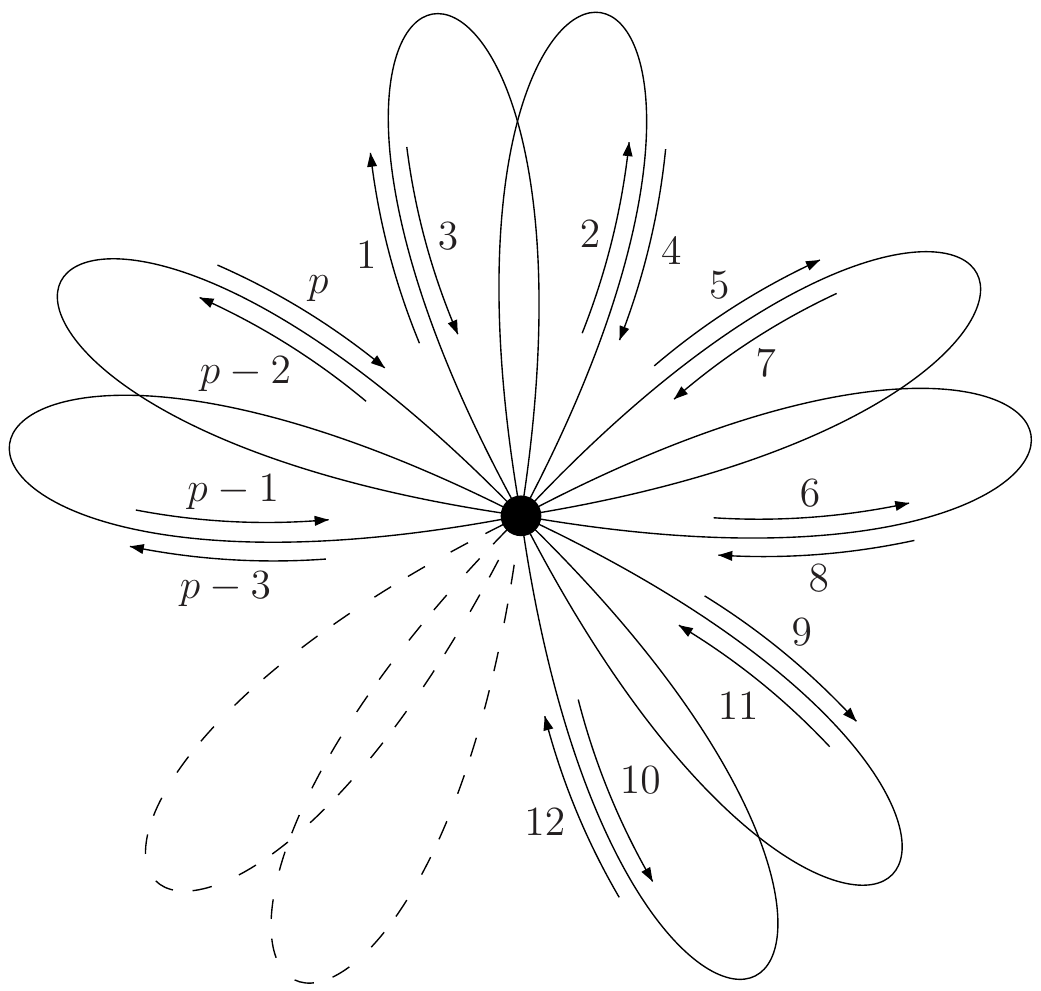}}
		\subfigure[]{\label{fig:graphe-4p-simple} \includegraphics[scale=0.5]{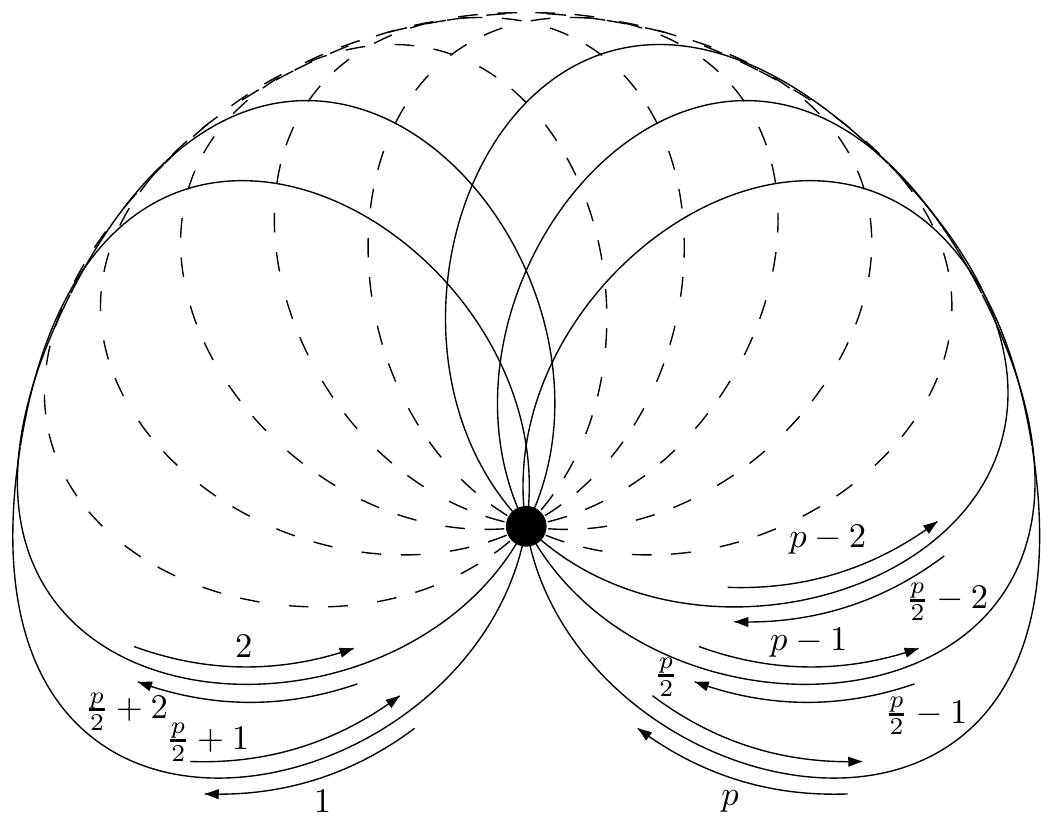}}
	\end{center}
	\caption{\label{fig:graphe-4p}Decorated graphs and associated closed walks of two ``families" of periodic boundary conditions in $H^2$ for a generic value of the genus $g\geqslant 2$. In both cases, the fundamental polygon is a $p$-gon with $p=4g$ (see \fref{fig:rep-4p}), and the associated tiling is the $\{4g,4g\}$ one. The graphs have $\frac{p}{2}$ edges that form $\frac{p}{2}$ interlaced closed loops (the loops are only labeled once in each direction).}
\end{figure}
Two graphs, however, are  simple to generate, each one with a unique possible closed walk on it: they are displayed in \fref{fig:graphe-4p}.
\begin{figure}
	\begin{center}
		\subfigure[]{\label{fig:rep-4p-can} \includegraphics[scale=0.45]{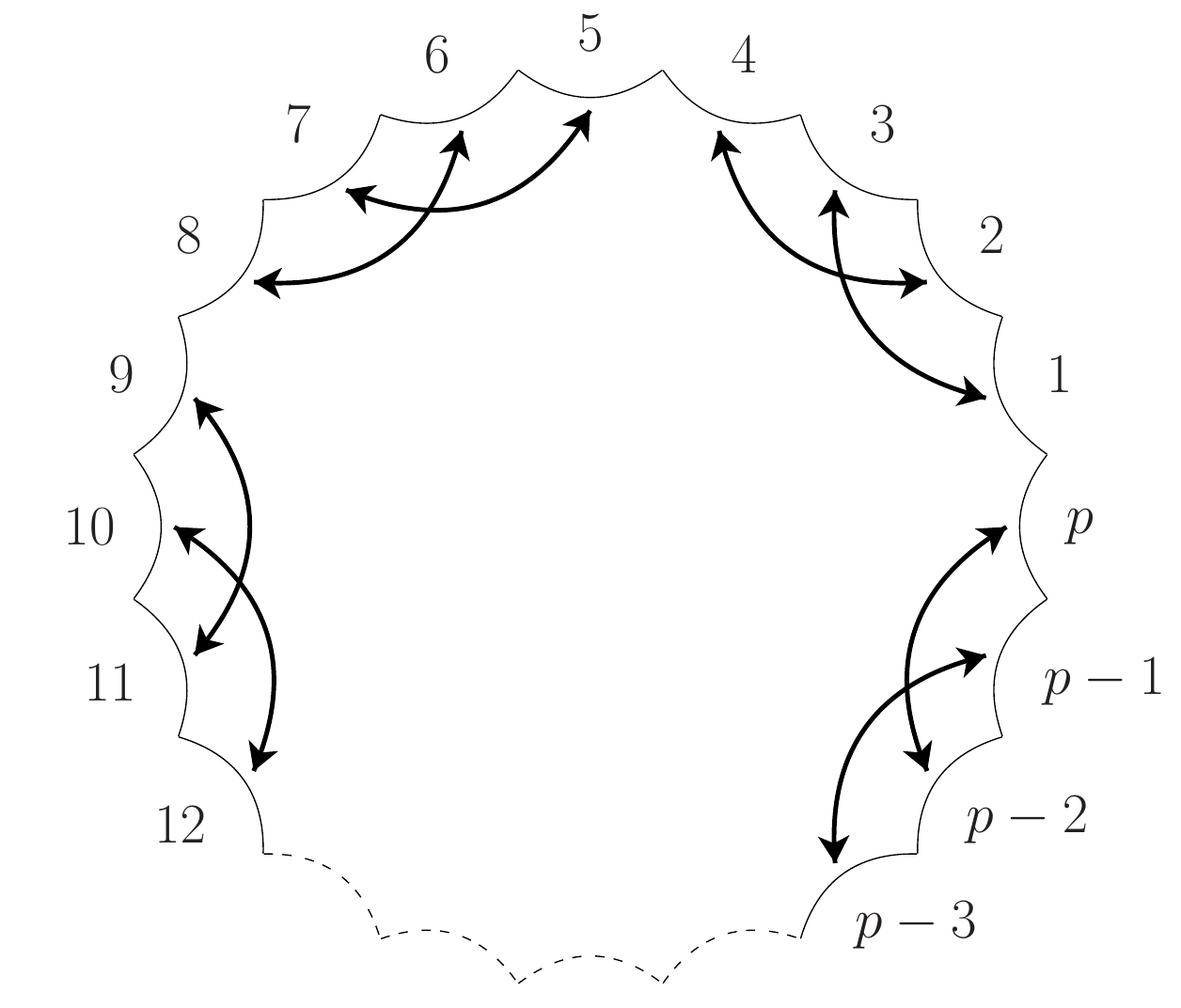}}
		\subfigure[]{\label{fig:rep-hex-gen} \includegraphics[scale=0.45]{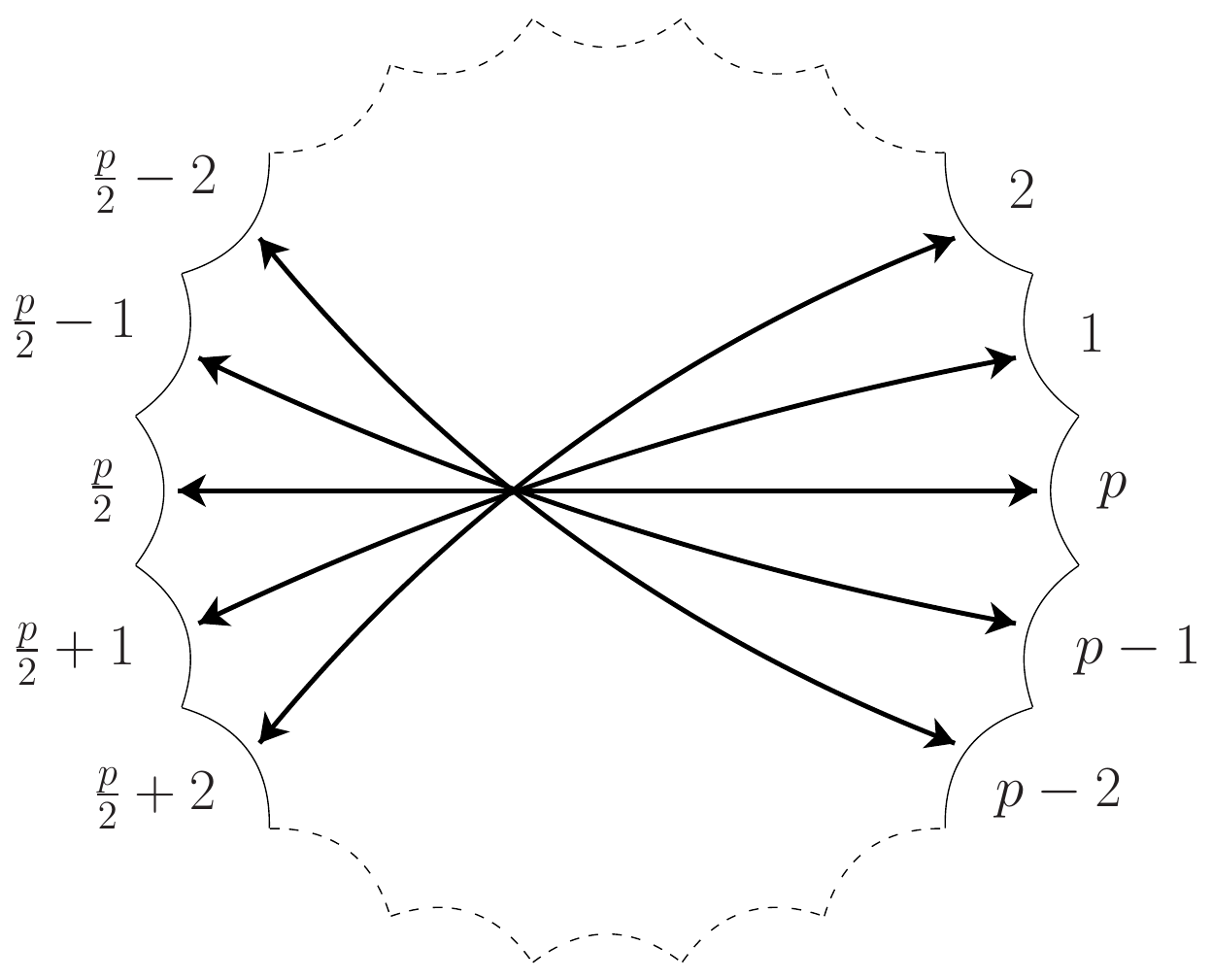}}
	\end{center}
	\caption{\label{fig:rep-4p}Fundamental polygons associated with the decorated graphs of \fref{fig:graphe-4p-can} (in (a)) and \fref{fig:graphe-4p-simple} (in (b)).}
\end{figure}
These two graphs lead to two different side pairings, and this is valid  for each value of $g$. As illustrated in \fref{fig:rep-4p}, the associated fundamental polygons are $4g$-gons which both represent the primitive cell of the same underlying $\{4g,4g\}$ tiling.

\begin{figure}
	\begin{center}
		\subfigure[]{\label{fig:tor-4p} \includegraphics[scale=0.25]{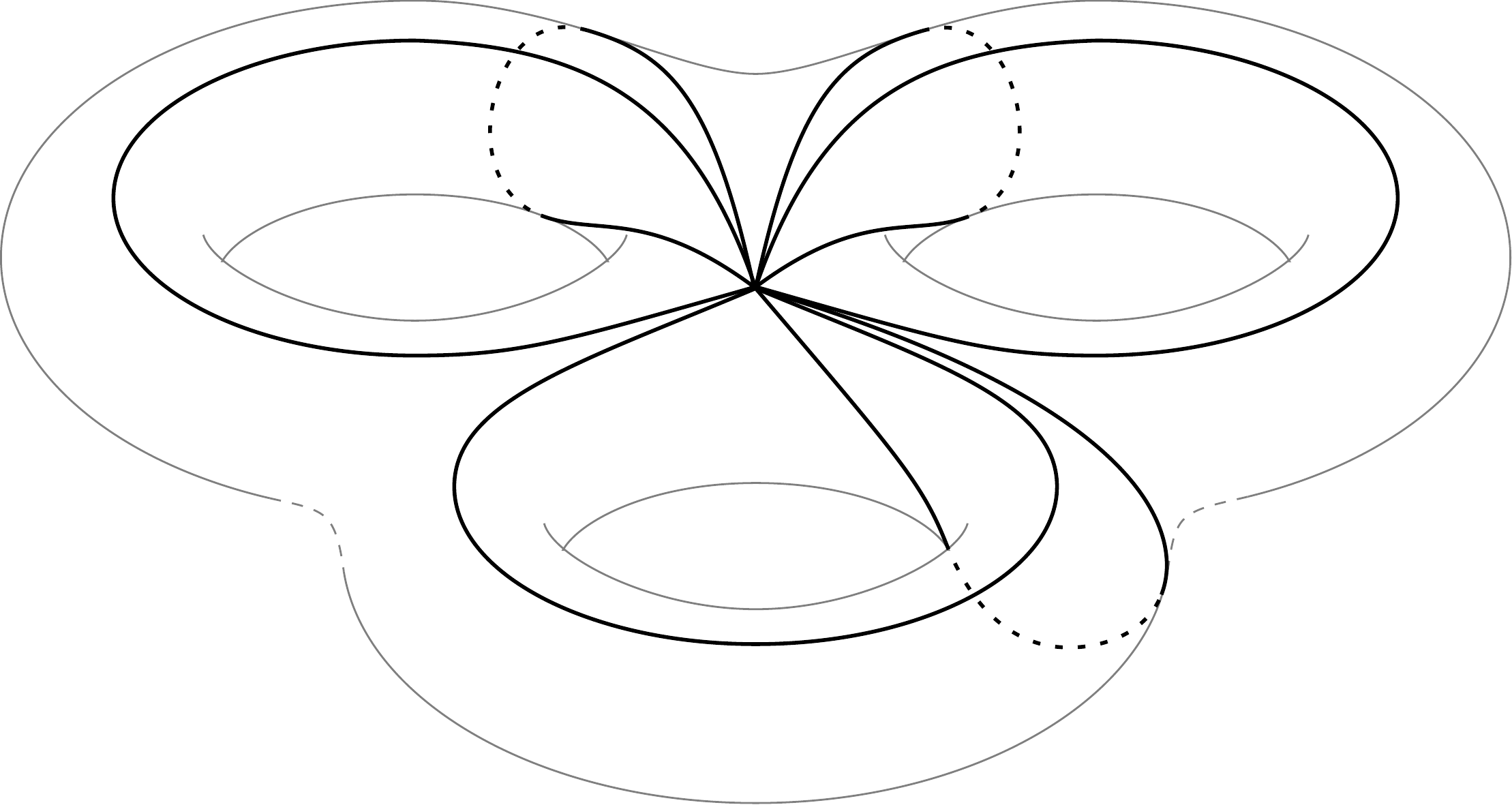}}
		\subfigure[]{\label{fig:tor-4p-simple} \includegraphics[scale=0.25]{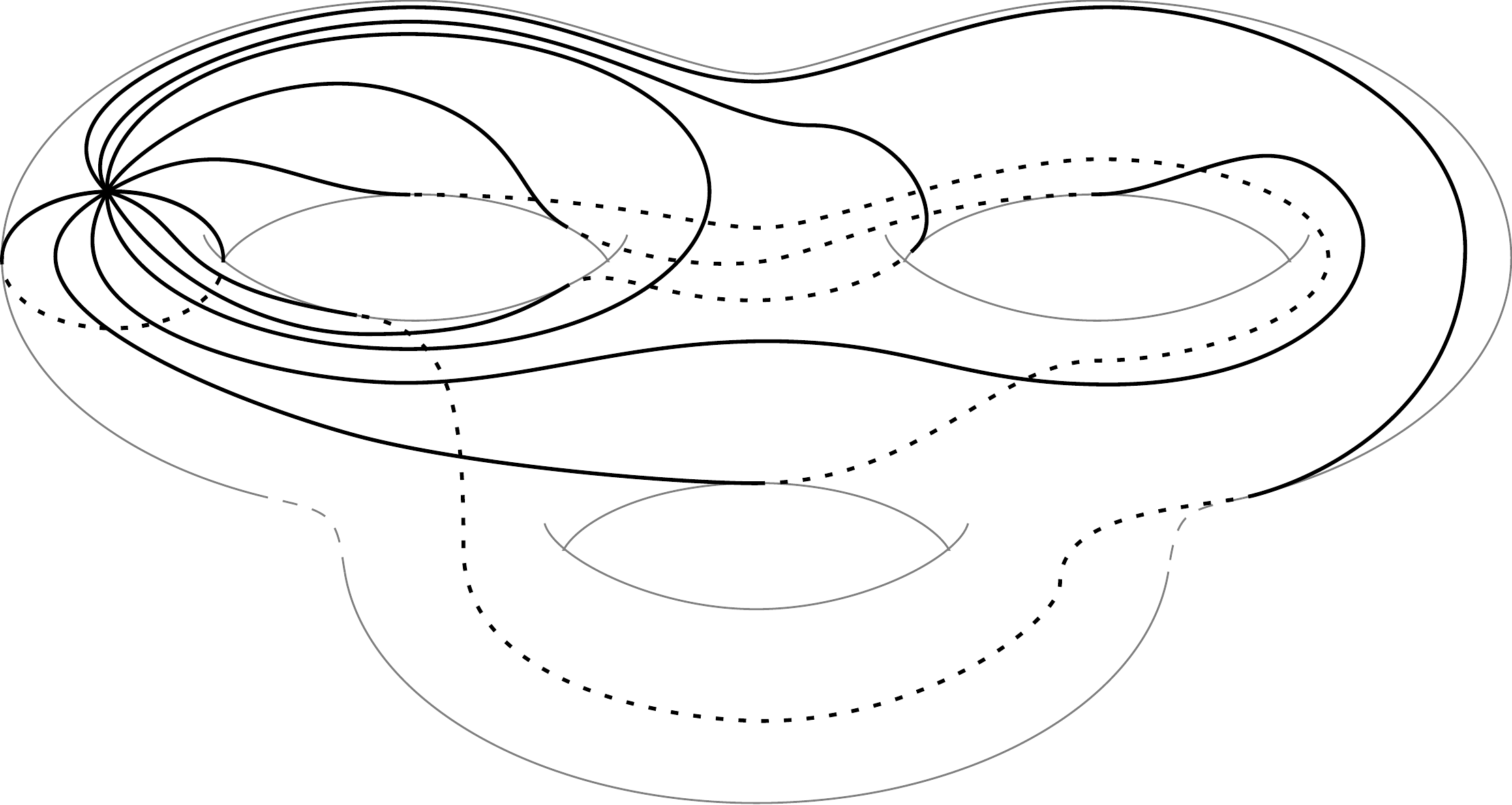}}
	\end{center}
	\caption{Compact manifold obtained by gluing the edges of the fundamental polygons of \fref{fig:rep-4p-can} (in (a)) and \fref{fig:rep-hex-gen} (in (b)). The manifold is generically a $g$-hole torus; in the present case, $g=3$. The graphs of \fref{fig:graphe-4p} are shown here embedded in the torus.}
\end{figure}

As seen previously, the connection between the ``graph" and the ``polygon" points of view is achieved by gluing polygon edges to obtain the quotient space where the edges now form the graph (see \fref{fig:tor-4p}).
The two families we have constructed in the above described manner can be considered as generalizations of the Euclidean square case, which we retrieve by taking $g=1$. In the case represented in \fref{fig:rep-4p-can}, it is noteworthy that the metric fundamental polygon is also a standard (or canonical) fundamental polygon, which is not true for the family illustrated in \fref{fig:rep-hex-gen}. However, from a practical point of view, there is no difference in the effort needed to implement these two types of periodic boundary conditions.

Another simple ``family" is made up of the hyperbolic equivalents of the Euclidean hexagonal periodic boundary conditions. This family corresponds to fundamental polygons of genus $g$ with $2N=4g+2$ sides (see \tref{tab:classification}). The main characteristic of this family is the relatively straightforward way to pair polygon sides. Indeed, in this case, pairing consists in linking opposite sides, regardless of the value of $g$. The corresponding graph has two vertices and $2g+1$ edges. Among all the possible graphs with such properties, the one corresponding to the ``hexagonal" family is such that when the two vertices have different ``rotations", edges do not intersect each others.
\begin{figure}
	\begin{center}
		\subfigure[]{\label{fig:graphe-hex-gen} \includegraphics[scale=0.5]{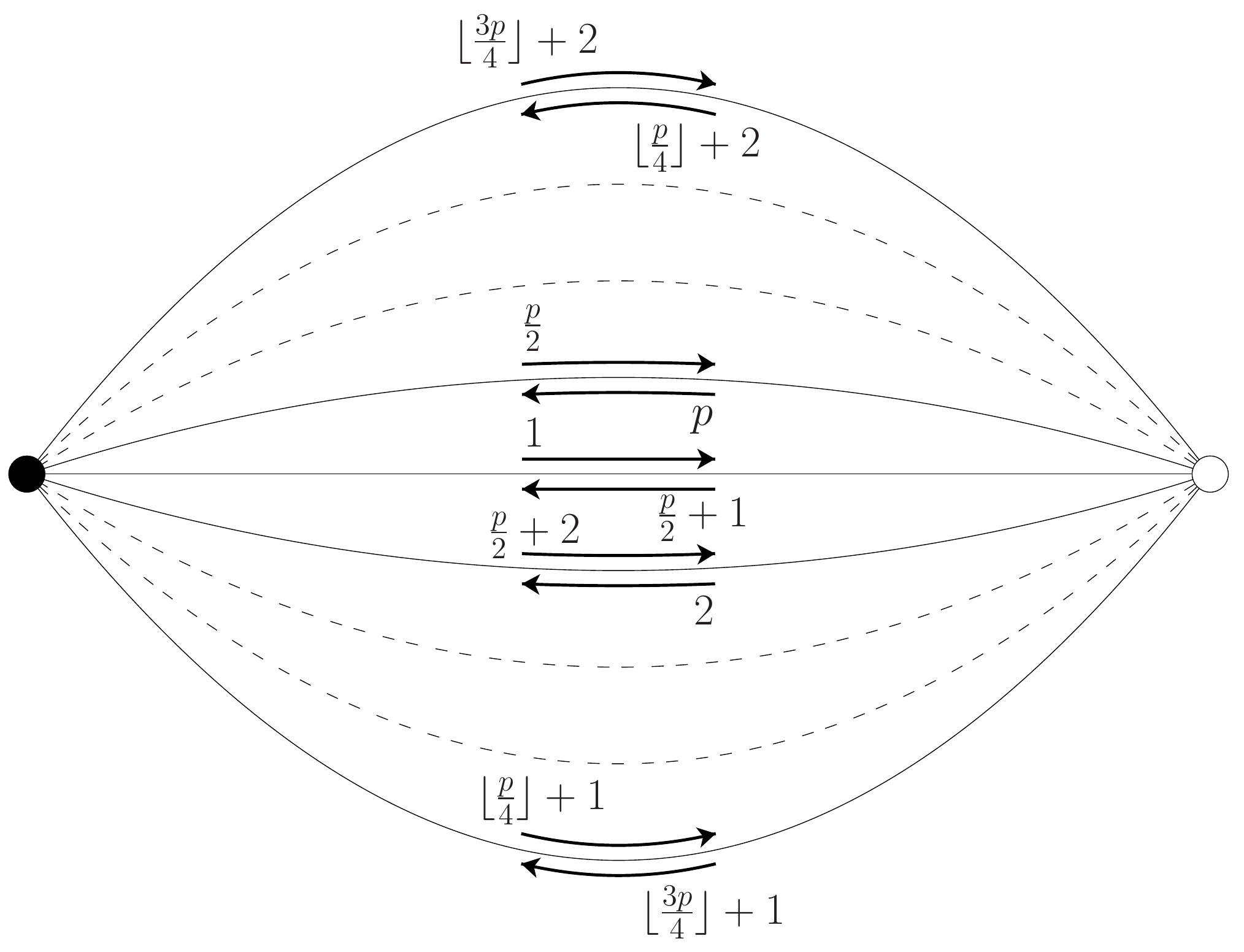}}
		\subfigure[]{\label{fig:graphe-hex-gen-twist} \includegraphics[scale=0.5]{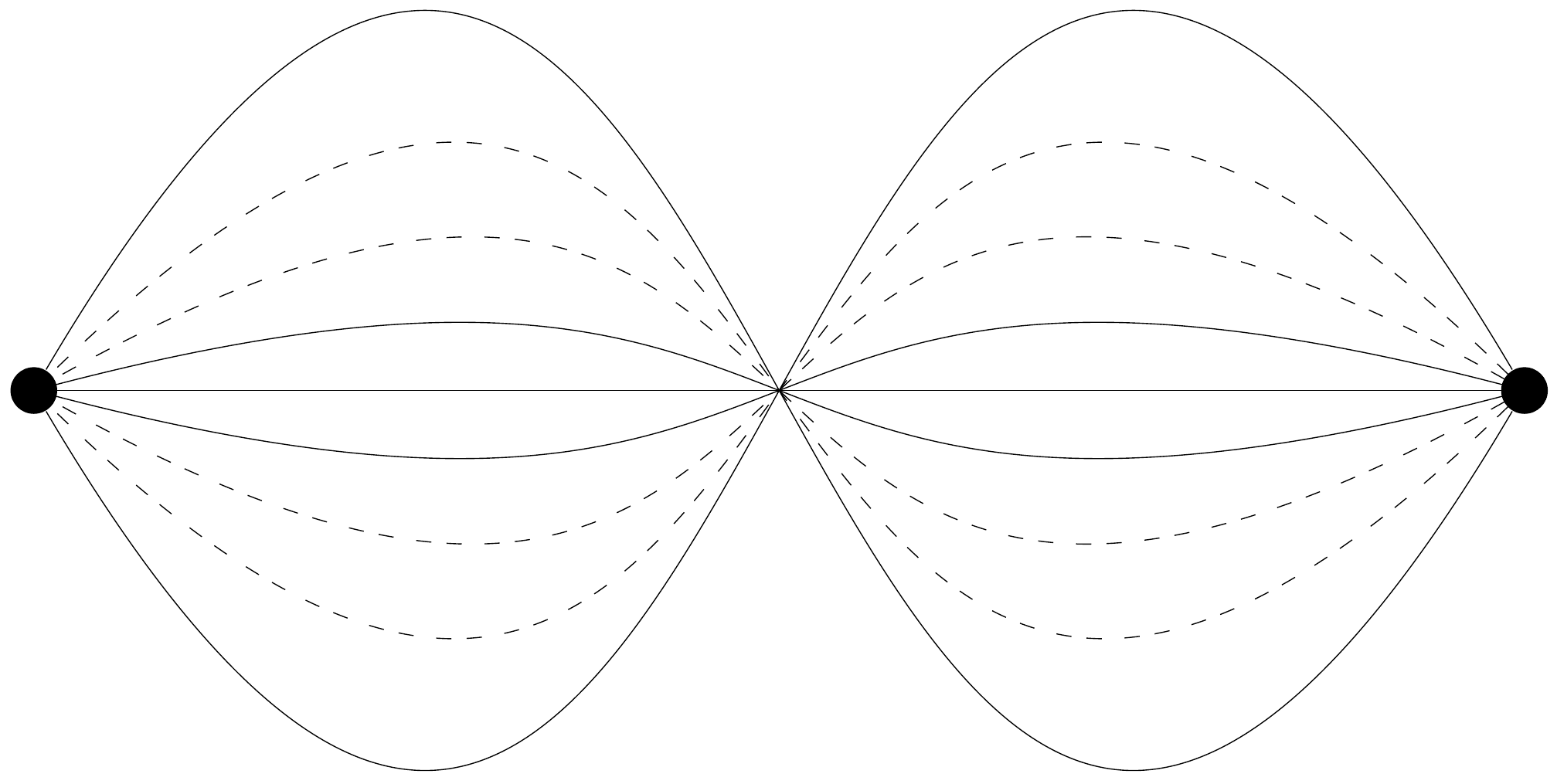}}
	\end{center}
	\caption{\label{fig:graphe-hex-gen-all}(a) Decorated graph of a third ``family" of periodic boundary conditions in $H^2$ for a generic value $g\geqslant 2$ of the genus. This family generalizes the hexagonal Euclidean case ($g=1$) shown in \fref{fig:graphe-hex}. The corresponding fundamental polygon is similar to that displayed in \fref{fig:rep-hex-gen} with $p=2(2g+1)$. The square brackets denote the integer part: here, $\lfloor p/4 \rfloor=g$, $\lfloor 3p/4 \rfloor=3g+1$, etc. The case illustrated here corresponds to an even value of $g$. (For an odd value, one simply interchanges $\lfloor p/4 \rfloor$ and $\lfloor 3p/4 \rfloor$ in the labels, so that odd labels always correspond to walks from left to right and opposite for even labels.) (b) Twisted version of the graph shown in (a), with two black vertices corresponding to clockwise rotations. Note that the two graphs in (a) and (b) are strictly equivalent in the sense that they lead to the same closed walk and the same embedded graph in the $g$-hole torus (see \fref{fig:tor-hex-gen}).}
\end{figure}
This leads to a unique closed walk illustrated in \fref{fig:graphe-hex-gen} and to a unique side pairing for each value of $g$. (Note that this graph can be twisted and is then equivalent to the graph with identical rotations of the two vertices but with crossing edges displayed in \fref{fig:graphe-hex-gen-twist}; the associated side pairing is however unchanged.)  The fundamental polygon characterizing this family is a $2(2g+1)$-gon which is associated with the underlying $\{2(2g+1),2g+1\}$ tiling and has paired opposite sides (see \fref{fig:rep-hex-gen}).
\begin{figure}
	\includegraphics[scale=0.25]{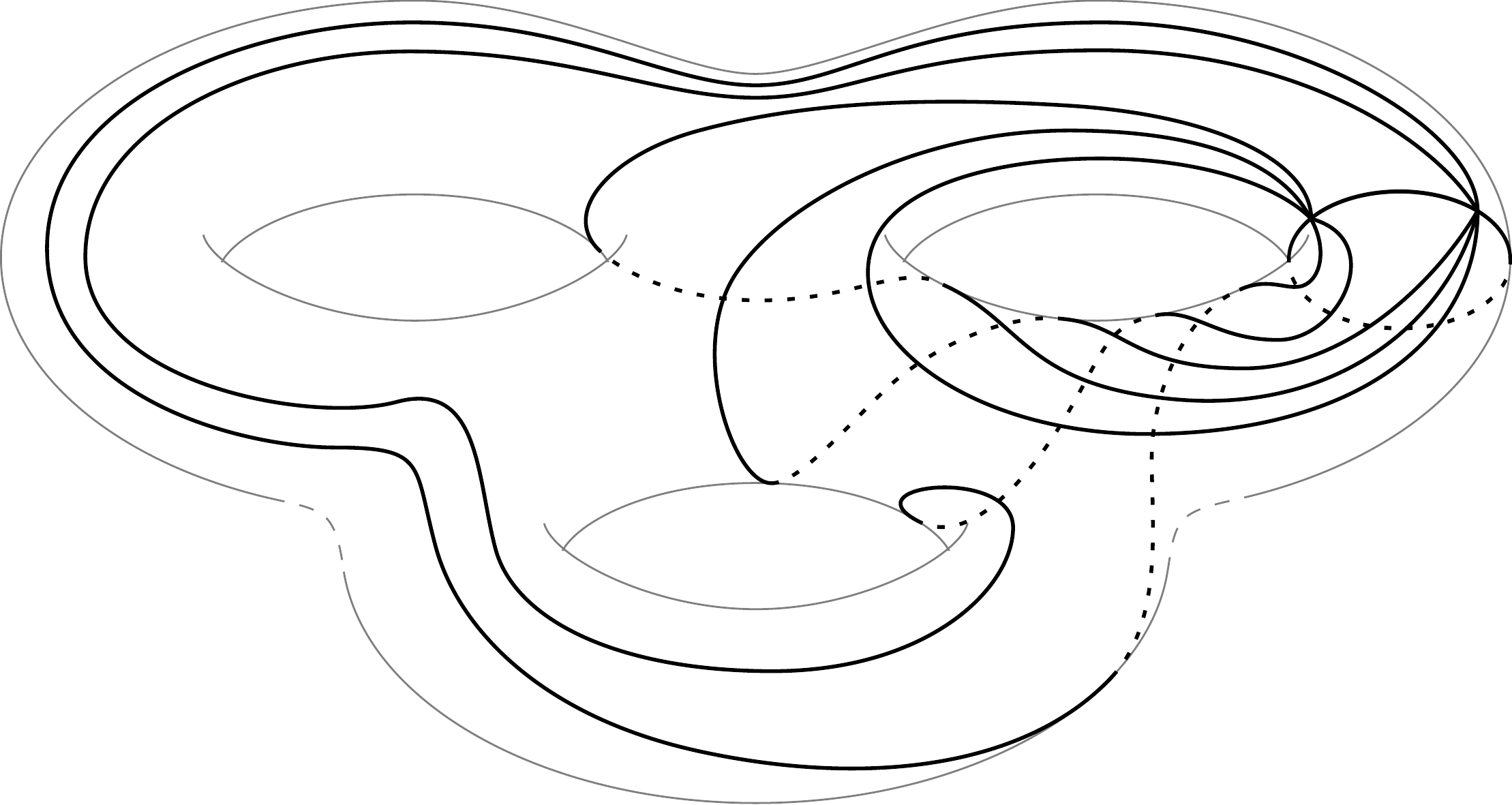}
	\caption{\label{fig:tor-hex-gen}Compact manifold obtained by gluing the edges of the fundamental polygon of \fref{fig:rep-hex-gen} with $p=2(2g+1)$, here with $g=3$. The planar representation of the graph is shown in \fref{fig:graphe-hex-gen-all}.}
\end{figure}
In this case, the quotient space is identical to that of the preceding $4g$-gons. It is a $g$-hole torus, but the embedding of the graph in the torus is different (see \fref{fig:tor-hex-gen}).

\begin{table}
	\caption{\label{tab:classification}Elements of classification of the periodic boundary conditions in the hyperbolic plane (conditions restricted to a regular primitive cell and a compact quotient space).}
\begin{tabular}{rl|cccc}
			\br
			\multicolumn{2}{|l|}{Genus of}& \centre{4}{\multirow{2}{*}{$g \geqslant 2^{\rm a}$}}\\
			\multicolumn{2}{|l|}{quotient space}& \\
			\multicolumn{2}{|l|}{} &\crule{4}\\
			\multicolumn{2}{|l|}{Number of sides}& $2N_{min}$ &  &  &$2N_{max}$ \\
			\multicolumn{2}{|l|}{of the}& $=$ &  &  &$=$ \\
			\multicolumn{2}{|l|}{primitive cell}& $4g$ & $4g+2$ & \ldots &$6(2g-1)$ \\
			\multicolumn{2}{|l|}{} & & & & \\
			\multicolumn{2}{|l|}{\multirow{2}{*}{Associated tiling}}& \multirow{2}{*}{$\{4g,4g\}$} &
			\multirow{2}{*}{$\{2(2g+1),2g+1\}$} & May not & \multirow{2}{*}{$\{6(2g-1),3\}$} \\
			\multicolumn{2}{|l|}{} & & & exist$^{\rm b}$ & \\
			\mr
			\multicolumn{1}{|l|}{ \multirow{22}{*}{\rotatebox{90}{Graph}}}&
			\multicolumn{1}{l|}{Number of edges} &
			$2g$ & $2g+1$ & \ldots & $3(2g-1)$ \\
			\multicolumn{1}{|l|}{}& & & & & \\
			\multicolumn{1}{|l|}{}& \multicolumn{1}{l|}{Number of} & \multirow{2}{*}{1} & \multirow{2}{*}{2} & \multirow{2}{*}{\ldots} & \multirow{2}{*}{$2(2g-1)$} \\
			\multicolumn{1}{|l|}{}& \multicolumn{1}{l|}{vertices} & & & & \\
			\multicolumn{1}{|l|}{}& & & & & \\
			\multicolumn{1}{|l|}{}& \multicolumn{1}{l|}{Coordinence} &
			\multirow{2}{*}{$4g$} & \multirow{2}{*}{$2g+1$} & May not & \multirow{2}{*}{3} \\
			\multicolumn{1}{|l|}{}& \multicolumn{1}{l|}{of vertices} & &  & exist$^{\rm b}$ & \\
			\multicolumn{1}{|l|}{}& & & & & \\
			\multicolumn{1}{|l|}{} & \multicolumn{1}{l|}{Number of} &
			\multirow{2}{*}{$\geqslant 2^{\rm c}$} & \multirow{2}{*}{$\geqslant 1$} & May not & \multirow{2}{*}{$\geqslant 5^{\rm d}$} \\
			\multicolumn{1}{|l|}{} & \multicolumn{1}{l|}{possible graphs} &  &  & exist$^{\rm b}$  & \\
			\multicolumn{1}{|l|}{}& & & & & \\
			\multicolumn{1}{|l|}{} & \multicolumn{1}{l|}{\multirow{5}{*}{Allowed}} &
			\includegraphics[scale=0.15]{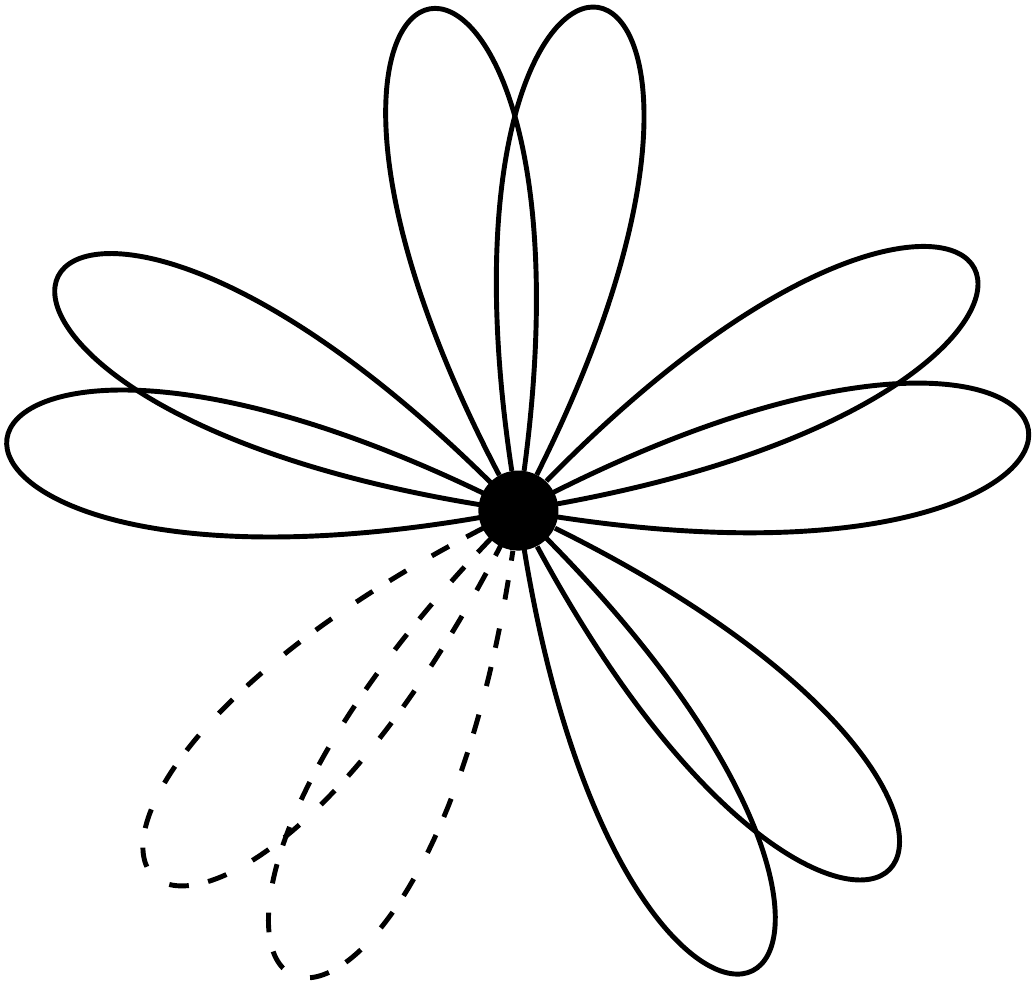} &
			\includegraphics[scale=0.1]{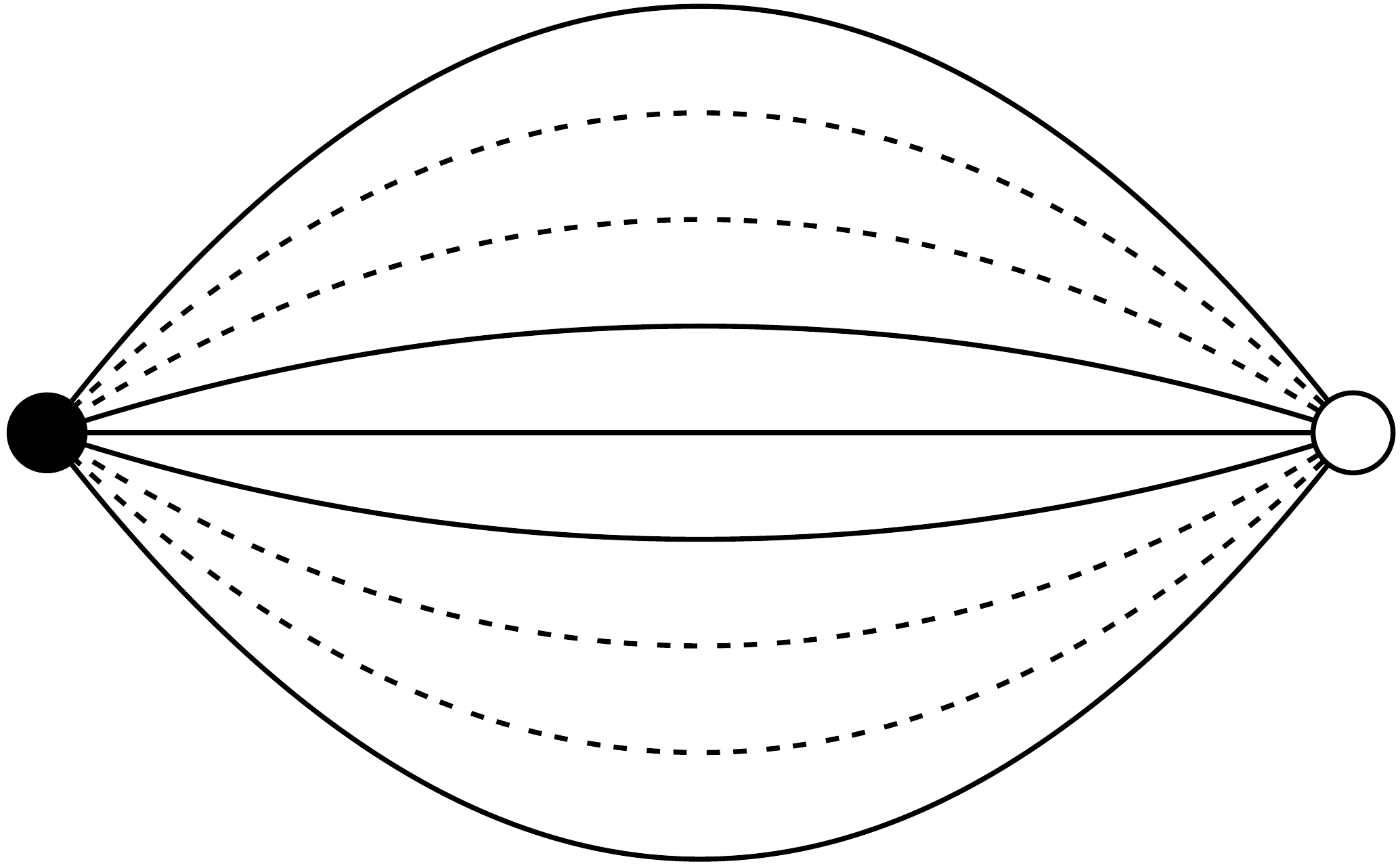} &
			\multirow{5}{*}{\ldots} & \multirow{5}{*}{?} \\
			\multicolumn{1}{|l|}{} & \multicolumn{1}{l|}{\multirow{5}{*}{closed walks}} & + & + & & \\
			\multicolumn{1}{|l|}{} & \multicolumn{1}{l|}{and decorations} & \includegraphics[scale=0.15]{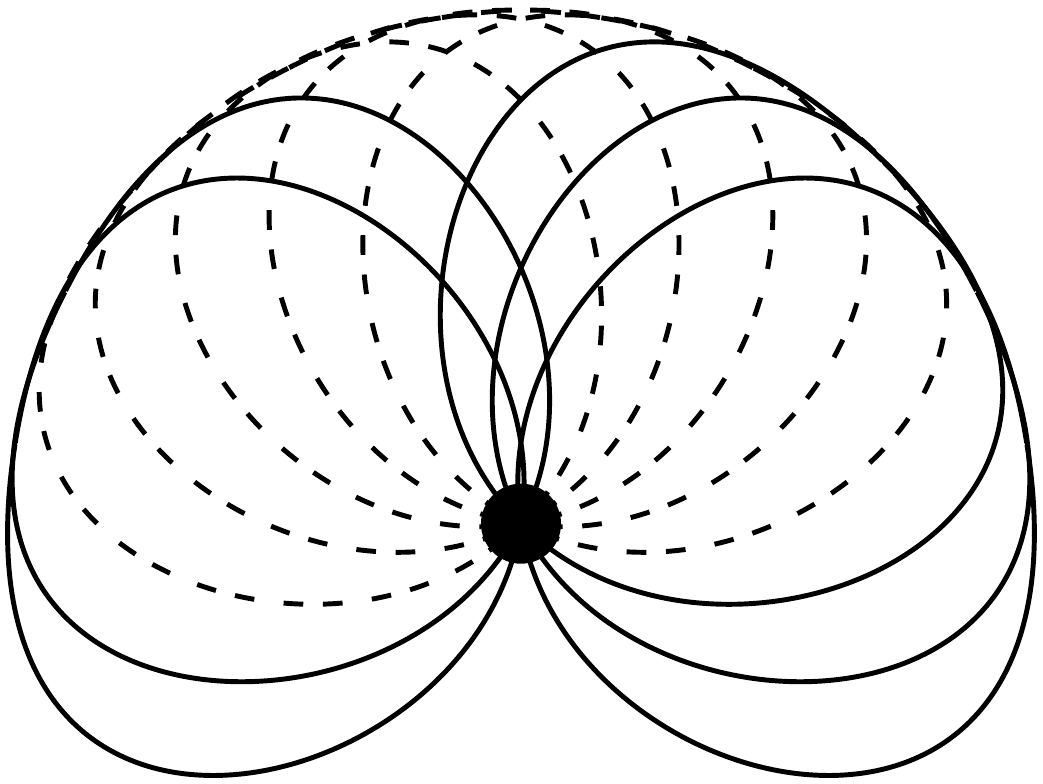} & \multirow{3}{*}[1em]{\vdots} & & \\
			\multicolumn{1}{|l|}{} & \multicolumn{1}{l|}{} & + &  & & \\
			\multicolumn{1}{|l|}{} & \multicolumn{1}{l|}{} & \vdots &  & & \\
			\mr
			\multicolumn{1}{|l|}{ \multirow{15}{*}{\rotatebox{90}{Fundamental polygon}}}& \multicolumn{1}{l|}{Number of} &
			\multirow{2}{*}{$\geqslant 2^{\rm c}$} & \multirow{2}{*}{$\geqslant 1$} & May not & \multirow{2}{*}{$\geqslant 8^{\rm d}$} \\
			\multicolumn{1}{|l|}{} & \multicolumn{1}{l|}{possible pairings} &  &  & exist$^{\rm b}$ & \\
			\multicolumn{1}{|l|}{}& & & & & \\
			\multicolumn{1}{|l|}{} & \multicolumn{1}{l|}{\multirow{5}{*}{Pairing families}} &
			\includegraphics[scale=0.15]{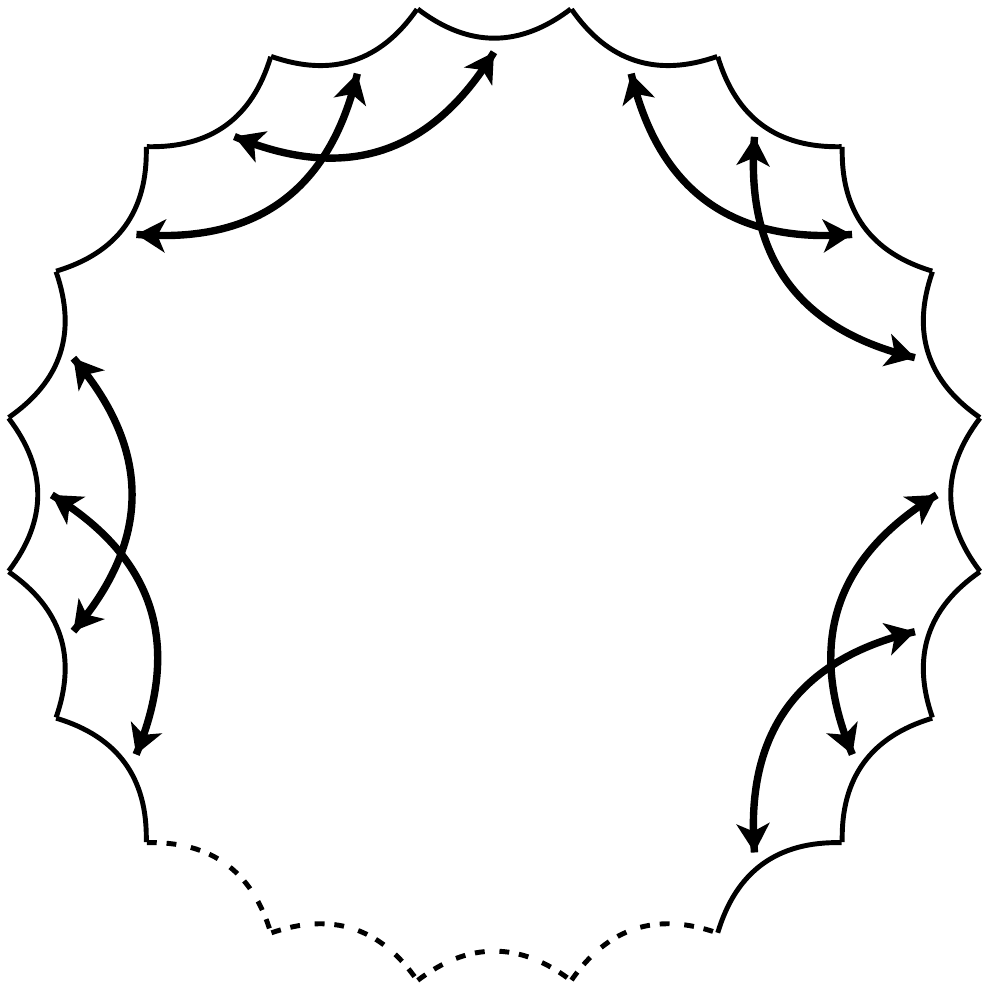} &
			\includegraphics[scale=0.15]{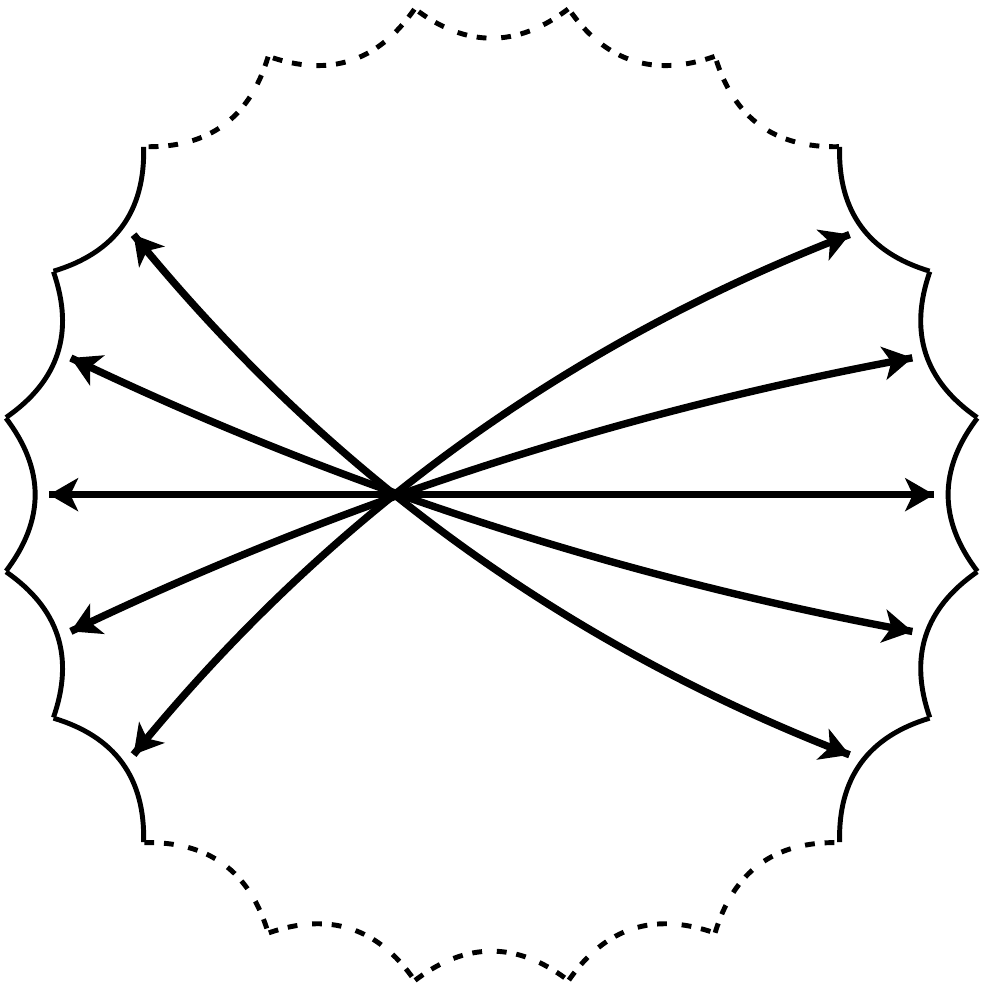} &
			\multirow{5}{*}{\ldots} & \multirow{5}{*}{?} \\
			\multicolumn{1}{|l|}{} & \multicolumn{1}{l|}{} & + & + & & \\
			\multicolumn{1}{|l|}{} & \multicolumn{1}{l|}{} & \includegraphics[scale=0.15]{Figures/Repliement-14-gone-simple-generique-tab} & \multirow{3}{*}[1em]{\vdots} & & \\
			\multicolumn{1}{|l|}{} & \multicolumn{1}{l|}{} & + &  & & \\
			\multicolumn{1}{|l|}{} & \multicolumn{1}{l|}{} & \vdots &  & & \\
			\br
		\end{tabular}

\noindent $^{\rm a}$	The Euclidean case is recovered by setting $g=1$.

\noindent $^{\rm b}$ The coordinence $q$ of the vertices must be an integer and satisfy $q=2\left( 1+(2g-1)/v \right)$, where $v$ is the number of vertices of the graph. If $q$ is not an integer, the corresponding polygon and the associated tiling are not regular, which is a case not studied in this paper.

\noindent $^{\rm c}$ Except in the Euclidean case where the number is equal to 1.

\noindent $^{\rm d}$ The case $g=2$ (see \cite{Jorgensen:1982}) is expected to provide a lower bound.
\end{table}

\subsection{Choosing the proper periodic boundary conditions}

Having sketched a classification of the possible periodic boundary conditions in the hyperbolic plane (see \tref{tab:classification}), we now address the way to build periodic boundary conditions adapted to a physical problem and, therefore, satisfying some prerequisite constraints. For example, the symmetry of the fundamental polygon may matter in some cases, such as the study of ordered, crystalline-like phases on the pseudosphere. The surface area is also an important ingredient, \textit{e.g.}, when envisaging a finite-size analysis to study the scaling behavior of a system near a critical point.

First, we note that due to equation \eref{eq:gauss-bonnet}, the area of a fundamental polygon is a multiple of $4 \pi \kappa^{-2}$, so that, contrary to the Euclidean case in which this area can vary continuously from zero to infinity, accessible areas form an infinite but countable set, \textit{for a given curvature}. Moreover, for two different values of the area, the corresponding fundamental polygons must be different, since the area of a fundamental polygon is fixed by its genus. This sets some constraints on the associated tiling of the hyperbolic plane. Generically, two fundamental polygons with different areas have different shapes and correspond to different tilings of the hyperbolic plane.

\begin{figure}
\begin{center}
	\subfigure[]{\label{fig:rep-14-gone-simple} \includegraphics[scale=0.4]{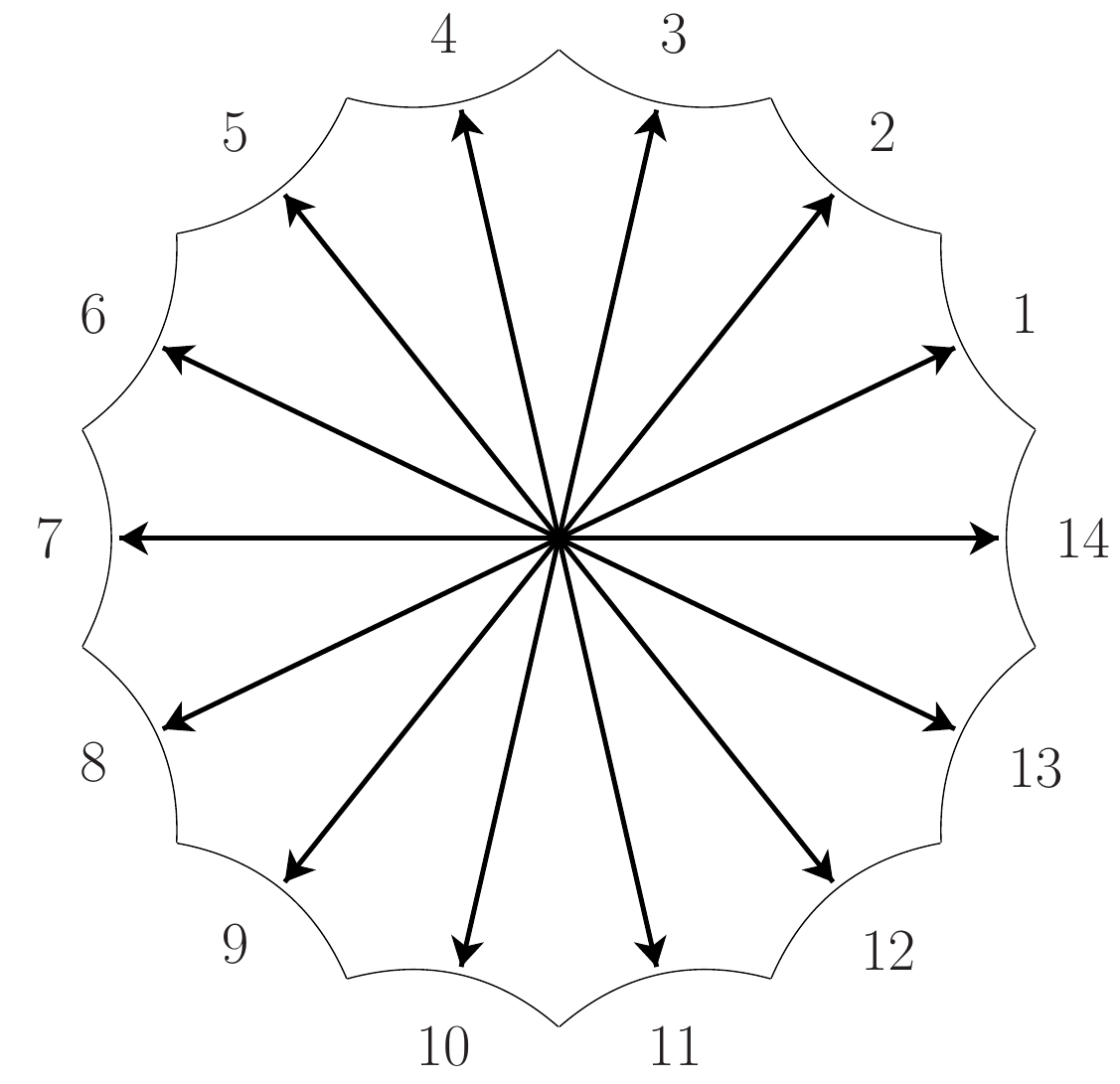}}
	\subfigure[]{\label{fig:tor-14-gone-simple} \includegraphics[scale=0.25]{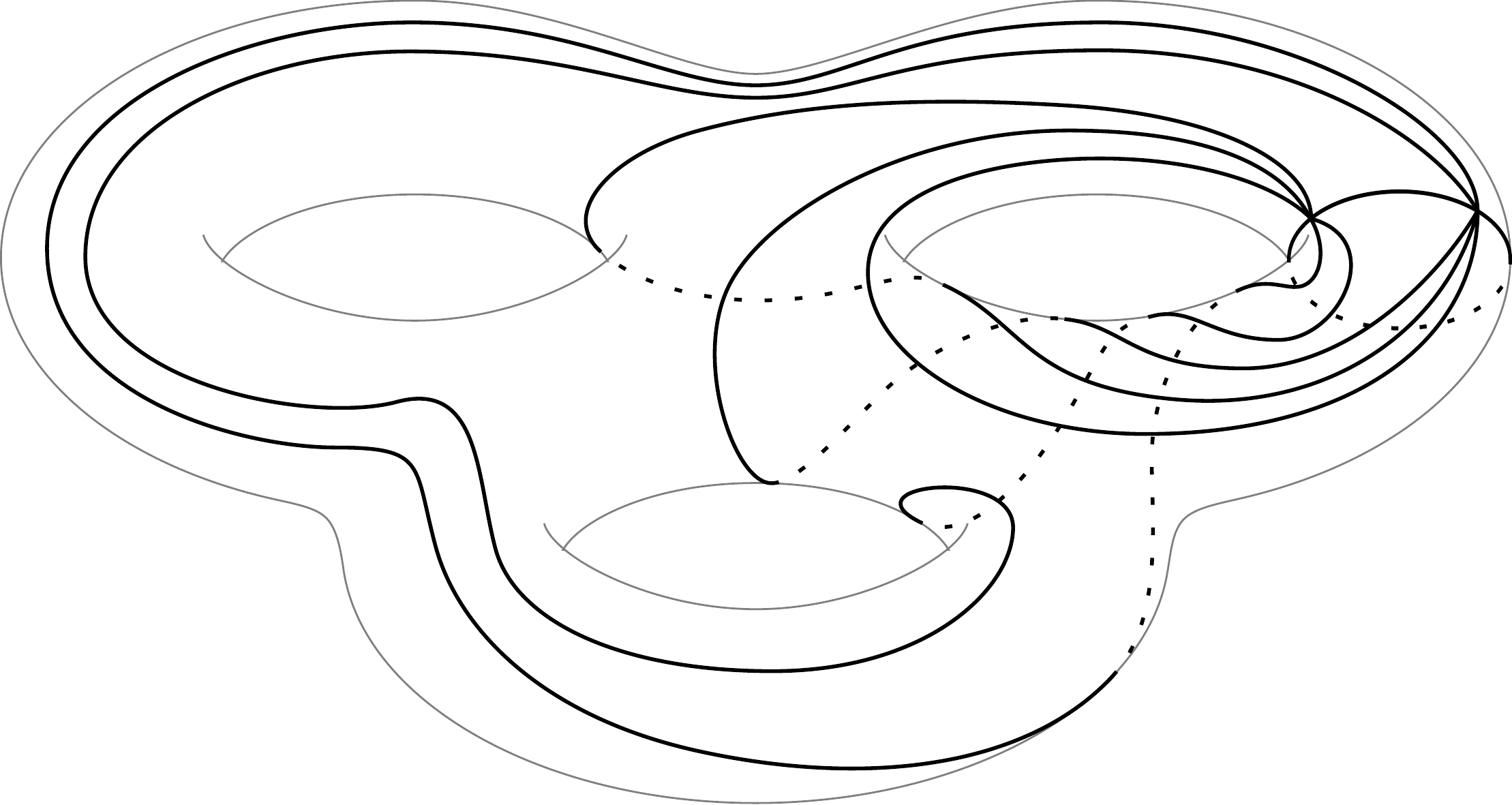}}
	\subfigure[]{\label{fig:graphe-14-gone-simple} \includegraphics[scale=0.45]{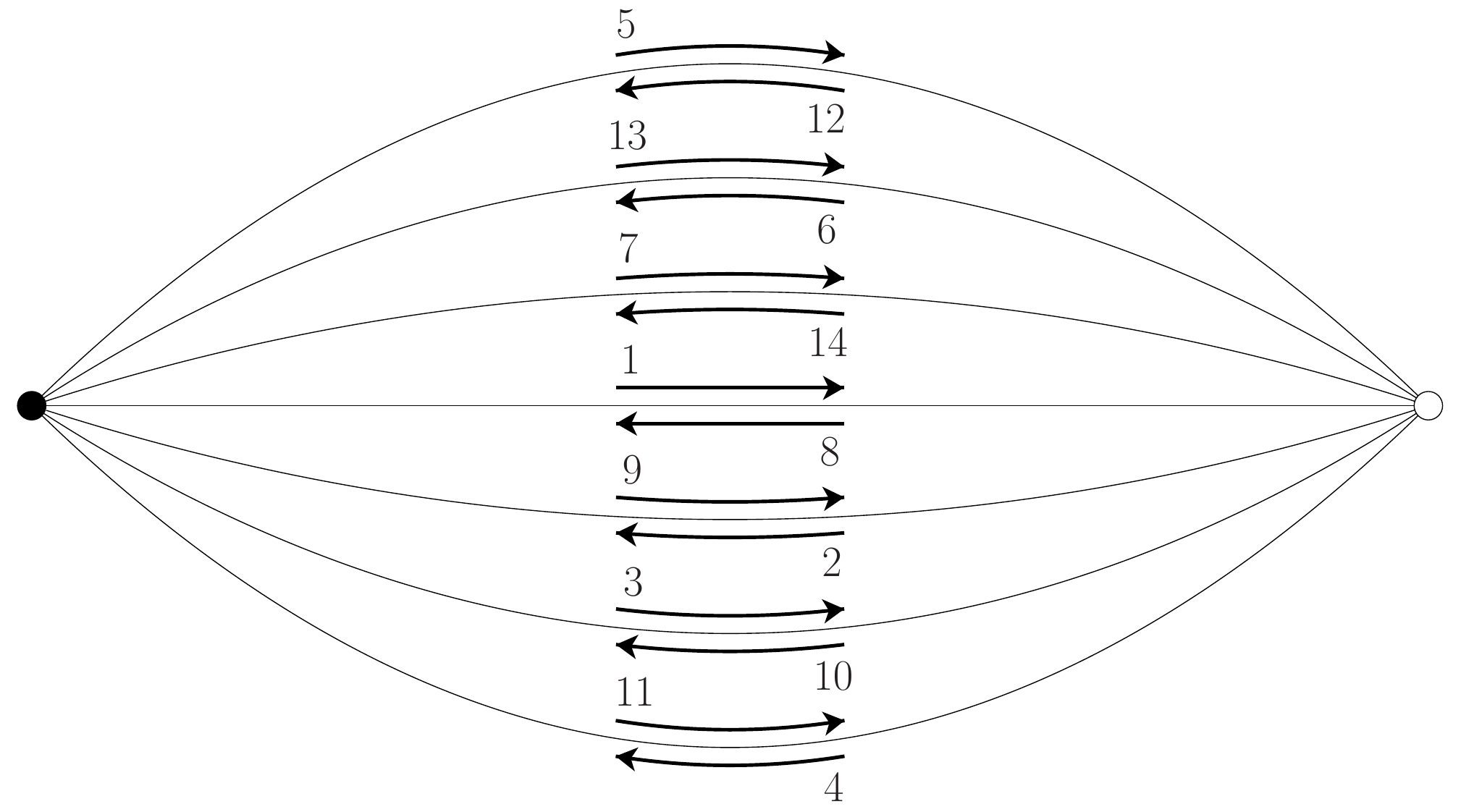}}
\end{center}
\caption{\label{fig:14-gone-simple}One possible periodic boundary condition in $H^2$ with $g=3$ and $2N=14$ leading to the $\{14,7\}$ tiling: (a) the fundamental polygon with the side pairing, (b) associated compact quotient space (three-hole torus) with the embedded graph, and (c) decorated graph with closed walk.}
\end{figure}

\begin{figure}
\begin{center}
	\subfigure[]{\label{fig:rep-14-gone} \includegraphics[scale=0.4]{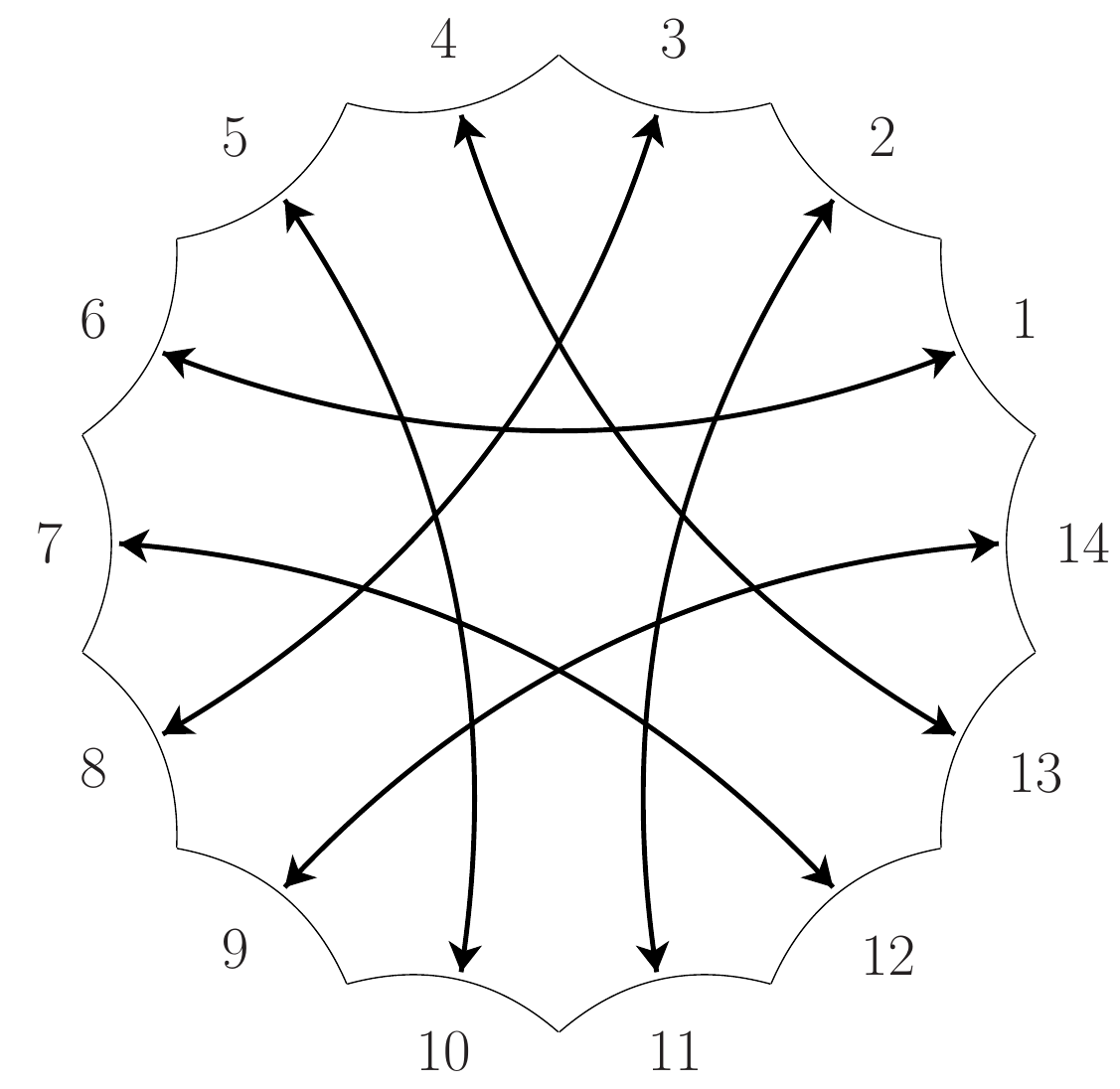}}
	\subfigure[]{\label{fig:tor-14-gone} \includegraphics[scale=0.25]{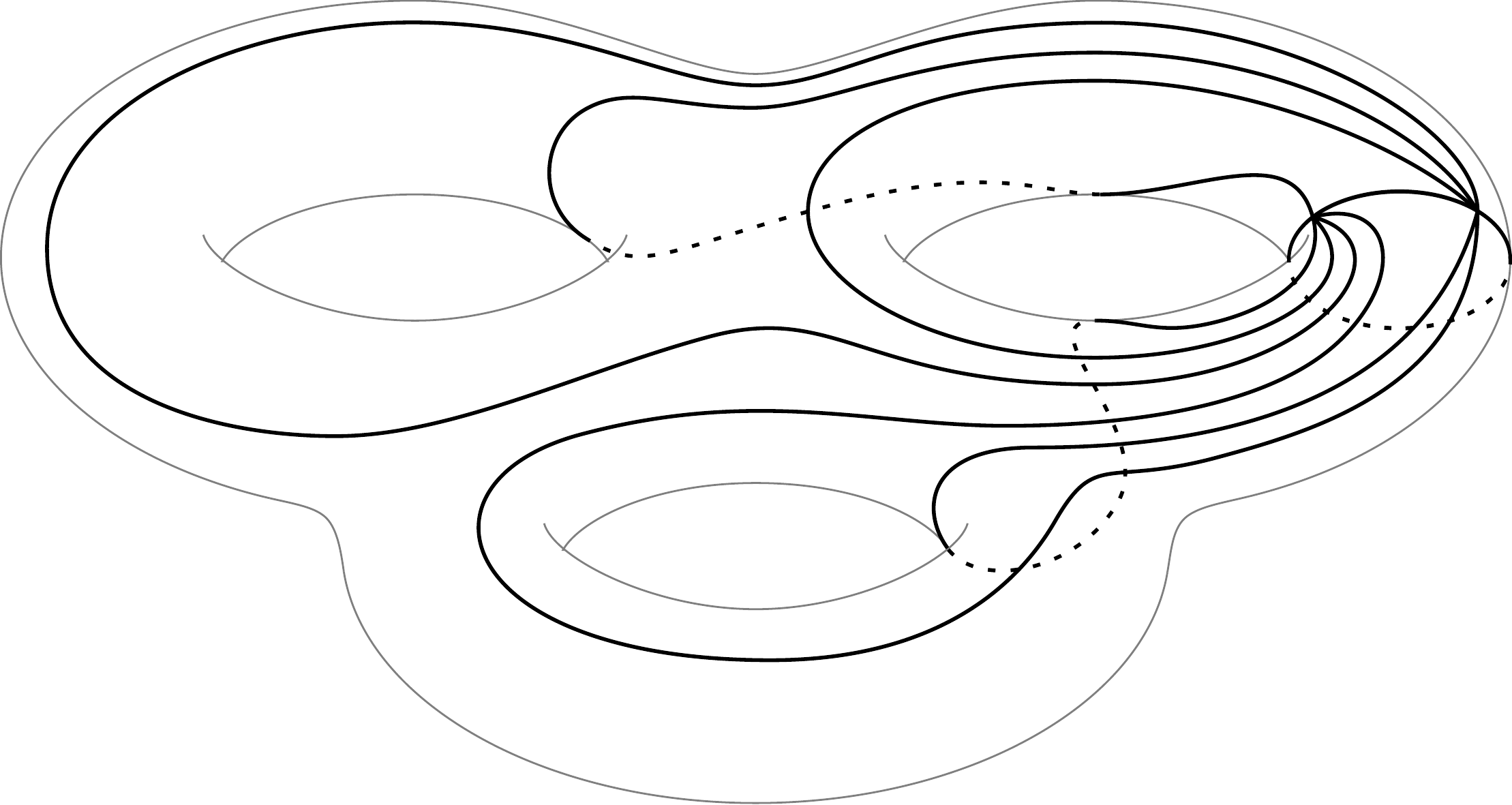}}
	\subfigure[]{\label{fig:graphe-14-gone} \includegraphics[scale=0.45]{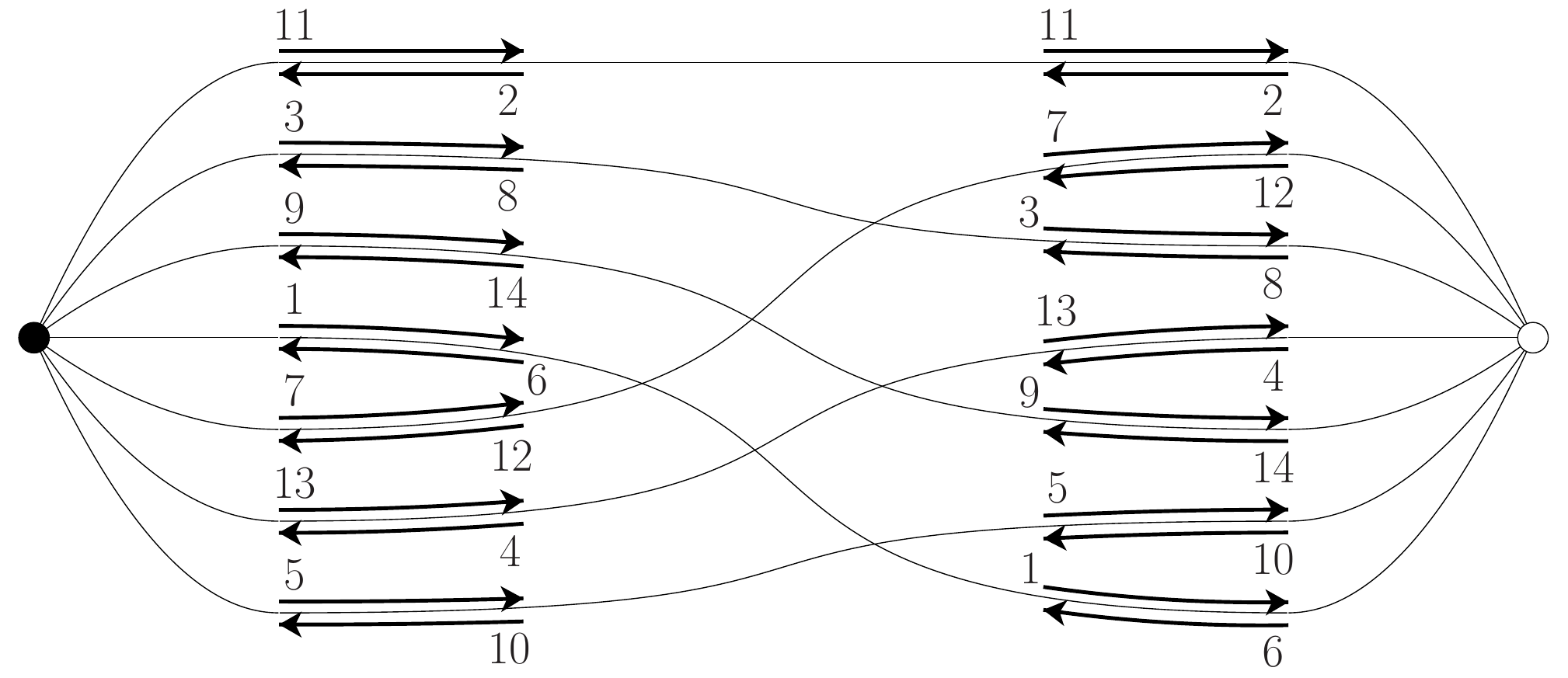}}
\end{center}
\caption{\label{fig:14-gone}Same as \fref{fig:14-gone-simple} for another periodic boundary condition scheme associated to the $\{14,7\}$ tiling. Only the edge pairing patterns differ between the cases shown in \fref{fig:14-gone-simple} and \fref{fig:14-gone}.}
\end{figure}

A fundamental polygon is characterized (in part) by its genus $g$ and the number of sides $2N$ (recall that we only consider regular metric fundamental polygons). Each couple $\{g,N\}$ exactly determines one tiling, $\{p,q\}$, of the hyperbolic plane, with
\begin{equation*}
	\{p,q\}=\left \{2N,\frac{2N}{N-2g+1} \right \}.
\end{equation*}
However, different (purely hyperbolic) Fuchsian groups $\Gamma$ of the hyperbolic plane $H^{2}$ can produce the same tiling. This is why for the same tiling and the same shape of the metric fundamental polygons, the side pairings may still differ. Consider for example a fundamental polygon with $g=3$ and $2N=14$. Different side pairings are possible. The simplest one is shown in \fref{fig:rep-14-gone-simple}, whereas a more intricate one, first described in \cite{Klein:1890}, is displayed in \fref{fig:rep-14-gone}. Yet, the two fundamental polygons are the primitive cell of the same $\{14,7\}$ tiling of $H^2$.

In summary, to build appropriate periodic boundary conditions on the pseudosphere (hyperbolic plane), one has to choose:
\begin{itemize}
	\item[-] the area of the fundamental polygon, which for a given curvature fixes its genus (and that of the associated quotient space),
	\item[-] the number of sides of the fundamental polygon, which for a given genus fixes the associated tiling (we only consider regular metric fundamental polygons and regular tilings),
	\item[-] the pairing of the sides, which can be achieved by using the graph formalism detailed above.
\end{itemize}

From a physical point of view, the choice of a pairing among all the possible ones is not obvious. It might depend on relevant symmetries of the problem, but how side pairing affects the physics of a system is still an open question for us. On one hand, once periodic boundary conditions have been imposed, the boundaries of the primitive cell have no physical meaning. On the other hand, different side pairings correspond to different ways of exploring a given tiling.

\section{Examples of applications}

We conclude this article by illustrating how to implement the preceding developments and build proper periodic boundary conditions in the case of two different statistical mechanical models defined on the pseudosphere (hyperbolic plane). As far as we know, this has never been achieved before.

\subsection{The dynamics of particles on the pseudosphere}

\begin{figure}
\begin{center}
	\subfigure[]{\includegraphics[scale=0.1]{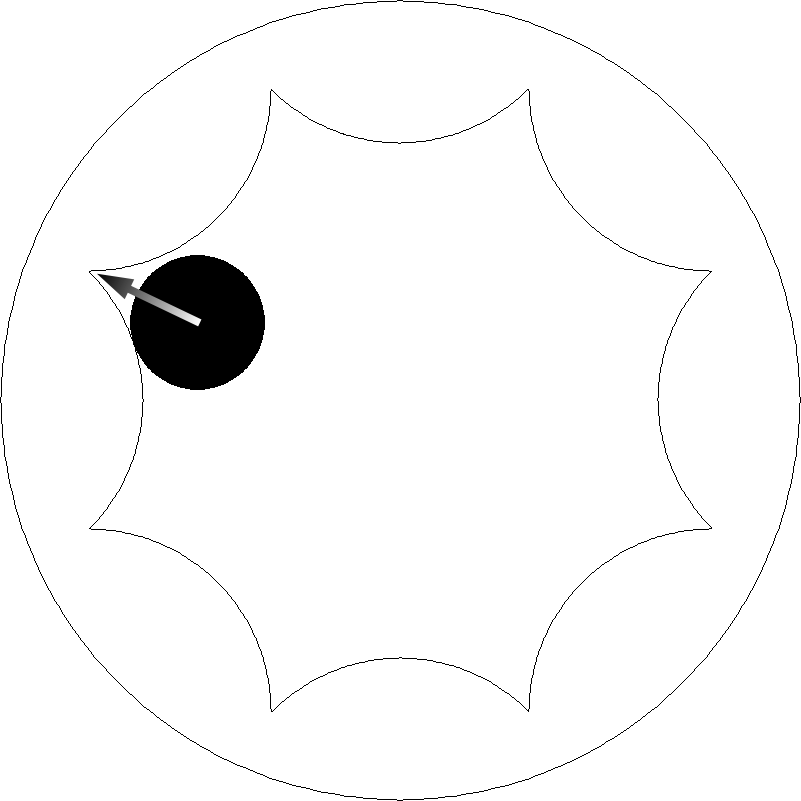}}
	\subfigure[]{\includegraphics[scale=0.1]{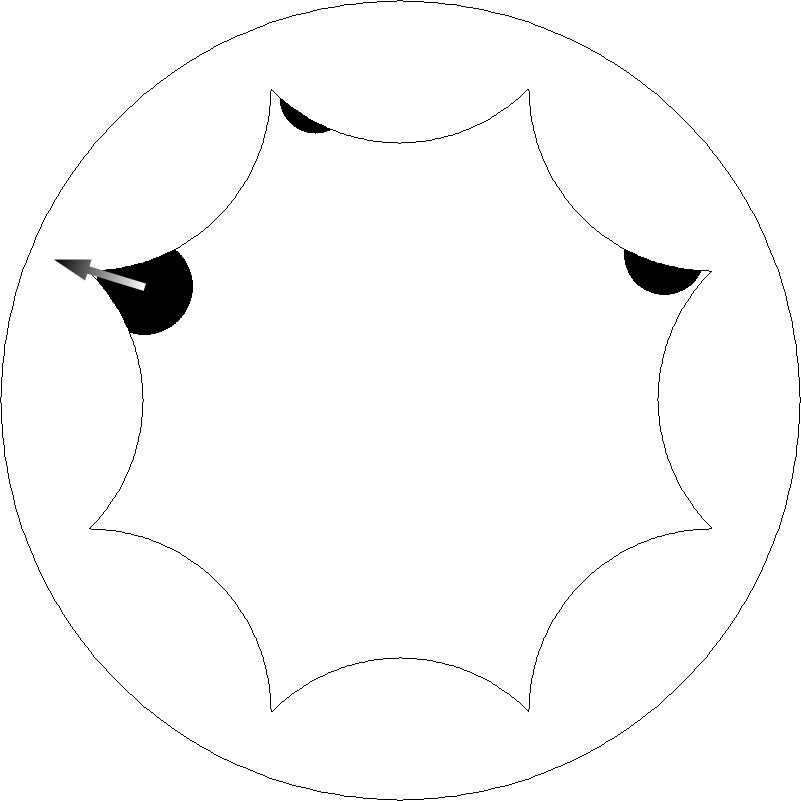}}
	\subfigure[]{\includegraphics[scale=0.1]{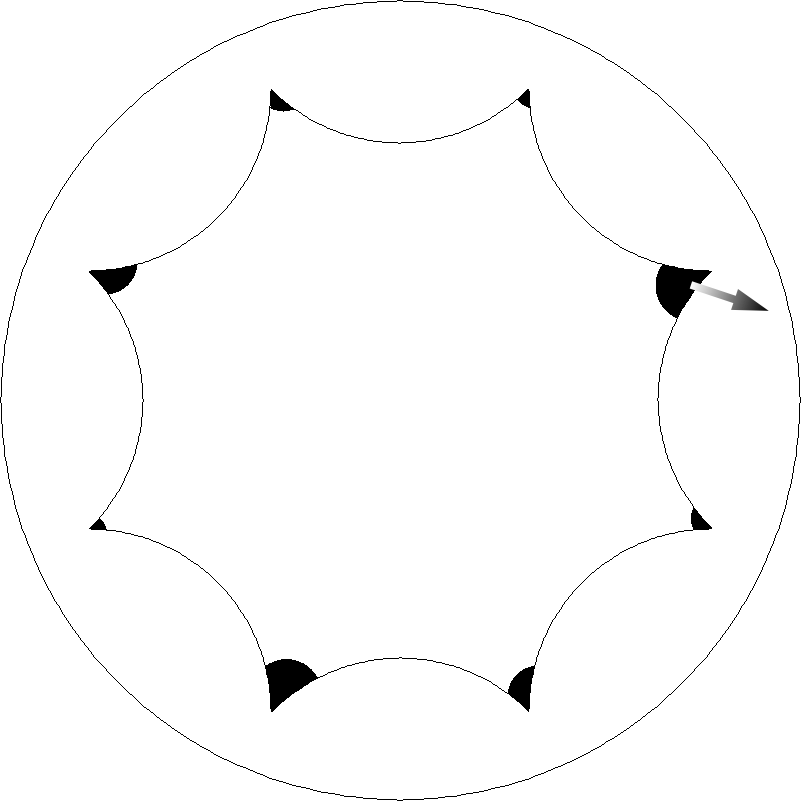}}
	\subfigure[]{\includegraphics[scale=0.1]{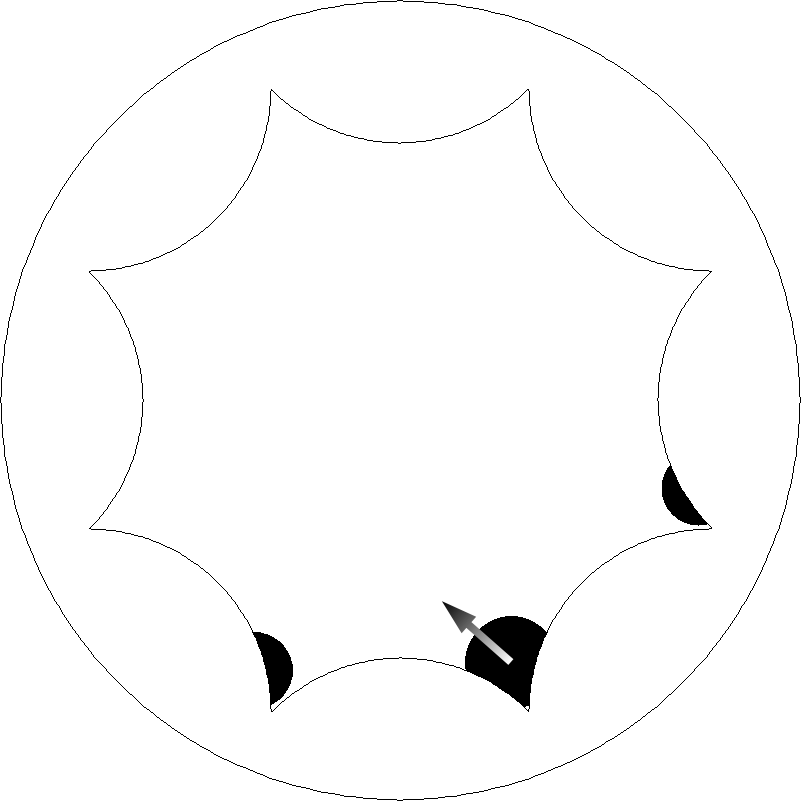}}
	\subfigure[]{\includegraphics[scale=0.1]{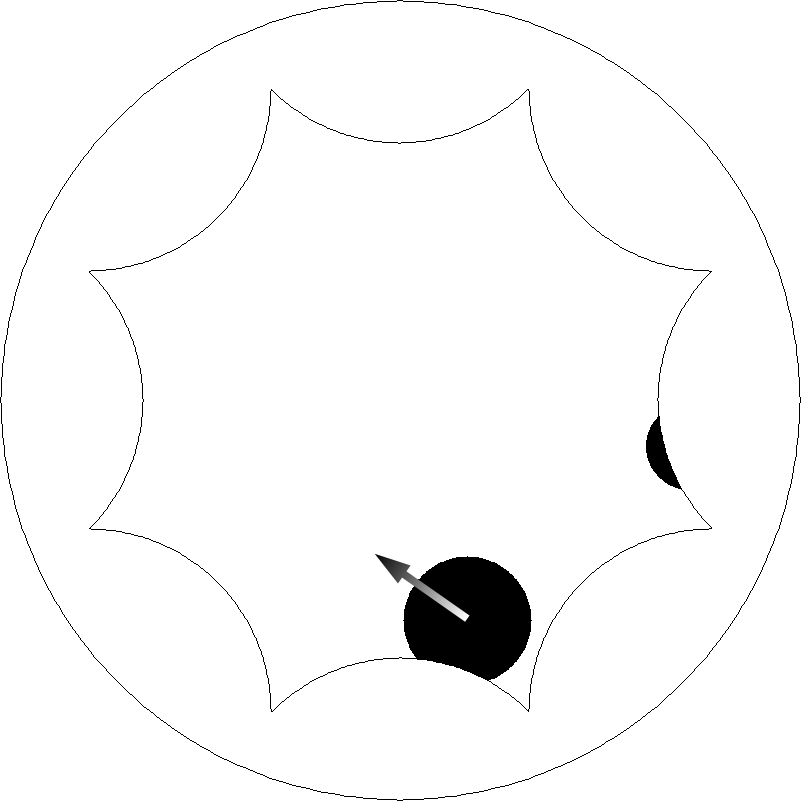}}
	\subfigure[]{\includegraphics[scale=0.1]{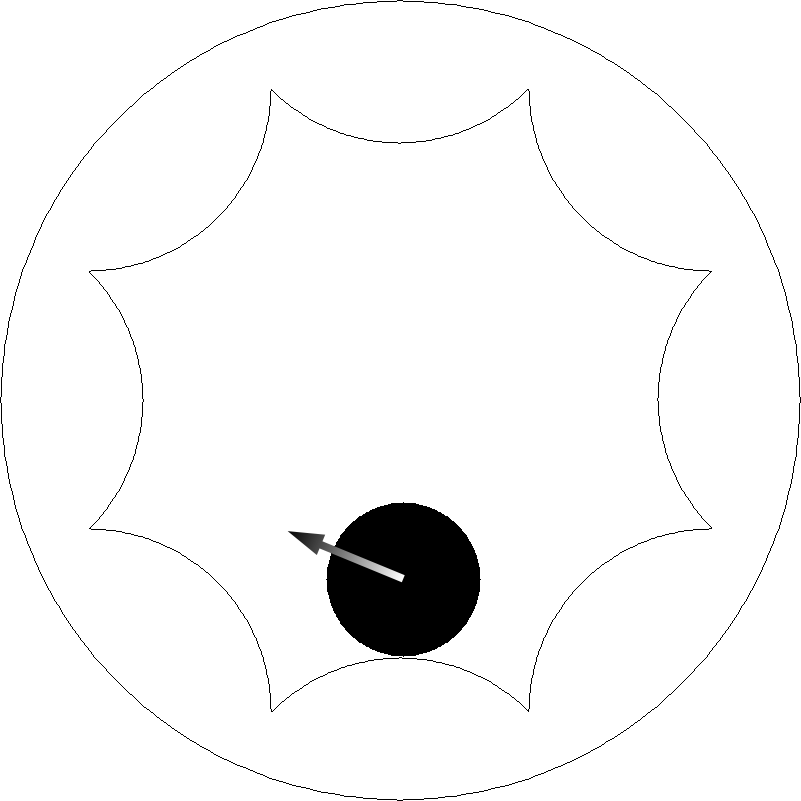}}
	\subfigure[]{\includegraphics[scale=0.1]{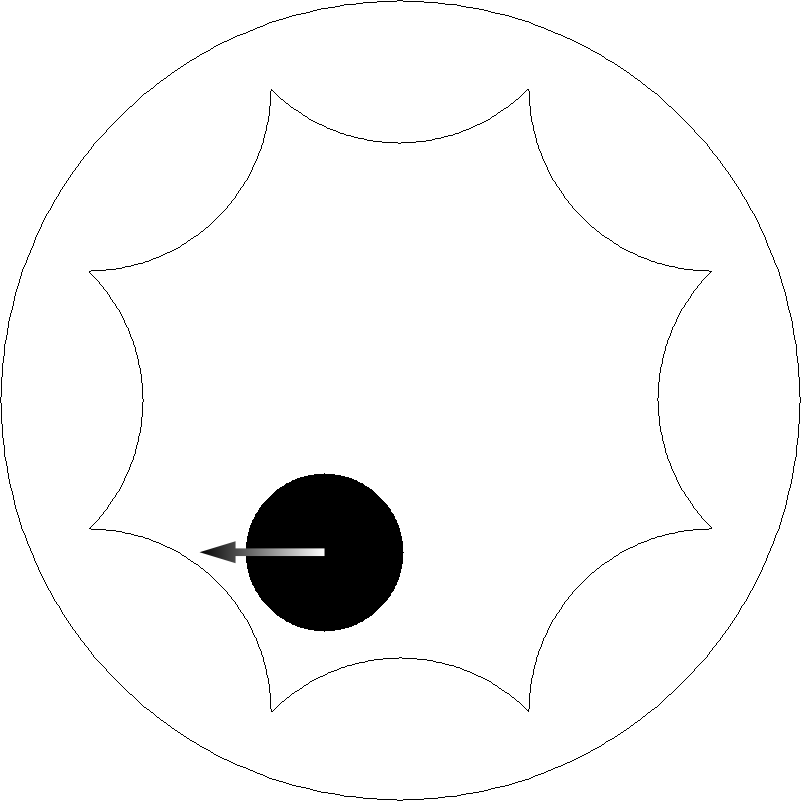}}
	\subfigure[]{\includegraphics[scale=0.1]{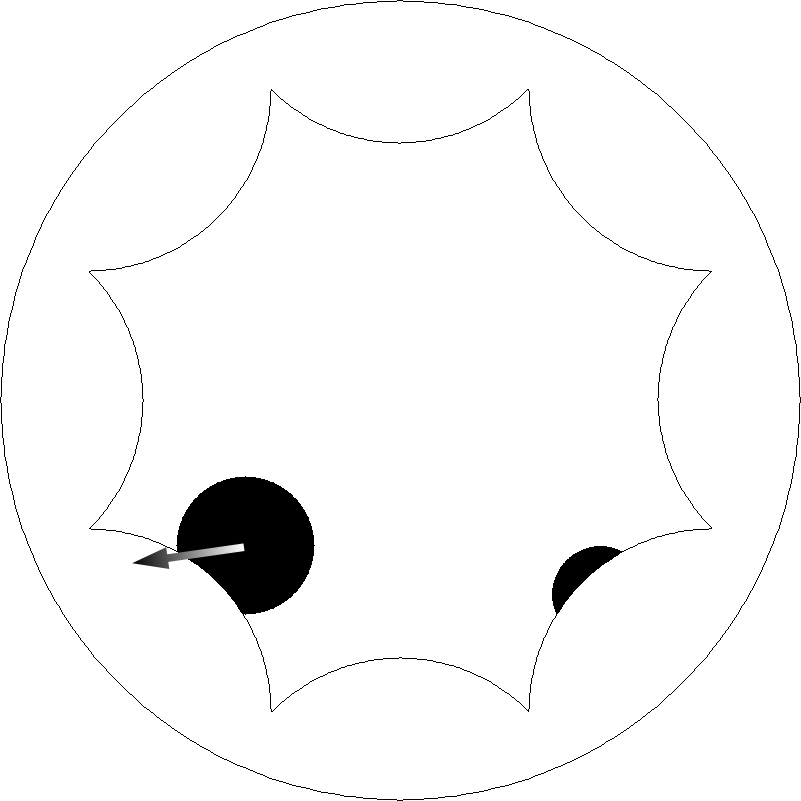}}
	\subfigure[]{\includegraphics[scale=0.1]{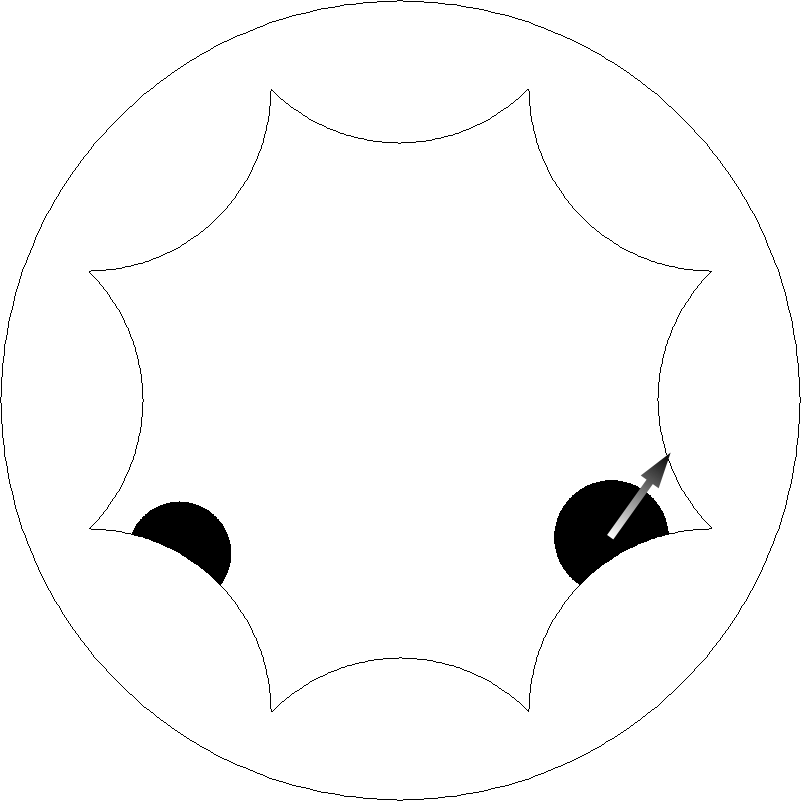}}
	\subfigure[]{\includegraphics[scale=0.1]{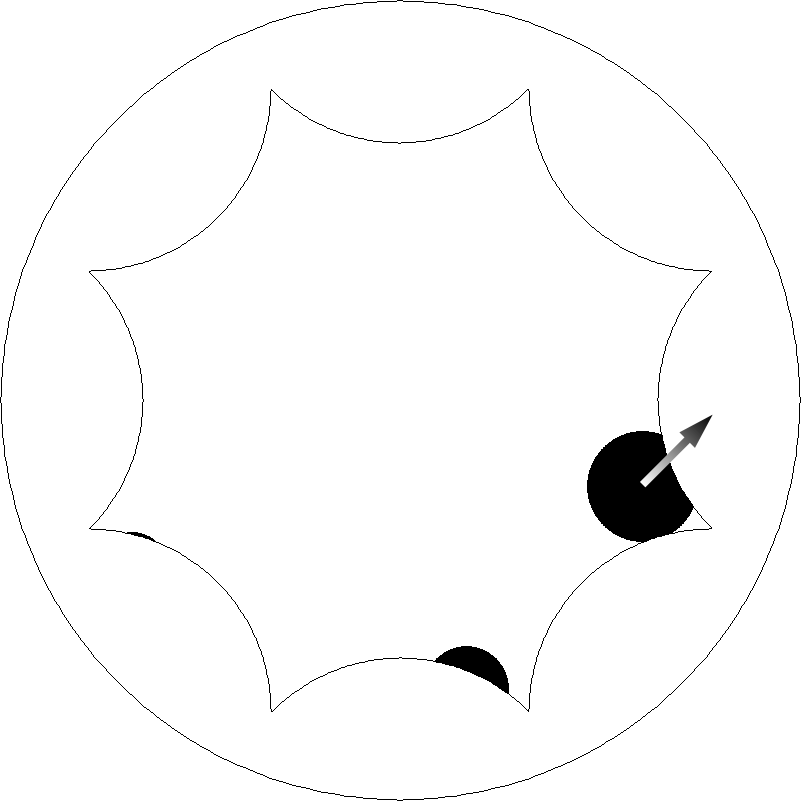}}
	\subfigure[]{\includegraphics[scale=0.1]{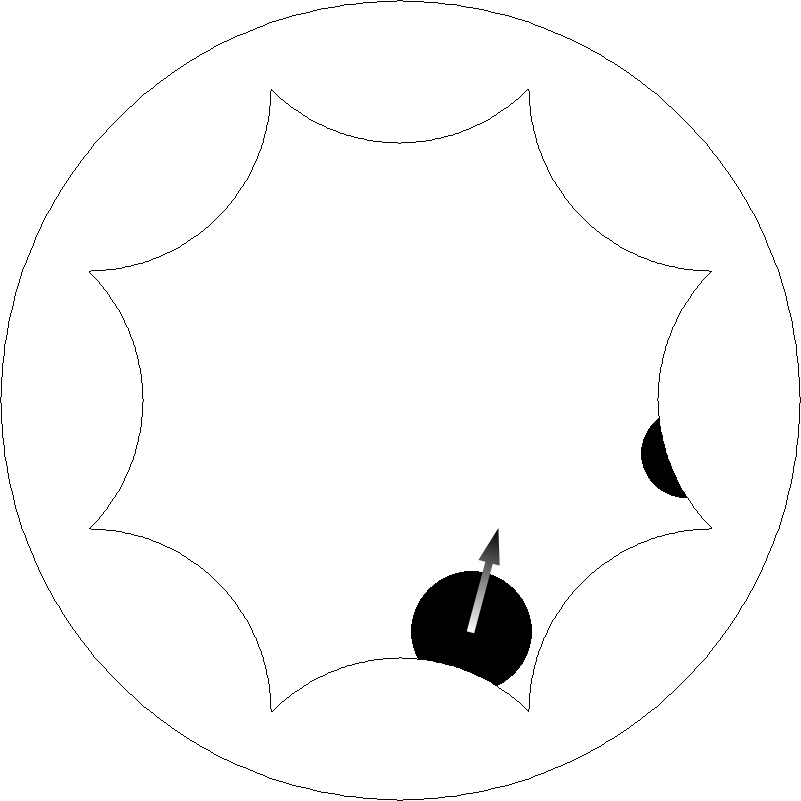}}
	\subfigure[]{\includegraphics[scale=0.1]{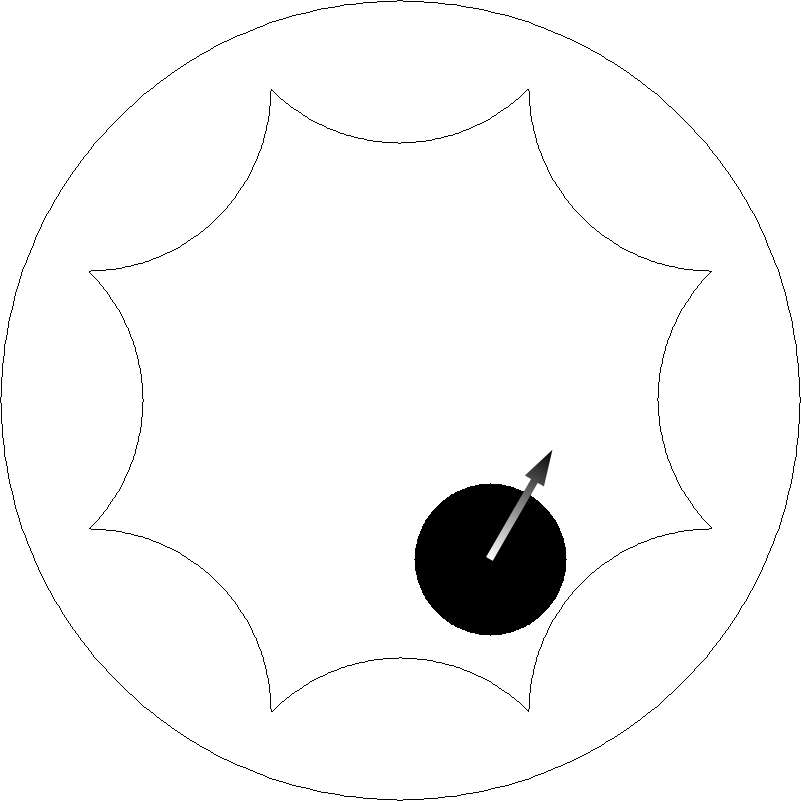}}
\end{center}
\caption{\label{fig:free-particle}Successive snapshots of the trajectory of a single disk-like particle freely evolving on the pseudosphere in the Poincaré representation, with the periodic boundary condition shown in \fref{fig:rep-8}. The trajectory follows geodesics, {\em i.e.}, circular arcs and the arrow indicates the direction of motion. Note the complex pattern of exits and reentries. The fact that the particle is not considered as a mere point but has a finite size makes the pattern even more intricate.}
\end{figure}

The first example concerns the dynamics of particles on the pseudosphere. The starting point for implementing periodic boundary conditions is to consider the free motion of one particle. We pick a simple example in which the primitive cell of the periodic boundary condition is an octagon (genus $g=2$). The trajectory of the particle is represented in the Poincaré disk model of $H^2$ in \fref{fig:free-particle}. Unlike in the Euclidean case, this trajectory is not rectilinear. The particle under free motion follows geodesics, which are arcs of circles in this representation of the hyperbolic plane (see \ref{ap:geo}). One can see in \fref{fig:rep-8} the complexity of the trajectory that results from the pairing of the octagon sides and the resulting ``folding" or ``compactification'' of space. (Alternatively, one may consider that the figure displays the trajectories of the particle and of all its images in the tessellation of $H^2$; the two points of view are strictly equivalent.) A detailed account of the free motion of a mass point on the pseudosphere with octagonal periodic boundary conditions has been given in \cite{Balazs:1986}.

Because we intend to generalize the study to assemblies of interacting particles, we do not simply consider a mass point, but a particle with a given shape (here, a disk) and a finite size (here, a radius $r_0$). Note that the apparent size of the particle depends on the distance to the center of the Poincaré disk. This stresses that the metric used in this representation is conformal (it preserves angles), but not isometric (it does not preserve lengths, which seem to contract when approaching the disk boundary): see \ref{ap:geo}.

Let us detail a little more the practical implementation of the periodic boundary conditions for a such system. First, since no force acts on it, the particle follows a geodesic at a constant velocity. In this representation, geodesics are arcs of circles which meet the disk boundary at a right angle (see \ref{ap:geo}), so that the trajectories of the particle are made of such arcs of circles. When the particle crosses a side $i$ of the octagon, the generator (a hyperbolic translation) $\gamma_i$ linking this side to its paired one, $\gamma_i(i)$, is applied to the particle. If the particle ends up in the interior of the octagon, the side crossing is achieved and the particle follows another geodesic. But, if after this first step, the particle is still outside the octagon, one has to iterate the preceding step until the particle reaches the interior of the octagon. In this case, each new step uses the generator $\gamma_{\rho(i)}$, where $\rho$ is the cyclic permutation $(1,2,\ldots,2N)$ of order $2N$. Here $N=4$ as we use an octagon. Note that this procedure is valid only for small displacements outside the polygon, typically less than the polygon side length, and that the maximum number of steps is $2N-1$.

To plot the neighborhood of the particle center, \textit{e.g.}, the disk of radius $r_0$ representing the whole particle, one duplicates the particle $2N-1$ times by using the preceding procedure when the disk (shown as black disk in \fref{fig:free-particle}) goes outside the polygon. This procedure will be necessary to study systems made of many moving and interacting particles.

Implementing such periodic boundary conditions is essential to study the statistical mechanics of assemblies of interacting particles on the pseudosphere by Molecular Dynamics or Monte Carlo simulation. The physical problem under investigation may be crystallization and glass formation of atomic liquids in hyperbolic geometry \cite{Nelson:1983,Nelson:2002,Rubinstein:1983}, with particles modeled as disks interacting with simple pair potentials such as a hard-core interaction or a Lennard-Jones interaction \cite{Hansen:1986}. It may also be the behavior of Coulomb systems living on a pseudosphere \cite{Jancovici:1998,Fantoni:2003,Jancovici:2004}, such as the one- and two-component plasmas with interactions given by the Green's function associated with the Laplace-Beltrami operator. In the latter case, and as shown in \cite{Callan:1990,Nelson:1983}, the curvature introduces an ``infrared cutoff" which screens the coulombic interactions between charged particles. As a result, the interaction is no longer logarithmic at large distance as it is on the Euclidean plane, but rather decays exponentially for distances larger than the radius of curvature $\kappa^{-1}$. A physical consequence of this curvature-induced screening is that the Kosterlitz-Thouless phase transition in which topological defects (vortices) all pair in dipoles is expected to be depressed to zero temperature. A technical consequence is that periodic boundary conditions can be used without recourse to specific methods such as the Ewald summation \cite{Frenkel:1996} for handling the coulombic interactions.

In all the above cases, one is interested in studying the influence of curvature on the behavior of the system. The area of the primitive cell in which is embedded the physical system is then fixed by the genus of the fundamental polygon. However, there is an additional length in the problem, which is the particle size, \textit{e.g.}, the radius $r_0$ of the disk formed by an atom. Working at fixed curvature means here keeping the product $\kappa r_0$ fixed. Varying the system size at constant curvature (when both the area and the curvature are made dimensionless by  relating them to the size of one particle) implies to change the genus and the fundamental polygon. Keeping the dimensionless particle density (\textit{i.e.}, the ``surface coverage") constant then requires to change the total number of particles in the primitive cell. We will not further describe how to implement in practice periodic boundary conditions coupled to a Molecular Dynamics or Monte Carlo algorithm. This will be detailed in a forthcoming publication \cite{Sausset:2007}.

\subsection{Spin models on hyperbolic lattices}

We finally consider models in which spins are fixed on the vertices of a regular lattice. The lattices correspond to  regular tilings of space, tilings which of course differ for the Euclidean and the hyperbolic planes. Taking as an example of statistical mechanical spin system the Ising model, the Hamiltonian on a hyperbolic lattice is given by
\begin{equation*}
\mathcal{H}=-J\sum_{\langle i,j\rangle }\sigma_i\sigma_j ,
\end{equation*}
where $\sigma_i=\pm 1$ and the sum runs on all distinct pairs of nearest-neighbor sites, \textit{i.e.}, vertices, on the chosen lattice.

The hyperbolic lattice Ising model has been studied in \cite{Rietman:1992,Wu:1996,Wu:2000,Angles-dAuriac:2001,Shima:2006a,Shima:2006} (a continuum field theoretical version has also been considered: see \cite{Doyon:2004}). We have already stressed in the introduction that when considering a finite-size system, surface effects do not become negligible compared to bulk effects when the size of the system increases. Indeed, the number of lattice sites around a given site grows exponentially with the distance, so that the contribution to any thermodynamic quantity of the sites at the boundary of a finite, open system is of the same order as that of the inside sites, even in the thermodynamic limit. Surface effects may be interesting \textit{per se} \cite{Angles-dAuriac:2001}, but one is nonetheless interested in ``bulk" behavior, that of the ``deep interior" of the system, because it is more directly comparable to the behavior in the absence of curvature. Until now, numerical simulation studies have considered finite systems with a free surface and studied how properties of the system evolve when neglecting an increasing amount of spins in the vicinity of the surface \cite{Angles-dAuriac:2001,Shima:2006a,Shima:2006}. It would seem more convenient to study the bulk behavior by using periodic boundary conditions, as such conditions minimize surface effects and more directly mimic the behavior of an infinite system.

Choosing the appropriate periodic boundary conditions compatible with a given hyperbolic lattice is however a highly nontrivial task. First, each periodic boundary condition corresponds to a given Fuchsian group, and so does a tiling (hence, a lattice), so that the Fuchsian group of the lattice and that of the periodic boundary condition must be compatible. Before giving a definition of what ``compatible" means, it is useful, as before, to revisit the Euclidean case. Consider the square lattice with lattice spacing $a$. The associated discrete group of isometries has for generators the two translations $T_{a \vec{x}}$ and $T_{a \vec{y}}$. It is clear in this case that one can choose as a primitive cell any rectangle of sides $m\times a$ and $n\times a$ with $m$ and $n$ two nonzero integers, or restricting ourselves to a regular cell, any square of side $n\times a$. The corresponding tessellation is also a $\{4,4\}$ square tiling, and the discrete group of isometries is generated by the two translations $T_{na \vec{x}}$ and $T_{na \vec{y}}$. This group is a subgroup of the group associated with the original square lattice of spacing $a$: the two groups are compatible. In the hyperbolic case, this has to be generalized in the following way: the Fuchsian group associated with the fundamental polygon corresponding to the periodic boundary conditions must be a \textit{normal subgroup} of the lattice (Fuchsian) group \cite{Coxeter:1965,Magnus:1974,Yuncken:2003}. We recall that a subgroup $\Gamma$ of a group $\Sigma$ is said ``normal" if it is invariant under conjugation, which means that for each element $\gamma \in \Gamma$ and each $\sigma \in \Sigma$, $\sigma\gamma\sigma^{-1}\in\Gamma$. (Normal subgroups are closely related to quotient groups, which can indeed be constructed from a given group by using a normal subgroup of this group.)
\begin{figure}
	\includegraphics[scale=0.4]{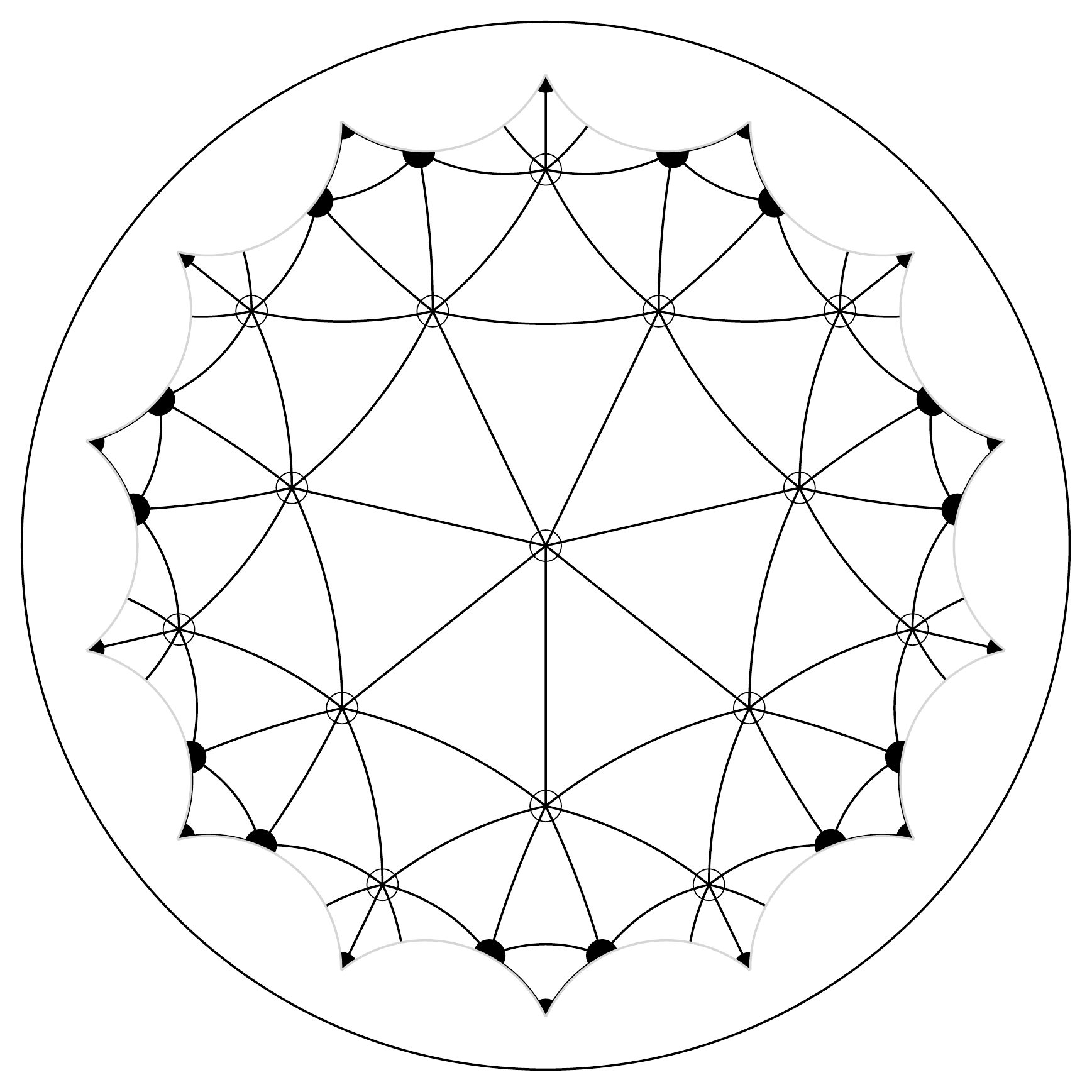}
	\caption{\label{fig:ising}Periodic boundary condition adapted for a spin model on a hyperbolic $\{3,7\}$ lattice in the Poincaré disk representation. The primitive cell in which the physical system is embedded is a 14-gon with the side pairing shown in \fref{fig:rep-14-gone}. Each spin on a vertex interacts with 7 nearest neighbors. The boundary sites shown with black circles are duplicated (see text).}
\end{figure}

For illustrative purpose,we consider the lattice based on the $\{3,7\}$ tiling, which has been used for instance in previous studies of the Ising model \cite{Angles-dAuriac:2001,Shima:2006a,Shima:2006}. The $\{3,7\}$ tiling is quite similar to the Euclidean triangular lattice. The normal subgroup compatible with the $\{3,7\}$ tiling whose fundamental polygon has the smallest genus, and consequently the smallest area, was first described in \cite{Klein:1890}. The metric fundamental polygon is a $14$-gon of genus $g=3$, with the side pairing displayed in \fref{fig:rep-14-gone}. As shown in \fref{fig:ising} in the Poincaré disk representation, it represents the unit cell of a $\{14,7\}$ tiling of $H^2$ and possesses $24$ vertices of the original $\{3,7\}$ tiling.

In \fref{fig:ising}, the spins which are on the boundary (black ones) are duplicated twice or more since they are shared between different sides of the fundamental polygon. To build the lattice, we have  used the procedure detailed in \ref{ap:tilings}. For implementing the periodic boundary conditions, the same procedure has been followed. We have first constructed one cell of the $\{14,7\}$ tiling associated with the normal subgroup of the lattice group. Then, the spins which are on the boundary of this primitive cell have been paired by following the fundamental polygon side pairing detailed in \fref{fig:rep-14-gone}. Note that paired spins are only replicas of the same spin and, in the course of a numerical simulation, will therefore flip together (out of the $28$ boundary spins, only $9$ are independent).

Increasing the system size, namely the size of the primitive cell, implies to choose another normal subgroup, and thus another fundamental polygon. The new subgroup corresponds to a larger value of the genus and to another tiling of $H^2$ (compatible of course with the original $\{3,7\}$ tiling). The enumeration of all possible normal subgroups of a given Fuchsian group is still, as far as we know, an open problem \footnote{Restrictions on the possible normal subgroups of a given Fuchsian group have been found: for instance in \cite{Yuncken:2003}, it has been shown that a $\{p,q\}$ tiling is itself tiled by fundamental polygons of some Fuchsian group if and only if $q$ has a prime divisor less than or equal to $p$.}, but results have been obtained in the case of the $\{3,7\}$ tiling, and the following normal subgroups with larger fundamental polygons than that detailed above are described in \cite{Kulkarni:1985}: their genus are equal to 7, 14, 118, 146, 411, 474, 2131 and 3404, which correspond, respectively, to systems with 56, 112, 944, 1168, 3288, 3792, 17048 and 27232 spins. To further implement the corresponding periodic boundary conditions, one should determine the tiling associated with these subgroups and the proper pairing pattern for boundary spins. Work in this direction is in progress.

\ack
We thank R. Mosseri for helpful discussions.
The LPTMC is CNRS UMR 7600.

\appendix

\section{Hyperbolic geometry} \label{ap:geo}

The hyperbolic plane $H^2$, also called pseudosphere or Bolyai-Lobachevsky plane, is a Riemannian surface of constant Gaussian negative curvature \cite{Coxeter:1961,Hilbert:1983}. Contrary to the sphere which is a surface of constant positive curvature, $H^2$ as a whole cannot be embedded in the 3-dimensional Euclidean space $E^3$. Actually, to embed $H^2$ in $E^3$, the corresponding surface should have a saddle point at every point of space, which is obviously not feasible. The hyperbolic plane is infinite (contrary to the sphere) and homogeneous.

The metric of $H^2$ is quite similar to that of the sphere (hence the name pseudosphere), particularly when expressed in polar coordinates $(r,\phi)$; the distance between two infinitesimally close points is indeed given by
\begin{equation*}
\rmd ^2s=\rmd ^2r+\left (  \frac{\sinh(\kappa r)}{\kappa } \right )^2 \rmd ^2\phi,
\end{equation*}
where $-\kappa^2$ is the negative Gaussian curvature of the plane.

Due to the impossibility of embedding $H^2$ in $E^3$, one has to use models to visualize $H^2$. The representation of $H^2$ which may be more familiar to physicists is the upper sheet of an hyperboloid embedded in $\mathbb{R}^3$ with a Minkowski metric. Other models also exist, like the Poincaré upper half-plane, the Klein model, and the Poincaré disk \cite{Coxeter:1961,Hilbert:1983}. In this work we primarily use the latter model, as it is convenient to visualize tilings of the hyperbolic plane.

The Poincaré disk model maps $H^2$ onto the open unit disk
\begin{equation*}
\Delta=\{ (x,y) \in \mathbb{R}^2, \left|x^2 + y^2 \right| <1\}.
\end{equation*}
In $\Delta$, the metric becomes
\begin{equation*}
\rmd s^2=\kappa^{-2} \frac{4\left(\rmd x^2 +\rmd y^2\right)}{\left(1-\left(x^2 + y^2\right)\right)^2 },
\end{equation*}
or expressed in terms of $z=x+\rmi y \in \mathbb{C}$ and its complex conjugate $\bar{z}$,
\begin{equation*}
\rmd s=\frac{4\;\rmd z\,\rmd \bar{z} }{\left(1-\left|z\right|^2\right)^2 }.
\end{equation*}

One can easily see that lengths in the disk are not conserved with this metric: the Euclidean distance between two points of $\Delta$ separated by a constant distance in $H^2$ shrinks to zero when the points approach the disk boundary. On the other hand, angles are strictly identical in $H^2$ and $\Delta$. This representation is therefore conformal, but not isometric.

The unit circle corresponding to the closure of $\Delta$ represents the set of points at infinity in $H^2$. In the Poincaré disk, geodesics are arcs of circles crossing orthogonally the limit (unit) circle.

The distance $\rho$ between two points $z_1$ and $z_2$ is given by
\begin{equation*}
\rho(z_1,z_2)=2 \kappa^{-1}\,\tanh^{-1}\left( \left| \frac{z_1-z_2}{z_1-\bar{z_2} } \right| \right).
\end{equation*}
The area of a hyperbolic disk of radius $r$ in $H^2$ is equal to $4 \pi \kappa^{-2} \,\sinh^2\left( \frac{\kappa r}{2} \right)$ whereas the length of a hyperbolic circle of radius $r$ is given by $2 \pi \kappa^{-1}\,\sinh(\kappa r)$. As a consequence, the ratio of the perimeter of a disk to its area goes to a nonzero value, $\kappa$, when $\kappa r \rightarrow \infty$.

\section{Isometries and Fuchsian groups} \label{ap:isometries}

Physically, isometries correspond to displacements that leave the metric invariant, so they are an important ingredient of any study of a Riemannian space.

As we mainly use the Poincaré disk model in all this paper, we only detail the isometries for this representation of $H^2$. The case of other representations will only be briefly touched on; it is well described for instance in \cite{Balazs:1986}.

The isometries of the Poincaré disk can be represented as maps from $\mathbb{C}$ to $\mathbb{C}$ of the following form,
\begin{equation*}
z \longmapsto \frac{az+\bar{c}}{cz+\bar{a}},\qquad z \longmapsto \frac{a\bar{z}+\bar{c}}{c\bar{z}+\bar{a}},
\end{equation*}
where $z,a,c \in\mathbb{C}$, the overbar denotes complex conjugation, and $|a|^2-|c|^2=1$ (implying that if $z$ is inside the unit disk $\Delta$, so is its image). Note that these two maps correspond, respectively, to orientation-preserving and orientation-reversing linear fractional transformations. The group formed by the orientation-preserving maps is isomorphic to the quotient group $SU(1,1)/\{\pm \mathbbm{1}\}$, where $SU(1,1)$ is the pseudo-unitary group. These transformations can indeed be represented by complex matrices,
\begin{equation*}
\left(
\begin{array}{ccc}
a & \bar{c}\\
c & \bar{a}\\
\end{array}
\right)
.
\end{equation*}
of unit determinant.

Note that for the Poincaré upper half-plane, the group of orientation-preserving isometries is isomorphic to the projective linear group $PSL(2,\mathbb{R})=SL(2,\mathbb{R})/\{\pm \mathbbm{1}\}$, where the special linear group $SL(2,\mathbb{R})$ consists of the set of $2\times 2$ real matrices whose determinant is equal to $+1$.
For the hyperboloid model, the group of orientation-preserving isometries is isomorphic to the restricted Lorentz group $SO^+(2,1)$. All these groups are isomorphic to each other. The representations of $H^2$ are indeed all strictly equivalent, each one having its own advantages to study properties of the hyperbolic plane. Finally, it is useful to notice that the group of isometries is generated by reflections in ``hyperbolic lines", \textit{i.e.}, geodesics. This property proves very helpful in practice.

A Fuchsian group $\Gamma$ is a discrete subgroup of the group of complex linear fractional transformations (also called Möbius transformations) with an invariant disc $D$, \textit{i.e.}, such that $\gamma(D)=D$ for any element $\gamma$ of $\Gamma$. If the unit disk $\Delta$ is $\Gamma$-invariant, then $\Gamma$ can be viewed as a discrete subgroup of the group of isometries of the hyperbolic plane. $\Gamma$ induces a tessellation of the plane by hyperbolic polygons (see \ref{ap:tilings}). Note that Fuchsian groups preserve orientation. In the present article, we further restrict ourselves to ``purely hyperbolic" Fuchsian groups that, except for the identity, do not contain transformations leaving points of invariant, such as rotations. (Such transformations with fixed points may be of two types, called ``elliptic" and ``parabolic"; in this context ``hyperbolic" denotes a transformation that has no fixed points, \textit{i.e.}, a ``translation".)

\section{Hyperbolic trigonometry and tilings} \label{ap:tilings}

A regular tessellation of the hyperbolic plane by congruent, regular hyperbolic polygons, can be built by using the above mentioned property of the group of isometries (see \ref{ap:isometries}), namely that this latter is generated by reflections in geodesics, which are inversions across circles in the Poincaré disk model. For discrete subgroups, these geodesics correspond to the edges of the tessellation. Therefore, to construct a tiling of the hyperbolic plane, one just has to construct one tile and to duplicate it by using reflections in tile edges.

\begin{figure}
	\includegraphics[scale=0.25]{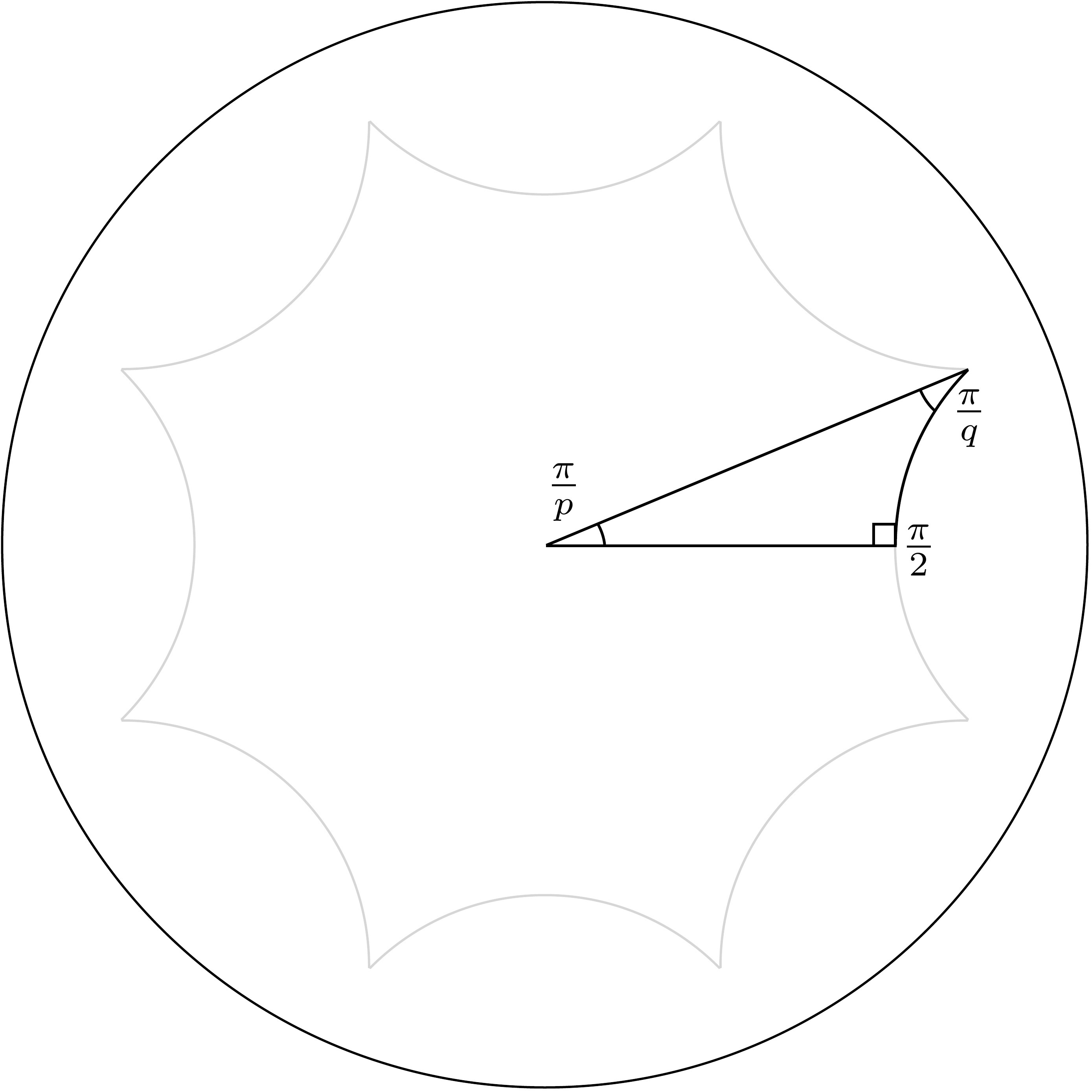}
	\caption{\label{fig:trigo}Hyperbolic ``orthoscheme" triangle used to build the $\{p,q\}$ tiling (Poincaré disk representation of $H^2$).}
\end{figure}

To build an elementary polygon associated with the tiling $\{p,q\}$, where $p$ denotes the number of sides of a tile and $q$ the coordinence of the tile vertices, one can decompose it (since it is regular) into copies of a single triangle called the ``orthoscheme" triangle \cite{Sadoc:1999} (see \fref{fig:trigo}). For any tiling $\{p,q\}$, one has to first construct the appropriate triangle and then to duplicate it by using the above described procedure, where just one vertex of the triangle is used to generate the vertices of the tessellation (see \fref{fig:trigo}). Such a triangle must have the following angles: $\frac{\pi }{2}$, $\frac{\pi }{p}$ and $\frac{\pi }{q}$, and the duplicable vertex generating the tiling is the vertex corresponding to the $\frac{\pi }{q}$ angle (see \fref{fig:trigo}). Constructing the triangle requires some elements of hyperbolic trigonometry, whose most important features are now summarized.

First, the area $A$ of any triangle with angles $\alpha$, $\beta$ and $\gamma$ is given by
\begin{equation*}
A=\kappa^{-2}\left(\pi-(\alpha + \beta + \gamma)\right),
\end{equation*}
where $-\kappa^2$, we recall, denotes the Gaussian curvature of the hyperbolic plane. In hyperbolic geometry, the sum of the angles of a triangle is always smaller than $\pi$ and depends on its area, contrary to the Euclidean case where the sum of the angles is always equal to $\pi$. (For the orthoscheme triangle introduced above, the sum of the angles is equal to $\pi(1-\frac{(p-2)(q-2)-4}{2pq})$, and is therefore less than $\pi$ due to the condition on hyperbolic tilings, equation \eref{eq:tiling}.)

Then, for a general hyperbolic triangle with sides $a$, $b$ and $c$ and opposite angles $\alpha$, $\beta$ and $\gamma$, the trigonometric rules are as follows:
\begin{eqnarray*}
\frac{\sinh (\kappa a) }{\mathrm{sin} (\alpha) }&=\frac{\sinh (\kappa b) }{\mathrm{sin} (\beta) }=\frac{\sinh (\kappa c) }{\mathrm{sin} (\gamma) },\\
\cosh(\kappa c)&=\cosh(\kappa a)\,\cosh(\kappa b)-\sinh(\kappa a)\,\sinh(\kappa b)\,\mathrm{cos}(\gamma),\\
\cosh(\kappa c)&=\frac{\mathrm{cos}(\alpha)\,\mathrm{cos}(\beta)+\mathrm{cos}(\gamma)}{\mathrm{sin}(\alpha)\,\mathrm{sin}(\beta)},
\end{eqnarray*}

Note that the first two relations have Euclidean analogues, which can be recovered by taking $\kappa \rightarrow 0$, whereas the last one has no Euclidean counterpart. This last relation implies that if two triangles have the same angles, then there is an isometry mapping one triangle onto the other. As a result, triangles are univocally determined by their angles. Finally, it is worth pointing out that the more familiar spherical trigonometry is recovered by making the replacement $\kappa \rightarrow i\kappa$.

\section*{References}

\bibliographystyle{unsrt}
\bibliography{CLP}

\end{document}